\documentclass[footinbib,a4paper,aps,superscriptaddress,reprint,twocolumn,preprintnumbers,amsmath,amssymb,nobalancelastpage,10pt]{revtex4-1}
\usepackage{bigints}
\usepackage{amsmath}
\usepackage{verbatim}
\usepackage{graphicx}
\usepackage{color}
\usepackage{tikz}
\usepackage{siunitx}
\usepackage{amssymb}
 \usepackage{relsize}

\usepackage{float}
\usepackage{slashed}
\usepackage{mathptmx}
\usepackage{amssymb}
\usepackage{amsmath}
\usepackage{amsfonts}
\usepackage{array}
\usepackage{multirow}
\usepackage{siunitx}

\usepackage{bm}
\usepackage{xcolor}
\graphicspath{{figures/}}

\usepackage{stmaryrd}

\usepackage{braket}

\definecolor{mygrey}{gray}{0.35}
\definecolor{myblue}{rgb}{0.1,0.0.1,1}
\definecolor{amber}{rgb}{1.0, 0.65, 0.0}
\definecolor{forestgreen}{rgb}{0.00, 0.45, 0.0}
\definecolor{mywhite}{rgb}{1,1,1}
\definecolor{myred}{rgb}{0.8,0.,0.3}

\usepackage[colorlinks=true,citecolor=myred,linkcolor=myred]{hyperref}

\usepackage{wasysym}

\usepackage[caption=false]{subfig}

 \def\ee{\mathord{\rm e}}
 
 \def\ii{\mathord{\rm i}}

\def\half{\textstyle\frac{1}{2}}

\def\fourth{\textstyle\frac{1}{4}}

\renewcommand{\ii}{{\rm i}}
\renewcommand{\ee}{{\rm e}}
\def\beq{\begin{equation}}
\def\eeq{\end{equation}}
\def\barray{\begin{eqnarray}}
\def\earray{\end{eqnarray}}

\begin{document}

\title{Towards quantum computing  Feynman diagrams in hybrid qubit-oscillator devices}

\author{S. Varona}
\affiliation{Instituto de F\'isica Teorica, UAM-CSIC, Universidad Autónoma de Madrid, Cantoblanco, 28049 Madrid, Spain. }

\author{S. Saner} 
\affiliation{Department of Physics, University of Oxford, Clarendon Laboratory, Parks Road, Oxford OX1 3PU, United Kingdom}

\author{O. Băzăvan}
\affiliation{Department of Physics, University of Oxford, Clarendon Laboratory, Parks Road, Oxford OX1 3PU, United Kingdom}

\author{G. Araneda}
\affiliation{Department of Physics, University of Oxford, Clarendon Laboratory, Parks Road, Oxford OX1 3PU, United Kingdom}

\author{G. Aarts}
\affiliation{Department of Physics, Swansea University, Singleton Campus, SA2 8PP Swansea, United Kingdom}
\author{A. Bermudez}
\affiliation{Instituto de F\'isica Teorica, UAM-CSIC, Universidad Autónoma de Madrid, Cantoblanco, 28049 Madrid, Spain. }

\begin{abstract}

We show that recent experiments in hybrid qubit-oscillator devices that measure the phase-space characteristic function of the oscillator via the qubit can be seen through the lens of functional calculus and path integrals, drawing a clear analogy with the generating functional of a quantum field theory. This connection suggests 
an expansion of the characteristic function in terms of Feynman diagrams, exposing the role of the real-time bosonic propagator, and identifying the external source functions with certain  time-dependent couplings that can be controlled experimentally. 
 By applying maximum-likelihood techniques, 
 we show that the ``measurement'' 
of these Feynman diagrams can be reformulated as a problem of multi-parameter point estimation that takes as input a set of Ramsey-type measurements of the qubit. By numerical simulations that consider leading imperfections in trapped-ion devices, we identify the optimal regimes in which Feynman diagrams could be reconstructed from measured data with low systematic and statistical errors. We discuss how these ideas can be generalized to finite temperatures
via the Schwinger-Keldysh formalism,
contributing to a bottom-up approach to probe quantum simulators of  lattice field theories by systematically increasing the qubit-oscillator number.

\end{abstract}

\maketitle

\setcounter{tocdepth}{2}
\begingroup
\hypersetup{linkcolor=black}
\tableofcontents
\endgroup

\section{\bf Introduction}

The generating functional is a central object of physical interest in any quantum field theory (QFT), for it encodes a vast amount of information about its real-time dynamics, including any possible $n$-particle scattering~\cite{Peskin:1995ev,Fradkin:2021zbi}.  Recent work~\cite{Jordan2018bqpcompletenessof} has shown that the accurate estimation of the full generating functional, even when restricting to a specific set of source functions and to a simple self-interacting scalar field theory~\cite{WILSON197475,PhysRevD.9.1686}, belongs to the complexity class of bounded-error quantum polynomial time (BQP) problems~\cite{nielsen00,doi:10.1137/S0097539796300921}. Paralleling the situation with Shor's algorithm for integer factorization~\cite{doi:10.1137/S0097539795293172}, which also falls into this complexity class,
it is not expected that a classical polynomial-time (P) algorithm for this problem will be found, as this would imply that BQP equals P.
Therefore, quantum computing the generating functional of an interacting QFT would constitute a rigorous demonstration of quantum advantage for a problem of practical relevance in physics. 
In contrast to integer factorization, however, one cannot straightforwardly verify that the result of this quantum computation is correct for arbitrary  couplings of the QFT. 

On the other hand, when the interactions of the QFT are small, the generating functional admits an expansion in terms of Feynman diagrams. In this regime,  the results of the quantum computation could be benchmarked against approximate perturbative predictions at increasing loop orders. In this way, confidence could be built
on the outcomes of the quantum computation of the generating functional in regimes that go beyond perturbation theory. 
 Feynman diagrams are not only a useful perturbative tool
but, together with the concept of renormalization, have revolutionized
the way in which we think about quantum many-body systems across various disciplines~\cite{Kaiser:2005tp}. It would thus be very interesting if
they could be directly estimated from experimentally measured data, serving as a useful benchmark for experimental quantum computers exploring QFTs.
We remark that such a Feynman diagram estimation is not the only validation route. Complementary non-perturbative benchmarks may include Monte Carlo lattice field theory calculations for equilibrium observables, variational tensor-network approaches, or exact results available in special integrable limits like large-$N$ or reduced dimensions. Each of these approaches has its own limitations. Respectively, these include problems where perturbation theory breaks down, where   fermions have finite densities,  where the regime of interest involves long real-time evolutions, or where the vacua withstand strong quantum correlations. The perturbative framework presented in this work should therefore be viewed as a complement to a wider benchmarking toolbox.
 
 Aside from its practical interest, using quantum computers to calculate Feynman diagrams has a foundational appeal, as it would bring together two of Feynman's revolutionary ideas to deepen our understanding of quantum many-body physics: drawing diagrams on a piece of paper~\cite{PhysRev.76.749} and drawing circuits on a quantum computer~\cite{Feynman_1982}. In this article, we take a first step in this direction, contributing to the prospects of using noisy intermediate scale quantum (NISQ) devices~\cite{Preskill2018quantumcomputingin} for problems in  relativistic QFTs, in particular those related to real-time evolution that  have  generated a significant interest recently (see, e.g.,~\cite{https://doi.org/10.1002/andp.201300104,Zohar,doi:10.1080/00107514.2016.1151199,Banuls2020,Carmen_Banuls_2020,doi:10.1098/rsta.2021.0064,Klco_2022, https://doi.org/10.48550/arxiv.2204.03381,Bauer:2023qgm,halimeh2023coldatom}). To connect this step with experiments that are already being realized in various laboratories, we shall remove the additional complexity associated with ultraviolet divergences and renormalization, and focus on  a (0+1)-dimensional quantum-mechanical version of the problem, identifying common features  shared with the higher-dimensional QFTs, as discussed at the end of this manuscript.

In a series of works~\cite{PhysRevX.7.041012,PRXQuantum.3.020352,martinez2023thermal}, we have argued that the generating functional of a QFT is intimately related to certain interferometric measurements in hybrid quantum simulators: special purpose quantum computers that are not only composed of qubits arranged in a static register, but also present additional continuous variables that can be exploited for the simulation of a QFT. Note that hybrid approaches sometimes refer to algorithms that combine both classical and quantum processing units, although we use hybrid in the context of quantum processors with discrete and continuous variables~\cite{liu2024hybrid}. In these schemes, one encodes the information about the QFT and its generating functional  in the bosonic degrees of freedom, and probe it with a collection of qubits coupled to them.
The qubits, after evolving in real time under the effect of the field, get projectively measured to extract the full generating functional describing the QFT, or other key quantities such as the renormalized couplings of the theory.
The results presented in this manuscript follow from the observation that an analogous interferometric setup has been recently implemented in experiments with a single trapped ion~\cite{Fl_hmann_2020,matsos2023robust,bazavan2024squeezing}, which are used to measure the characteristic function~\cite{https://doi.org/10.1002/qute.202100016,barnett_1998} of a (bosonic) harmonic oscillator in different motional states, including higher-order non-Gaussian versions of the squeezed states, so-called generalized squeezed states~\cite{braunsteain1987generalized}. 
In these experiments, the real-time dynamics of the problem is somewhat hidden but, once unveiled, it allows us to  connect to functional techniques and a $D=0+1$ dimensional version of a QFT. 

\begin{figure*}
\includegraphics[width=0.95\textwidth]{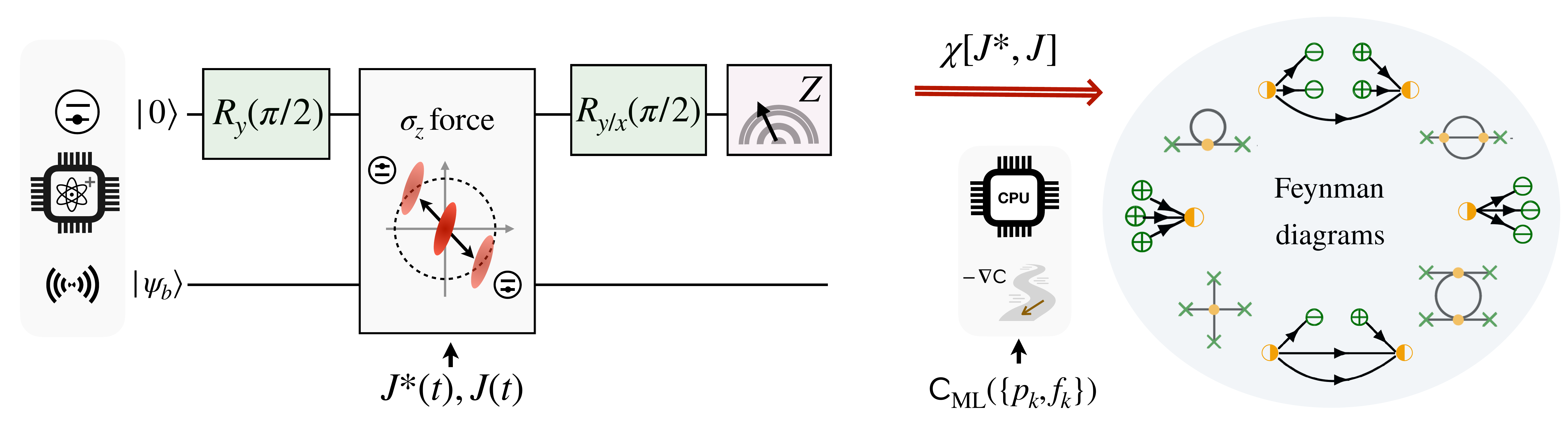}
  \caption{{\bf Quantum computing Feynman diagrams:} In the left panel, we present a circuit description of a hybrid quantum processor composed of both discrete qubit levels and continuum bosonic modes. The circuit describes the unitary operations that are used to implement a Ramsey-interferometry protocol that measures the characteristic functional $\chi[J^*,J]$ of a certain bosonic state $\ket{\psi_b}$, evaluated for a particular set of source/sink and vertex functions $J^*\!(t),J(t),\lambda^*\!(t),\lambda(t)$ that are inputs in the circuit. The circuit maps the characteristic functional information to the coherences of a qubit, which is initialized in $\ket{\psi_q}=R_y(\pi/2)\ket{0}=(\ket{0}+\ket{1})/\sqrt{2}$. The qubit-oscillator system is then subjected to a joint entangling unitary generated by a state-dependent linear potential $V_{I}(t)=-J(t)a^\dagger\!(t)-J^*(t)a(t)$. After a given time evolution $t\in[t_0,t_{\rm f}]$, the qubit is projectively measured in the $X(Y)$ basis by applying $R_y(\pi/2)(R_x(\pi/2))$ prior to a projective measurement in the $Z$ basis. In this way, one collects information about  ${\rm Re}\{\chi[J^*,J]\} ({\rm Im}\{\chi[J^*,J]\})$, which is imprinted in the measurement relative frequencies $f_k(\chi)$. In the right panel, we depict  a second step, in which one uses classical hardware to estimate  Feynman diagrams. This is performed by minimizing a maximum likelihood cost function $\mathsf{C}_{\mathrm{ML}}(\{p_k, f_k\})$ that uses the different orders of each contribution with source/sink and vertex parameters. We use  $\nabla\mathsf{C}$ to depict gradient-descent methods for non-linear minimization  of this cost function.} 
\label{fig:scheme}
\end{figure*}

In particular, we discuss how to formulate the single-oscillator characteristic function as a functional path integral, and how to compute it applying Feynman-diagram techniques similar to those commonly used in QFTs (see Fig.~\ref{fig:scheme}). This will allow us to understand the connections of the characteristic function with variants of the vacuum persistence amplitude in QFTs, which describe the vacuum-to-vacuum transition amplitude subject to the so-called Schwinger sources~\cite{schwinger_sources}, and lead to the notion of the generating functional~\cite{Peskin:1995ev,Fradkin:2021zbi}. In the present context, we will need to exchange the vacuum for a different, possibly non-Gaussian, state and study the persistence amplitude in that state after its evolution under the microscopic Hamiltonian.
We will show that such modified persistence amplitudes can be written in terms of a functional, provided that one allows for the interaction vertices of the QFT to have a specific time dependence. We borrow techniques from higher-dimensional QFTs to derive a perturbative series of the characteristic functional. This series can be graphically represented using Feynman diagrams, which involve time integrals of various products of the bosonic propagator, source, and vertex functions. We then argue that a restricted type of quantum process tomography for the effective dynamics of a probe qubit coupled to the bosonic oscillator could be used to estimate the various Feynman diagrams in future experiments. Performing  realistic numerical simulations of a trapped-ion device, we present  a detailed study of the maximum-likelihood inference of the different Feynman diagrams. In particular, we simulate experimental errors such as imperfect state preparation and decoherence of the oscillator state as a result of motional heating, and  analyze the interplay of stochastic and systematic error sources, and identifying the parameter regimes in which the estimation can be more accurate and precise. We also incorporate thermal effects in the bosonic mode, and show that the estimations agree with finite-temperature real-time diagrammatic predictions of the Schwinger-Keldysh formalism.

This article is organized as follows. In Sec.~\ref{sec:qubit_probe}, we connect the characteristic function to a functional, highlighting the role of the boson propagator and the Schwinger sources, and how this can be measured in real time using a qubit probe. Sec.~\ref{sec:feynman} contains the core of our results, including a path integral formulation for characteristic functionals, the diagrammatic expansion for both Gaussian and non-Gaussian squeezed states, the maximum-likelihood estimation of the Feynman diagrams and the incorporation of thermal effects via the Schwinger-Keldysh formalism. We present an outlook and  future directions in Sec.~\ref{sec:glimpse}.

\section{\bf Qubit probes for the characteristic function}
\label{sec:qubit_probe}
\subsection{From the characteristic function to a functional}
We consider a single quantum-mechanical oscillator of frequency $\omega_{b}$, the excitations of which are created and annihilated by the bosonic operators $a^{\dagger}$ and  $a$, respectively, such that the state with no excitations fulfils $a\ket{0}=0$. The {\it characteristic or Weyl's function} for an arbitrary pure state $\ket{\psi_{ b}}$ evaluated at phase-space coordinates ($\xi$, $\xi^\ast$) can be written as 
\begin{equation}
\label{eq:charact_func}
\chi(\xi^{\!*}\!,\xi)=\bra{\psi_b}D(\xi)\!\ket{\psi_b},\hspace{2ex} D(\xi)=\ee^{\xi a^\dagger-\xi^{\!*}\! a}
\end{equation}
where $D(\xi)$ is Glauber's displacement operator~\cite{PhysRev.131.2766,PhysRev.177.1857}
with complex displacement parameter $\xi\in\mathbb{C}$ ~\cite{Weyl1927,https://doi.org/10.1002/qute.202100016}, which is related to the position-momentum coordinate pair or to the  conjugate quadratures of the bosonic mode. 
The characteristic function plays an important role in the statistical phase-space formulation of quantum mechanics~\cite{Groenewold:1946kp,Moyal_1949}, and acts as the generating function of any symmetric correlation $\langle a^{n}(a^{\dagger})^m\rangle_S=\partial^n_{\xi^{\!*}}\partial^m_\xi\chi(\xi^{\!*}\!,\xi)|_{\xi=\xi^{\!*}=0}$, which is expressed as the sum of expectation values over all possible orderings of the bosonic operators~\cite{barnett_1998}. Moreover, a straightforward generalization $\chi_{s}(\xi^{\!*}\!,\xi)=\chi(\xi^{\!*}\!,\xi){\rm exp}\{\frac{s}{2}\xi^*\xi\}$ can lead to generators of normally ($s=+1$) or anti-normally ($s=-1$) ordered correlators. This function can be used to reconstruct the full density matrix of the mode, providing in this way a practical alternative to other quantum tomography methods~\cite{PhysRevLett.75.2932,PhysRevA.53.R1966,PhysRevLett.78.2547} for continuous-variable systems~\cite{RevModPhys.77.513,RevModPhys.81.299}. We note that Weyl's characteristic function~\cite{PhysRev.40.749,PhysRev.131.2766} is the (inverse) Fourier transform of the Wigner function~\cite{PhysRev.177.1857,HILLERY1984121}, and can thus provide the same information about the non-classicality of the quantum state. Likewise, the $s=\pm1$ characteristic function connects to other quasi-probability distributions~\cite{1940264,PhysRev.131.2766,PhysRevLett.10.277}.

So far, the discussion has been completely static. The time-evolution of the state paralleling the Schr\"{o}dinger equation in phase-space can be entirely formulated in phase space as a set of equations of motion for the Wigner function~\cite{https://doi.org/10.1002/qute.202100016}. In this work, we are interested in a different type of time-dependence that arises from generalizing the phase-space coordinates of the characteristic function to generic functions of time $\xi,\xi^*\!\mapsto J(t),J^*\!(t)$, which connect to Schwinger's source approach to QFT in which they are classical spacetime fields $J(t,\boldsymbol{x}),J^*\!(t,\boldsymbol{x})$~\cite{schwinger_sources}. These external classical fields can create/annihilate excitation of a quantum field, and will be referred to   as Schwinger sources/sinks in this manuscript, and are used to estimate the time-dependent correlation functions in QFTs~\cite{Peskin:1995ev,Fradkin:2021zbi}. Following this analogy, we promote the characteristic function~\eqref{eq:charact_func}  to a  functional, called the {\it characteristic functional} in this work
\begin{equation}
\label{eq:charac_time}  \chi[J^*,J]=\bra{\psi_{b}}U_J(t_{\rm f},t_0)\ket{\psi_b}, \hspace{1ex} U_J(t_{\rm f},t_0)= \mathcal{T}\!\Bigl\{\ee^{-\ii\bigintssss_{\,\,t_0}^{t_{\rm f}}\!{\rm d}t V_{J}(t)}\Bigr\},
\end{equation}
where $V_J(t)=-J(t)a^{\dagger}\!(t)-J^*\!(t)a(t)$ is the source potential, and $\mathcal{T}\{\cdot\}$ is the time-ordering operator. Here, we have set $\hbar=1$ and, as customary, used square brackets to indicate that the arguments $J^*\!,J$ are functions of time. Generalizing to  many modes per spatial point  and using spatial-dependent sources $J(t,\boldsymbol{x}),J^*\!(t,\boldsymbol{x})$ connects the characteristic functional directly to the standard definition of the generating functional of a scalar QFT~\cite{Peskin:1995ev,Fradkin:2021zbi}.

Let us now
add boson-boson interactions, which will allow us to make connections to path integrals and Feynman diagrams in QFTs. In this context, one is typically interested in a zero-temperature regime, such that $\ket{\psi_b}\mapsto\ket{\Omega_b}$ is the groundstate of the interacting system typically referred to as the vacuum,
and one focuses on the Feynman propagator 
$G (t_i-t_j)=\bra{\Omega_b} \mathcal{T}\{a(t_i)a^\dagger\!(t_j)\}\ket{\Omega_b}$. In a QFT context, the times $t_0$ and $t_{\rm f}$ are arbitrary as far as they correspond to the remote past and distant future, and thus far  from the time lapse during which  particles represented by excitations of the field collide with each other, feeling in this way  the interactions mediated by the field non-linearities  and thus scattering. In our work, these times have instead an explicit meaning, and are connected to the time during which the quantum system is driven, as will be clarified below.
The evolution of the operators
is no longer trivial, and depends on the interaction potential $V_{\lambda}$, which is usually expressed as a polynomial of the creation and annihilation operators
\beq
\label{eq:int_ham}
H_\lambda=\omega_b\,a^\dagger a+V_{\lambda}(a^\dagger\!,a),\hspace{2ex} a(t)=\ee^{\ii H_\lambda t}a\,\ee^{-\ii H_\lambda t}.
\eeq  
As a consequence, the propagator $G(t_i-t_j)$ will deviate from the non-interacting Green's function 
\beq
\label{eq:real_time_vcuum_prop}
G_0(t_i-t_j)=%
\theta(t_i-t_j)\ee^{-\ii\omega_b(t_i-t_j)}
\eeq
which, after a  Fourier transform, has a simple pole at $\omega=\omega_b$ 
\beq
\label{eq:free_prop}
G_0(t)=\int_{-\infty}^{\infty}\!\frac{{\rm d\omega}}{2\pi}G_0(\omega)\ee^{-\ii\omega t},\hspace{1ex}G_0(\omega)=\frac{\ii }{(\omega-\omega_b)+\ii\epsilon},
\eeq
where we have introduced $\epsilon\to 0^+$.

As discussed in Appendix~\ref{app:vpa}, the full propagator $G(t_i-t_j)$ can be depicted as the sum of various Feynman diagrams that can be obtained by taking functional derivatives on a lower-dimensional version of the {\it normalised generating functional} in QFTs~\cite{Peskin:1995ev,ryder_1996}. In the present context, the generating functional $Z[[J^*,J]/{Z}[0,0]]$ is in one-to-one correspondence to the aforementioned characteristic functional~\eqref{eq:charac_time}, i.e.,   $\chi[J^*,J]\leftrightarrow{Z}[J^*,J]/{Z}[0,0]$, provided one considers that the initial state is the vacuum $\ket{\psi_b}=\ket{\Omega_b}$ and the operators evolve under Eq.~\eqref{eq:int_ham} yielding 
\beq
\label{eq:gen_func}
  \frac{Z[J^*,J]}{Z[0,0]}=\big\langle\Omega_b\big |\mathcal{T}\!\Bigl\{{\rm exp}\Bigl(\ii\int_t \big( J(t)a^{\dagger}\!(t)+J^*\!(t)a(t)\big)\Bigr)\Bigr\}\big|\Omega_b\big\rangle.
\eeq
This  has a neat interpretation in terms of the quantum persistence amplitude, the square modulus of which is also known as the Loschmidt echo~\cite{GORIN200633},  and represents the probability amplitude with which a quantum system prepared in a given state remains in that same state after some time evolution $\mathcal{A}_{\rm pers}(t-t_0)=\bra{\Psi_0}U_J(t,t_0)\ket{\Psi_0}$. Setting the initial state as $\ket{\Psi_0}=\ket{\Omega_b}$ shows that the above functional~\eqref{eq:gen_func} is  the vacuum persistence amplitude, quantifying the  probability  amplitude with which  the system would remain in the vacuum state after a certain time evolution.
The sourced evolution can create excitations
via the Schwinger sources, which subsequently propagate in time including the non-trivial scattering due to  $V_{\lambda}$ and, finally, get absorbed  by other Schwinger sources such that the system returns with some probability to the initial vacuum state. 

In order to make the connection to a full QFT in $D=d+1$ spacetime dimensions, one would need to upgrade the operators and source functions to quantum and  source fields, e.g. $a(t),a^\dagger\!(t),J(t),J^*(t)\mapsto \phi(t,\boldsymbol{x}),\phi^\dagger\!(t,\boldsymbol{x}),J(t,\boldsymbol{x}),J^*(t,\boldsymbol{x})$, substituting the above integrals over time by spacetime integrals. Likewise, the Green's function would not only represent how the field excitations evolve in time, but how they propagate between two distinct spatial points $G(t_i-t_j)\mapsto G(t_i-t_j,\boldsymbol{x}_i-\boldsymbol{x}_j)$ which, interestingly, can include the time propagation of excitations and holes, leading to the notion of particles and antiparticles as excitations of the field. As noted in the introduction, this functional contains all the relevant information about the real-time dynamics of an interacting QFT~\cite{Peskin:1995ev,Fradkin:2021zbi}.

From this perspective, if one can upgrade the phase-space coordinates to source functions, the original characteristic function in Eq.~\eqref{eq:charact_func} can be interpreted as a $D=0+1$ dimensional version of the generating functional. As discussed in the following subsection, this is indeed the case for the recent trapped-ion experiments~\cite{Fl_hmann_2020,matsos2023robust,bazavan2024squeezing} in which $\xi^*$ and $\xi$ grow linearly in time. The final ingredient, as we discuss in more detail below, is that the state $\ket{\psi_{ b}}$ in Eq.~\eqref{eq:charact_func} can be considered as the result of the non-linear potential in Eq.~\eqref{eq:int_ham}, provided one allows for the vertex to have a specific time dependence $\lambda(t)$. Depending on the nature of the vertex function, the non-equilibrium initial state will  contain excitations that can propagate in time and interact with the Schwinger sources. As detailed in the following subsection, motivated by the recent experimental results~\cite{bazavan2024squeezing}, we will focus on the {\it generalized squeezed persistence amplitude} (GSPA) in which $\ket{\psi_{ b}}$ arises from applying generalized squeezing~\cite{braunsteain1987generalized} of order $n$ to the initial vacuum state $\ket{0}$. For $n>2$ these squeezed states are non-Gaussian \cite{PRXQuantum.2.030204} and the excitations always appear in bundles of $n$ quanta. We show below that, benefiting from the connection to the generating functional of QFTs, it is possible to find a path-integral representation for this characteristic functional, and how a specific expansion in Feynman diagrams arises from this formulation. Before turning to this, however, let us provide the details of how time-dependence appears in the experimental setup~\cite{Fl_hmann_2020,matsos2023robust,bazavan2024squeezing}, and how the structure of the boson propagator and the Schwinger sources is actually responsible for the specific time dependence found in the experiments.%

\subsection{ Hybrid qubit-oscillator interferometry}\label{eq:squeezing_rwa}
Let us now discuss a qubit-oscillator interferometric scheme for the measurement of the characteristic function $\chi$, which falls under the class of quantum tomography methods for continuous variable systems (see~\cite{RevModPhys.81.299} and references therein). There are several methods that directly reconstruct the Wigner function via homodyne detection, photon counting, or measurement of the parity operator.
In this work, however, we focus on the direct tomography of the characteristic function of a continuous-variable state by coupling the boson to a discrete-variable probe, in particular, a qubit and performing essentially a Ramsey interferometric scheme (see Fig.~\ref{fig:scheme}) where each of the qubit states acquires a residual phase that depends on the bosonic state. It turns out that the Ramsey interferometric scheme can be simplified while simultaneously allowing for arbitrary evolution times. Moreover, it also allows for a more flexible regime that can interpolate between resonant and off-resonant couplings. The common underlying idea of the schemes discussed in Refs.~\cite{PhysRevA.54.R25,PhysRevA.59.R950,PhysRevA.83.062120,PhysRevA.85.032334,Gerritsma:2010bpn,PhysRevLett.104.100503,PhysRevLett.115.213001,PhysRevX.7.041012,Fl_hmann_2020,matsos2023robust,bazavan2024squeezing} is that, during the interferometric evolution, the qubit-oscillator system is subjected to a linear force that induces a state-dependent displacement in phase space. This effectively rotates the qubit populations in a certain measurement basis such that the information about the characteristic function gets imprinted onto the probe qubit.

From now on, we focus on trapped-ion systems~\cite{Wineland1998experimental,3981}, and present the details of this interferometric scheme.
For the oscillator~\eqref{eq:charac_time}, it suffices to consider a single trapped ion, although we anticipate that the complexity of the simulated system can be increased by considering larger trapped ion crystals. Increasing the ion number sequentially allows for a bottom-up approach towards measuring the full generating functional of an interacting QFT, offering an alternative to the direction explored in~\cite{PhysRevX.7.041012,PRXQuantum.3.020352,martinez2023thermal} as outlined in Sec.~\ref{sec:glimpse}. Focusing on the single-ion case for the moment, the bosonic oscillator will correspond to one of the secular vibrational modes of an atomic ion of mass $m$ confined in a radio-frequency trap. We consider the axial mode $\omega_{b}=\omega_z$ in a linear Paul trap, the excitations of which will be referred to as phonons. Using state-dependent forces, one can map the information of the characteristic functional~\eqref{eq:charac_time} onto the populations of the qubit, which can then be measured through the fluorescence of the ion (see Fig.~\ref{fig:scheme}). Spin-dependent forces can be generated in various ways by illuminating the ion with different configurations of electromagnetic fields~\cite{molmer1999multiparticle, sorensen2000entanglement, milburn2000ion, cai2023entangling} and are described as follows
\beq
H\approx\frac{\omega_0}{2}\sigma_z+\omega_b\,a^\dagger a+2\Omega\eta\sin\big((\omega_b+\Delta) t-\Delta\varphi\big)\sigma_z(a+a^\dagger),
\eeq
without loss of generality, we choose the spin conditioning to be $\sigma_z=\ket{0}\!\bra{0}-\ket{1}\!\bra{1}$, where the qubit states $\ket{0},\ket{1}$ are separated by a transition frequency $\omega_0 \gg \omega_b$. Here, $\Omega\ll\omega_0$ is the Rabi frequency, $\eta=\Delta {k}/\sqrt{2m\omega_b}$ is the Lamb-Dicke parameter and $\Delta$ and $\Delta\varphi$ are the detuning and the phase, respectively, of the force relative to the motional mode frequency $\omega_b$. 
It is customary in the trapped-ion literature to consider a further rotating-wave-approximation for $|\Omega\eta|\ll\omega_b+\Delta$, such that 
\beq
\label{eq:sd_force_sources}
H\approx\frac{\omega_0}{2}\sigma_z+\omega_b\,a^\dagger a-\big(J(t)a^\dagger+J^{*}\!(t)a\big)\sigma_z,
\eeq
where we have introduced the functions that will play the role of our previous Schwinger sources and sinks
\beq
\label{eq:rwa_sources}
J(t)=J_0\ee^{-\ii\Delta t},\hspace{1ex}J^{*}\!(t)=J_0^*\ee^{\ii\Delta t},\hspace{2ex} J_0=-\ii\Omega\eta\ee^{\ii\Delta\varphi}.
\eeq

We consider that the initial state of the system is $\ket{\psi_0}=\ket{\psi_b}\otimes\ket{0}$, where we note that the vibrational state $\ket{\psi_b}$ can be prepared using various possible techniques~\cite{3981}. In this work, we consider a method recently demonstrated in experiments~\cite{bazavan2024squeezing} to prepare higher-order squeezed states~\cite{PhysRevA.104.032609} (see Sec.~\ref{sec:bosonic_states_sim}). Once the bosonic state $\ket{\psi_b}$ is prepared, we apply a $\pi/2$ rotation along the $\sigma_y$ axis on the qubit, expressed as \( R_y(\pi/2)\ket{0} = {\rm exp}\{-\ii\frac{\pi}{4}\sigma_y\}\ket{0} \). This unitary operation is equivalent to a Hadamard gate in quantum computing~\cite{nielsen00}. Finally, we apply the $\sigma_z$-dependent force~\eqref{eq:sd_force_sources}, introducing terms that function as Schwinger sources and sinks. The time evolution under this spin-dependent force, after tracing over the phonon degrees of freedom, leads to the qubit density matrix 
\beq \label{eq:density_matrix_qubit}
\rho_{\rm I}(t_{\rm f},t_0)=\half\bigg(\mathbb{I}_2+\chi[J^*,J]\ket{0}\!\!\bra{1}+\chi^*[J^*,J]\ket{1}\!\!\bra{0}\bigg).
\eeq
Hence, the information about the characteristic functional~\eqref{eq:charac_time}, evaluated at the specific source/sink harmonic functions in Eq.~\eqref{eq:rwa_sources}, has been mapped into the qubit coherences. Let us note that this expression assumes the interaction picture with respect to $H_0=\omega_ba^\dagger\!a+\frac{\omega_0}{2}\sigma_z$. The remaining step in the Ramsey scheme is to measure the qubit.
At $t=t_{\rm f}$, the state-dependent force in Eq.~\eqref{eq:sd_force_sources} is switched off, and one applies a  resonant $\pi/2$-pulse with the same laser beams used for the initialization, but now driving a qubit rotation along the $y$ or $x$ axis in the Bloch sphere, after which the state of the qubit is measured by state-selective fluorescence in a cycling transition~\cite{Wineland1998experimental,3981}. Using phase-coherent laser pulses allows to measure  $\sigma_x$ or $\sigma_y$, and to infer the real and imaginary parts of the characteristic function 
\beq
\label{eq:sx_sy}
\langle \sigma_x(t_{\rm f})\rangle={\rm Re}\,\chi[J^*,J],\hspace{1ex} \langle \sigma_y(t_{\rm f})\rangle={\rm Im}\,\chi[J^*,J].
\eeq
We note that this Ramsey interferometry scheme can be reinterpreted as a so-called Hadamard test to measure a unitary operator in quantum computation~\cite{10.1145/1132516.1132579,childs_qc}. We also note that the scheme just discussed is also valid for any source and sink functions different from Eq.~\eqref{eq:rwa_sources}, which could be designed, for instance,  by amplitude or phase modulation of the driving laser fields.
One can also consider the instantaneous switching on/off of these sink/source functions, which would allow us to reconstruct functional derivatives and calculate any $n$-point full propagators, as considered in the QFT proposal~\cite{PhysRevX.7.041012}. 

As previously noted, the time dependence and the role of the source/sink functions is somewhat hidden in the recent experimental accounts~\cite{Fl_hmann_2020,matsos2023robust,bazavan2024squeezing}. As it will become more clear when we derive closed analytical expressions based on the Feynman-diagram expansion, the dynamics manifest through a linear time dependence of the phase-space coordinates, which is a result of using resonant forces $\Delta\approx0$, such that
\beq
\label{eq_resonant_source}
\xi(t)=-\ii\int_{0}^{t}\!{\rm d}t'J(t')G_0(t')\approx\xi_0+\Omega\eta\ee^{\ii\Delta\varphi}\,t.
\eeq
Once this connection is clear, we can move to the following section, where we will present the functional path-integral formulation of the problem.

\section{\bf Path integrals and Feynman diagrams}
\label{sec:feynman}
 
\subsection{\bf Squeezed persistence amplitude}

\begin{figure*}
\subfloat[\label{fig:chi_squeezing}]{%
 \includegraphics[width=0.33\linewidth]{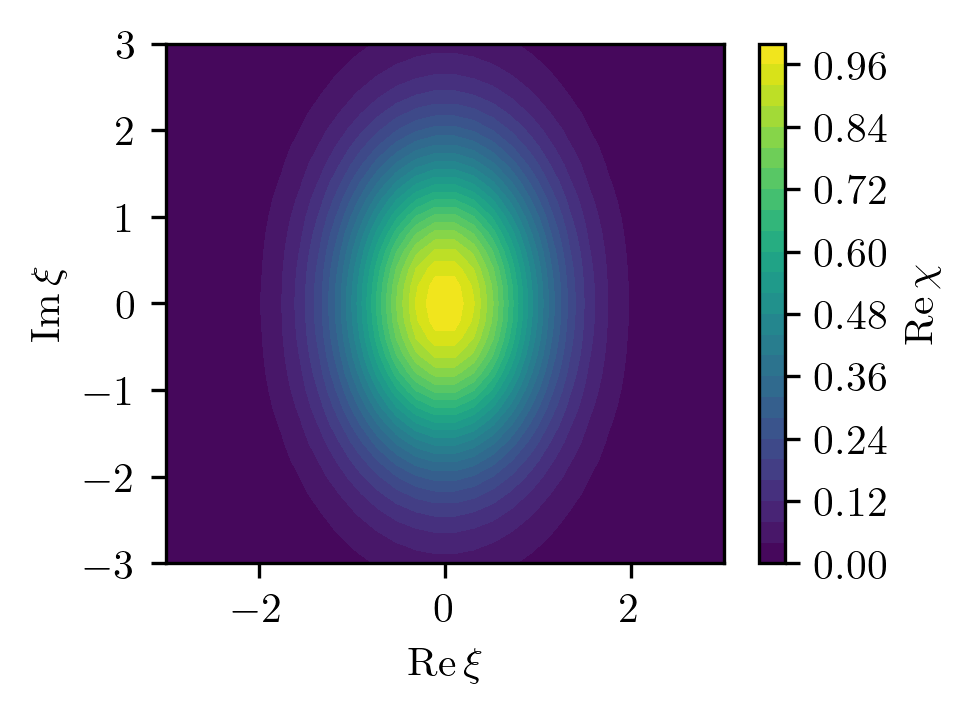}%
}\hfill
\subfloat[\label{fig:chi_real_trisqueezing}]{%
 \includegraphics[width=0.33\linewidth]{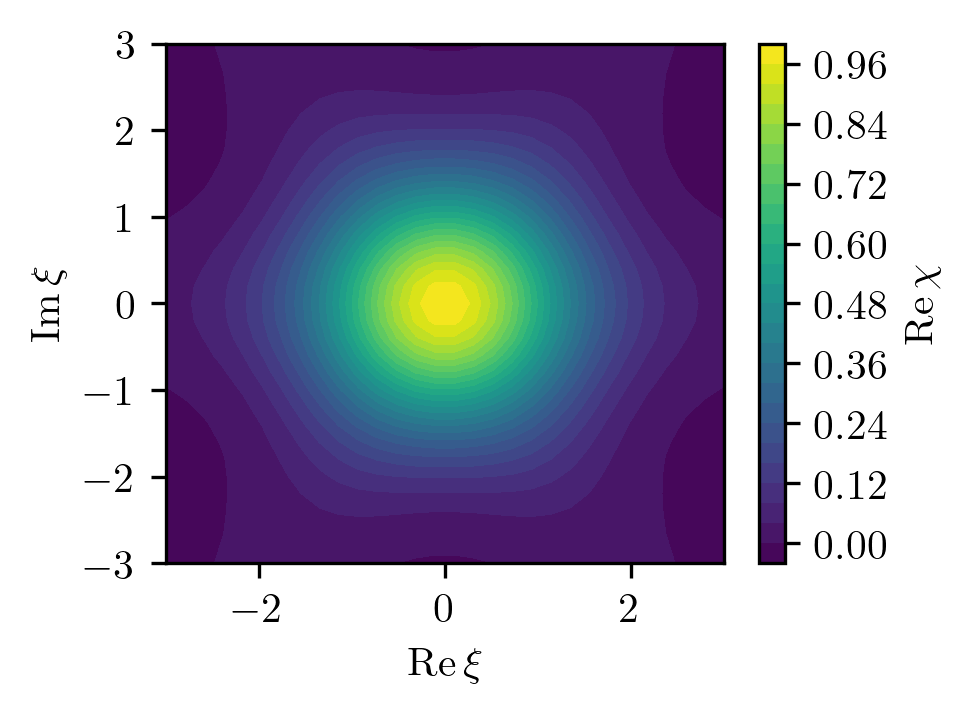}%
}\hfill
\subfloat[\label{fig:chi_imag_trisqueezing}]{%
 \includegraphics[width=0.33\linewidth]{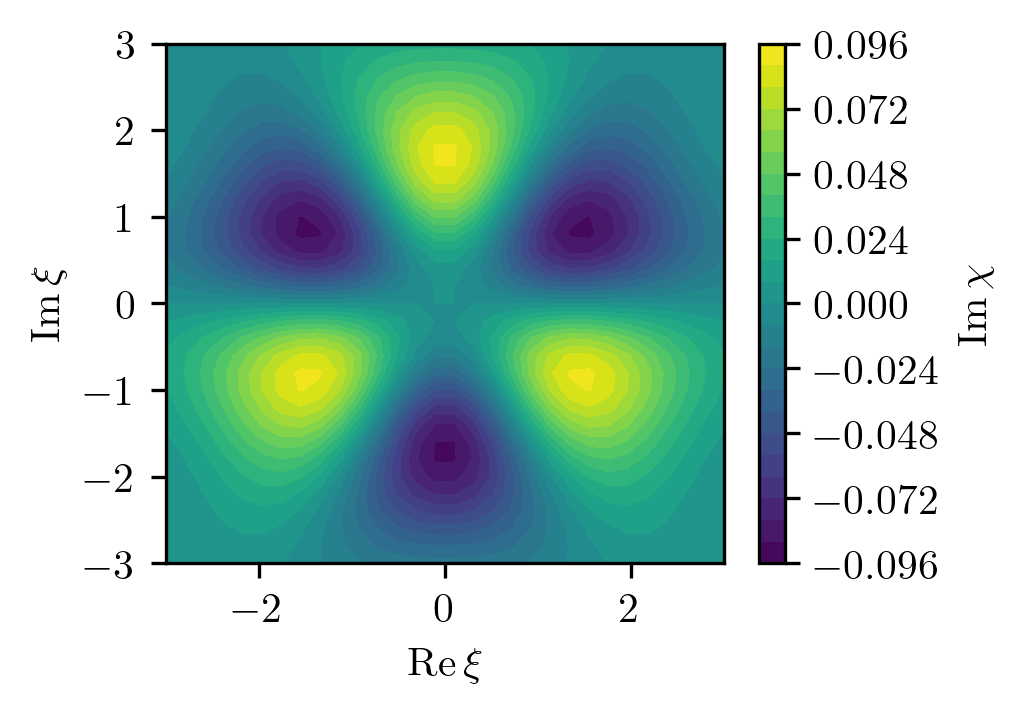}%
}
\caption{{\bf Characteristic functions of squeezed states:} (a) Standard $n=2$ squeezed state with amplitude $r=0.25$ and phase $\theta=0$. The contour of the characteristic function, which is always real in this case, displays the typical ellipsoidal contour levels that are compressed along one phase-space axis at the expense of the other. (b,c) Non-Gaussian $n=3$ squeezed state with amplitude $r=0.25$ and phase $\theta=0$. The real part (b) displays concentric contour lines which, upon close inspection, have a $C_6$ rotational symmetry. In this case, the imaginary part (c) no longer vanishes, and is actually an odd function under $\xi\mapsto-\xi$ in contrast to the real part, which is even. As a result, the $C_6$ symmetry is reduced to $C_3$ rotations. Whereas the Weyl characteristic function of the $n=2$ squeezed has an exact analytical expression~\eqref{eq:char_squeezed}, the $n=3$ squeezed state is non-Gaussian and requires a numerical computation.}.
\label{fig:char_functions}
\end{figure*}

We start by recalling that the path integral approach to the vacuum persistence amplitude~\eqref{eq:gen_func} reviewed in Appendix~\ref{app:vpa} results in  a closed functional expression~\eqref{eq:gen_functional_series}, which can be expanded to the desired power of the vertex and sources. This leads, for example, to the standard Feynman diagrams of a quartic oscillator displayed in~\eqref{eq:Z_interacting}. The underlying assumption shared with
 QFTs is that, while the
 Schwinger sources can be instantaneously switched on/off, the remaining microscopic parameters can only vary slowly, such that the system follows the groundstate adiabatically.
 This would be the case of the microscopic interaction strength $\lambda(t):0\mapsto\lambda\mapsto 0$, which is adiabatically switched on from the remote past to the present and then off towards the distant future, connecting $\ket{0_b}\mapsto\ket{\Omega_b}\mapsto\ket{0_b}$.
The reason why the Schwinger sources stand on a different footing
is that they should be able to create and absorb excitations instantaneously which, mathematically, translates into the possibility of extracting the desired propagator by functional differentiation, e.g., 
\beq
G(t_i-t_j)=-\frac{1}{Z[0,0]}\frac{\delta^2{Z}[J^*,J]}{\delta J^*(t_i)\delta J(t_j)}\bigg|_{J=0}.
\eeq
We recall that the functional derivatives involve an instantaneous delta-type switching of the sources/sinks $\delta F/\delta J(t_i)|_{J=0}=\lim_{\epsilon\to 0}\frac{1}{\epsilon}(F[J(t)+\epsilon\delta(t-t_i)]-F[J(t)])|_{J=0}$.

In order to use the path-integral formalism for the squeezed persistence amplitude \eqref{eq:charac_time}, we  consider that also the interactions $\lambda(t)$ can be switched on/off impulsively, generating an off-equilibrium initial state.
We thus need to consider that the interacting Hamiltonian has now an explicit time dependence,
such that 
\beq
\label{eq:int_ham_time} H(t)=\omega_ba^\dagger a+V_{\lambda(t)}(a^\dagger,a),\hspace{1ex} a(t)=U_\lambda^\dagger(t)a\, U_\lambda(t),
\eeq 
where we have introduced the time-ordered unitary $U_\lambda(t)=\mathcal{T}\{{\rm exp}(-\ii\int_{t_0}^t{\rm d}t' H(t'))\}$. One then proceeds with the path integral
as in Appendix~\ref{app:vpa}
and, by selecting the appropriate $V_{\lambda(t)}$, the characteristic functional can be expressed as
\beq
\label{eq:char_functional}
\chi[J^{*},J]=\frac{\ee^{-\ii\!\bigintssss_{t}V_{\lambda(t)}\!\!\big(\!-\ii\delta_ {J(t)},-\ii\delta_ {J^*\!(t)}\!\big)}\!\ee^{-\bigintssss_{t_1}\!\!\!\bigintssss_{t_2}\!\! J^{*}\!(t_1)G_0(t_1-t_2)J(t_2)}\phantom{\big|_{\!J=0}}}{\ee^{-\ii\!\bigintssss_{t}V_{\lambda(t)}\!\!\big(\!-\ii\delta_ {J(t)},-\ii\delta_ {J^*\!(t)}\!\big)}\!\ee^{-\bigintssss_{t_1}\!\!\!\bigintssss_{t_2}\!\! J^{*}\!(t_1)G_0(t_1-t_2)J(t_2)}\big|_{\!J=0}},
\eeq
which will admit an expansion in terms of
Feynman diagrams to any arbitrary order, paralleling Eq.~\eqref{eq:Z_interacting}. 
 
 From now on, we will be interested in interactions that create/annihilate excitations in bundles of $n$ quanta
\beq
\label{eq:n_squeezing}
V_{\lambda(t)}\big(a^\dagger,a\big)=\frac{1}{n!}\bigr(\lambda\!(t)\,a^{\dagger n}+\lambda^{\!*}\!(t)\,a^n\bigl),
\eeq
in contrast to the more standard QFT-like situation with a quartic potential~\eqref{eq:pol_int} in which the potential only depends on the position of the oscillator.
As 
these types of potentials are no longer functions of a single quadrature of the bosonic mode, we generally need to work with complex-valued sinks and sources.
By instantaneously switching on/off the potential~\eqref{eq:n_squeezing}  with a certain strength, one prepares the initial state
\beq
\label{eq:squeezed}
\lambda_0=-\ii\zeta=-\ii r\ee^{\ii\theta},\hspace{2ex} \ket{\psi_{ b}}=S_n({\zeta})\!\ket{0}=\ee^{\frac{1}{n!}\left(\zeta^*a^n-{\zeta}a^{\dagger n}\right)}\ket{0},%
\eeq
 where we have introduced the squeezing amplitude $r>0$ and phase $\theta\in[0,2\pi)$.  Here $S_n$ denotes the generalized squeezing operator.

For $n=2$, this corresponds to a single-mode squeezed state~\cite{PhysRevD.1.3217,PhysRevA.13.2226,PhysRevD.19.1669}
which belongs to the family of Gaussian states~\cite{WANG20071} and, by definition, has a characteristic function~\eqref{eq:charact_func} that can be expressed as a Gaussian, and is thus purely real (see Fig.~\ref{fig:chi_squeezing}). 
For $n=3$ this corresponds to a tri-squeezed state~\cite{fisher1984impossibility,braunsteain1987generalized,PhysRevA.55.2368,PhysRevA.104.032609}, which has been recently generated in superconducting-qubit~\cite{PhysRevX.10.011011} and trapped-ion~\cite{bazavan2024squeezing} experiments. In this case, the Weyl characteristic function is no longer a Gaussian, and indeed has a non-vanishing imaginary part (see Figs.~\ref{fig:chi_real_trisqueezing} and \ref{fig:chi_imag_trisqueezing}).
We note that non-Gaussian states cannot be efficiently simulated on a classical computer and that to achieve universality for continuous-variable quantum computation, one possibility is to include this type of interaction with $n=3$ ~\cite{PhysRevLett.82.1784,PhysRevA.64.012310}. For $n> 3$, one can obtain even higher-order squeezing~\cite{fisher1984impossibility, braunsteain1987generalized}, such as the quad-squeezed states that have been recently observed in trapped ions~\cite{bazavan2024squeezing}. 

In any of these examples, we see that the  coupling~\eqref{eq:n_squeezing} must be complex, requiring a pair of time-dependent vertices
that are  connected to either two outgoing or two incoming lines, respectively, as they can only emit or absorb bundles of $n=2$ or $n=3$ bosonic excitations, respectively. This will change the form of the Feynman diagrams one can find with respect to those of the quartic potential~\eqref{eq:Z_interacting}. In order to find these diagrams, in addition to the initial impulsive switching that sets the squeezed state at $t=t_0$~\eqref{eq:squeezed}, we note that a path-integral formulation will also require a vertex at the final time $t=t_{\rm f}$ that sets the correct boundary condition for the GSPA. This can be achieved by considering the following $n$-squeezed vertex 
\beq
\label{eq:vertices}
\begin{split}
\lambda(t)&=\lambda_+(t)\delta(t-t_0)+\lambda_{-}(t)\delta(t-t_{\rm f}), \hspace{1ex}\lambda_\pm(t)=\pm\lambda_0\ee^{-\ii n\omega_{b}t}.%
\end{split}
\eeq
  Here, the complex exponentials keep track of the phase reference, and connect directly to the type of qubit measurement discussed in the previous section~\eqref{eq:sx_sy}, where we considered using the same phase-referenced radiation to implement the initial and final qubit rotations. In addition, the overall opposite sign of the interaction vertex at the final time is required to set the boundary condition of the path-integral formulation $\bra{\psi_{b}}=\bra{0}S^\dagger_n\!({\zeta})=\bra{0}S_n({-\zeta})$. Let us remark that the squeezing operation is only applied once for each experimental run. It is only in the path-integral formulation where we need to allow for the initial and final action of these interaction vertices to set the boundary conditions and connect to the GSPA~\eqref{eq:charac_time}.

Let us emphasise that, with this choice~\eqref{eq:vertices},
the interaction vertices $\lambda(t_i),\lambda^{\!*}(t_j)$
can only act at the initial and final instants of time, where they create/annihilate arbitrary numbers of bosons in bundles of $n$ quanta. The excitations created at $t_0$ will propagate according to the free Green's function $G_0(t_j-t_0)$ for all intermediate times ${t_j}$, until they either get absorbed by the action of the sinks $J^*\!(t_j)$, or otherwise propagate all the way towards $t_{\rm f}$ where they can connect to the final interaction vertex.
This type of processes can be combined at all possible orders and all intermediate times until, at the end $t_{\rm f}$, the vertices act by annihilating the remaining excitations again in bundles of $n$ quanta, such that the resulting state is brought back to the vacuum with some probability amplitude. As shown in the following subsections, by allowing for vertex functions that can be switched on/off impulsively, we can associate each of these terms to a Feynman diagram, and connect this to current experiments that measure the characteristic functional~\eqref{eq:charac_time} for resonant sources~\eqref{eq_resonant_source}.

\subsubsection{Diagrams for standard squeezing}

Let us now present the specific Feynman diagrams from Eq.~\eqref{eq:char_functional} with the interaction potential~\eqref{eq:n_squeezing} and the vertices~\eqref{eq:vertices}, focusing on the standard $n=2$ squeezed state~\eqref{eq:squeezed}. This falls under the class of Gaussian states~\cite{WANG20071} which, as discussed in detail in Sec.~\ref{eq:squeezing_rwa}, will allow for an exact expression of the characteristic function $\chi(\xi^*,\xi)$ in Eq.~\eqref{eq:charact_func} in the resonant regime
$\xi\propto (t_{\rm f}-t_0)$. In this section, we derive the different Feynman diagrams with the correct symmetry factors, preparing the ground for a quantitative check of the validity of our path integral formalism in the resonant regime. 

In our closed expression of the characteristic functional~\eqref{eq:char_functional}, we need to Taylor expand the exponential of
\beq
\label{eq:2_squeezing_v}
-\ii\int_{t_0}^{t_{\rm f}}\!\!{\rm d}tV_{\lambda(t)}=+\ii\frac{\lambda(t_0)}{2}\delta^2_{ J(t_0)}+\ii\frac{\lambda^{\!*}\!(t_{\rm f})}{2}\delta^2_{ J^*\!(t_{\rm f})}- {\rm c.c.},
\eeq
where we have  used a short-hand notation for the functional derivatives $\delta^n_{J(t_j)}=\delta^n/\delta J^n(t_j),\delta^n_{J^*(t_j)}=\delta^n/\delta J^{*n}(t_j)$. One observes in this equation that the interaction vertices can only be inserted at the initial and final times. Moreover, for a zero-temperature regime, the terms in the above complex conjugation shall not give any contribution, as they would require backwards hole propagation. Finally, another simplification that is also usual in QFTs is that all the Feynman diagrams that do not depend on source functions, the so-called vacuum diagrams, do not contribute due to the normalization in Eq.~\eqref{eq:char_functional}. 

The problem has thus been reduced the calculation of two types of functional derivatives over the free characteristic functional, which is nothing but a Gaussian in the sources
\beq
\label{eq:free_charact}
\chi_0[J^*,J]={\rm exp}\left\{- \int_{t_1}\!\int_{t_2}\, J^*\!(t_1)G_0(t_1-t_2)J(t_2)\right\}.
\eeq
The leading orders of the expansion in $J,J^*,\lambda,\lambda^{\!*}$ are
 \begin{widetext}
 \beq
  \label{eq:chi_squeezed} 
 \chi[J^*,J]=\bigg(\, 1\hspace{-7.ex}
 \setlength{\unitlength}{1cm}
\thicklines
\begin{picture}(19,0)
\put(1.2,0.0){$+\,\,\frac{\ii}{2}$}
\put(2.1,.05){\line(2,1){0.5}}
\put(2.1,0.05){\vector(2,1){0.4}}
\put(2.1,0.05){\vector(2,-1){0.4}}
\put(2.1,.05){\line(2,-1){0.5}}
\put(1.84,-0.05){${\color{amber}\boldsymbol{\circledcirc}_{\!\!_{_{0}}}}$}
\put(2.55,.27){${\color{forestgreen}\boldsymbol{\ominus}}$}
\put(2.55,-.35){${\color{forestgreen}\boldsymbol{\ominus}}$}
\put(2.9,0.){$+\,\,\frac{\ii}{2}$}
\put(3.55,0.27){${\color{forestgreen}\boldsymbol{\oplus}}$}
\put(3.55,-0.35){${\color{forestgreen}\boldsymbol{\oplus}}$}
\put(3.8,.26){\line(2,-1){0.4}}
\put(3.8,-.19){\line(2,1){0.4}}
\put(3.8,0.26){\vector(2,-1){0.3}}
\put(3.8,-0.19){\vector(2,1){0.3}}
\put(4.18,-0.05){${\color{amber}\boldsymbol{\circledast}_{\!\!_{_{\rm f}}}}$}
\put(4.8,0.0){$+$}
\put(5.5,.05){\line(2,1){0.5}}
\put(5.5,0.05){\vector(2,1){0.4}}
\put(6.32,-0.265){\vector(1,0){0.1}}
\qbezier(5.5,-.075)(6.23,-0.46)(7.2,-0.075)
\put(5.24,-0.1){${\color{amber}\boldsymbol{\circledcirc}_{\!\!_{_{0}}}}$}
\put(5.95,.27){${\color{forestgreen}\boldsymbol{\ominus}}$}
\put(6.5,.27){${\color{forestgreen}\boldsymbol{\oplus}}$}
\put(6.75,.3){\line(2,-1){0.44}}
\put(6.75,0.3){\vector(2,-1){0.3}}
\put(7.15,-0.1){${\color{amber}\boldsymbol{\circledast}_{\!\!_{_{\rm f}}}}$}
\put(7.75,0.){$-\hspace{1ex}\fourth $}
\put(8.75,.05){\line(2,1){0.5}}
\put(8.75,0.05){\vector(2,1){0.4}}
\put(8.75,0.05){\vector(2,-1){0.4}}
\put(8.75,.05){\line(2,-1){0.5}}
\put(8.49,-0.05){${\color{amber}\boldsymbol{\circledcirc}_{\!\!_{_{0}}}}$}
\put(9.2,.27){${\color{forestgreen}\boldsymbol{\ominus}}$}
\put(9.2,-.35){${\color{forestgreen}\boldsymbol{\ominus}}$}
\put(9.7,0.27){${\color{forestgreen}\boldsymbol{\oplus}}$}
\put(9.7,-0.35){${\color{forestgreen}\boldsymbol{\oplus}}$}
\put(9.93,.25){\line(2,-1){0.4}}
\put(9.93,-.18){\line(2,1){0.4}}
\put(9.95,0.235){\vector(2,-1){0.3}}
\put(9.95,-0.175){\vector(2,1){0.3}}
\put(10.3,-0.05)
{${\color{amber}\boldsymbol{\circledast}_{\!\!_{_{\rm f}}}}$}
\put(10.75,0.){$-\hspace{1ex}\textstyle{\frac{1}{8}} $}
\put(11.75,.15){\line(4,3){0.5}}
\put(11.75,0.15){\vector(4,3){0.4}}
\put(11.75,0.15){\vector(2,0){0.4}}
\put(11.75,.15){\line(2,0){0.5}}
\put(11.49,0.1){${\color{amber}\boldsymbol{\circledcirc}_{\!\!_{_{0}}}}$}
\put(12.2,.45){${\color{forestgreen}\boldsymbol{\ominus}}$}
\put(12.2,.07){${\color{forestgreen}\boldsymbol{\ominus}}$}
\put(11.75,-.15){\line(4,-3){0.5}}
\put(11.75,-0.15){\vector(4,-3){0.4}}
\put(11.75,-0.15){\vector(2,0){0.4}}
\put(11.75,-.15){\line(2,0){0.5}}
\put(11.49,-0.28){${\color{amber}\boldsymbol{\circledcirc}_{\!\!_{_{0}}}}$}
\put(12.2,-.6){${\color{forestgreen}\boldsymbol{\ominus}}$}
\put(12.2,-.25){${\color{forestgreen}\boldsymbol{\ominus}}$}
\put(12.75,0.){$-\hspace{1ex}\textstyle{\frac{1}{8}} $}
\put(13.75,.56){\line(4,-3){0.5}}
\put(13.75,0.56){\vector(4,-3){0.4}}
\put(13.75,0.18){\vector(2,0){0.4}}
\put(13.75,.18){\line(2,0){0.5}}
\put(14.23,0.12){${\color{amber}\boldsymbol{\circledast}_{\!\!_{_{\rm f}}}}$}
\put(13.5,.45){${\color{forestgreen}\boldsymbol{\oplus}}$}
\put(13.5,.1){${\color{forestgreen}\boldsymbol{\oplus}}$}
\put(13.75,-.55){\line(4,3){0.5}}
\put(13.75,-0.55){\vector(4,3){0.4}}
\put(13.75,-0.15){\vector(2,0){0.4}}
\put(13.75,-.15){\line(2,0){0.5}}
\put(14.23,-0.25){${\color{amber}\boldsymbol{\circledast}_{\!\!_{_{\rm f}}}}$}
\put(13.5,-.6){${\color{forestgreen}\boldsymbol{\oplus}}$}
\put(13.5,-.25){${\color{forestgreen}\boldsymbol{\oplus}}$}
\put(14.65,0.01){$+\,\,\cdots\bigg) \mathlarger{\chi_0[J^*,J]}$.}
\end{picture}
\eeq
 \end{widetext}
 In order to translate each diagram to a mathematical expression, we introduce  the conventions $ {\color{forestgreen}\boldsymbol{\oplus}}=J(t_j)$, ${\color{forestgreen}\boldsymbol{\ominus}}=J^*\!(t_i)$, $ {\color{amber}\boldsymbol{\circledcirc}_{_{0}}}\!=\lambda_+(t_0)$,${\color{amber}\boldsymbol{\circledast}_{_{{\rm f}}}}\!=\lambda_-^*\!(t_{\rm f})$, and $\boldsymbol{\rightarrow}\hspace{-1.2ex}\boldsymbol{-}=G_0(t_i-t_j)=\theta(t_i-t_j)\ee^{-\ii\omega_{b}(t_i-t_j)}$, and integrate over all intermediate times in which the source/sink functions are inserted. 
 We remark that, in spite of the representation of the diagrams in a plane, all the sources and vertices only account for different instants of time, and there is no spatial propagation involved at all.
 As a specific example, the first diagram amounts to the integral
 \beq
 \label{eq:feyn_diag_example}
 \setlength{\unitlength}{1cm}
\thicklines
\begin{picture}(19,0)
\put(0.2,.05){\line(2,1){0.5}}
\put(0.2,0.05){\vector(2,1){0.4}}
\put(0.2,0.05){\vector(2,-1){0.4}}
\put(0.2,.05){\line(2,-1){0.5}}
\put(-0.05,-0.05){${\color{amber}\boldsymbol{\circledcirc}_{\!\!_{_{0}}}}$}
\put(0.65,.27){${\color{forestgreen}\boldsymbol{\ominus}}$}
\put(0.65,-.35){${\color{forestgreen}\boldsymbol{\ominus}}$}
\put(1.2,0.01){$\!\!\!\!\!\!\!=\lambda_+(t_0)\!\!\bigintsss_{\,\,t_0}^{ t_{\rm f}}\!{\rm d}t_1J^*\!(t_1)G_0(t_1-t_0)\!\!\bigintsss_{\,\,t_0}^{ t_{\rm f}}\!{\rm d}t_2J^*\!(t_2)G_0(t_2-t_0)$.}
\end{picture}
 \eeq
 Since the interaction vertices cannot be inserted at intermediate times during the propagation of the bosonic excitations, we find
diagrams that differ
from those of the $D=0+1$ dimensional version of the $\lambda\phi^4$ QFT~\eqref{eq:Z_interacting}. Nonetheless, the underlying ingredients are the same and, by changing the corresponding vertices, one would approach this paradigmatic QFT.

\subsubsection{ Diagrams for non-Gaussian squeezing}

The potential interest of the diagrammatic techniques is that they can be applied to non-Gaussian regimes~\cite{PRXQuantum.2.030204}, which would connect to the more challenging cases of interacting QFTs~\cite{Peskin:1995ev,Fradkin:2021zbi,Jordan2018bqpcompletenessof} when going beyond this single-mode scenario.
 In our present context, the previous path-integral approach can be adapted to any $n$-order squeezing interactions~\eqref{eq:n_squeezing}, which is no longer exactly solvable when $n\geq 3$, by simply changing Eq.~\eqref{eq:2_squeezing_v} for 
\beq
\label{eq:n_squeezing_v}
-\ii\int_{t_0}^{t_{\rm f}}\!\!{\rm d}tV_{\lambda(t)}=-\ii^{n+1}\frac{\lambda(t_0)}{n!}\delta^n_{ J(t_0)}-\ii^{n+1}\frac{\lambda^{\!*}\!(t_{\rm f})}{n!}\delta^n_{ J^*\!(t_{\rm f})}- {\rm c.c.},
\eeq
Substituting this equation in our closed functional expression~\eqref{eq:char_functional}, we can carry a diagrammatic power expansion to desired order of the microscopic couplings. For instance, for $n=3$ squeezing, we obtain the Feynman diagrams 
 \begin{widetext}
 \beq
  \label{eq:chi_squeezed_3} 
 \chi[J^*,J]=\bigg(\, 1\hspace{-7.ex}
 \setlength{\unitlength}{1cm}
\thicklines
\begin{picture}(19,0)
\put(1.2,0.0){$-\,\,\frac{1}{6}$}
\put(2.1,.05){\line(2,1){0.5}}
\put(2.1,0.05){\vector(2,1){0.4}}
\put(2.1,.05){\line(2,0){0.6}}
\put(2.1,0.05){\vector(2,0){0.55}}
\put(2.1,.05){\line(2,-1){0.5}}
\put(2.1,0.05){\vector(2,-1){0.4}}
\put(2.1,.05){\line(2,-1){0.5}}
\put(1.84,-0.05){${\color{amber}\boldsymbol{\circledcirc}_{\!\!_{_{0}}}}$}
\put(2.55,.27){${\color{forestgreen}\boldsymbol{\ominus}}$}
\put(2.65,-.04){${\color{forestgreen}\boldsymbol{\ominus}}$}
\put(2.55,-.35){${\color{forestgreen}\boldsymbol{\ominus}}$}
\put(3.,0.){$-\,\frac{1}{6}$}
\put(3.65,0.27){${\color{forestgreen}\boldsymbol{\oplus}}$}
\put(3.57,-0.04){${\color{forestgreen}\boldsymbol{\oplus}}$}
\put(3.65,-0.35){${\color{forestgreen}\boldsymbol{\oplus}}$}
\put(3.9,.26){\line(2,-1){0.4}}
\put(3.8,0.035){\line(2,0){0.5}}
\put(3.9,-.19){\line(2,1){0.4}}
\put(3.9,0.25){\vector(2,-1){0.3}}
\put(3.85,0.035){\vector(2,0){0.3}}
\put(3.9,-0.18){\vector(2,1){0.3}}
\put(4.28,-0.05){${\color{amber}\boldsymbol{\circledast}_{\!\!_{_{\rm f}}}\boldsymbol{}}$}
\put(4.7,0.0){$-\frac{1}{6}$}
\put(5.5,.05){\line(1,1){0.4}}
\put(5.5,0.04){\vector(1,1){0.3}}
\put(5.5,-.02){\line(1,0){0.4}}
\put(5.5,-0.02){\vector(1,0){0.3}}
\put(6.32,-0.41){\vector(1,0){0.1}}
\qbezier(5.5,-.1)(6.25,-0.75)(7.15,-0.07)
\put(5.21,-0.1){${\color{amber}\boldsymbol{\circledcirc}_{\!\!_{_{0}}}}$}
\put(5.86,.4){${\color{forestgreen}\boldsymbol{\ominus}}$}
\put(5.86,-.1){${\color{forestgreen}\boldsymbol{\ominus}}$}
\put(6.5,.4){${\color{forestgreen}\boldsymbol{\oplus}}$}
\put(6.5,-.1){${\color{forestgreen}\boldsymbol{\oplus}}$}
\put(6.75,.41){\line(1,-1){0.4}}
\put(6.75,0.41){\vector(1,-1){0.3}}
\put(6.75,-.025){\line(1,0){0.4}}
\put(6.75,-0.025){\vector(1,0){0.3}}
\put(7.125,-0.1){${\color{amber}\boldsymbol{\circledast}_{\!\!_{_{\rm f}}}}$}
\put(7.75,0.){$+\hspace{1ex}\half $}
\put(8.75,.05){\line(1,1){0.4}}
\put(8.75,0.04){\vector(1,1){0.3}}
\put(8.75,-.02){\line(1,0){1.6}}
\put(9.0,-0.02){\vector(1,0){0.65}}
\put(9.55,-0.41){\vector(1,0){0.1}}
\qbezier(8.75,-.1)(9.5,-0.75)(10.4,-0.07)
\put(8.45,-0.1){${\color{amber}\boldsymbol{\circledcirc}_{\!\!_{_{0}}}}$}
\put(9.11,.4){${\color{forestgreen}\boldsymbol{\ominus}}$}
\put(9.75,.4){${\color{forestgreen}\boldsymbol{\oplus}}$}
\put(10.,.41){\line(1,-1){0.4}}
\put(10.,0.41){\vector(1,-1){0.3}}

\put(10.4,-0.1){${\color{amber}\boldsymbol{\circledast}_{\!\!_{_{\rm f}}}}$}
\put(10.8,0.0){$+\half\bigg(\frac{1}{6}$}
\put(12.1,.05){\line(2,1){0.5}}
\put(12.1,0.05){\vector(2,1){0.4}}
\put(12.1,.05){\line(2,0){0.6}}
\put(12.1,0.05){\vector(2,0){0.55}}
\put(12.1,.05){\line(2,-1){0.5}}
\put(12.1,0.05){\vector(2,-1){0.4}}
\put(12.1,.05){\line(2,-1){0.5}}
\put(11.84,-0.05){${\color{amber}\boldsymbol{\circledcirc}_{\!\!_{_{0}}}}$}
\put(12.55,.27){${\color{forestgreen}\boldsymbol{\ominus}}$}
\put(12.65,-.04){${\color{forestgreen}\boldsymbol{\ominus}}$}
\put(12.55,-.35){${\color{forestgreen}\boldsymbol{\ominus}}$}
\put(13.,0.){$+\,\frac{1}{6}$}
\put(13.65,0.27){${\color{forestgreen}\boldsymbol{\oplus}}$}
\put(13.57,-0.04){${\color{forestgreen}\boldsymbol{\oplus}}$}
\put(13.65,-0.35){${\color{forestgreen}\boldsymbol{\oplus}}$}
\put(13.9,.26){\line(2,-1){0.4}}
\put(13.8,0.035){\line(2,0){0.5}}
\put(13.9,-.19){\line(2,1){0.4}}
\put(13.9,0.25){\vector(2,-1){0.3}}
\put(13.85,0.035){\vector(2,0){0.3}}
\put(13.9,-0.18){\vector(2,1){0.3}}
\put(14.28,-0.05){${\color{amber}\boldsymbol{\circledast}{\!_{_{\rm f}}}\boldsymbol{}}\!\!\bigg)^{\!\!\!2}$}

\put(15.,0.01){$\!\!\!\!\!+\,\cdots\!\bigg) \mathlarger{\chi_0[J^*,J]}$,}
\end{picture}
\eeq
 \end{widetext}
 where, once more, the vacuum diagrams
with no sinks and sources disappear due to the normalization in Eq.~\eqref{eq:char_functional}. To simplify the notation, the last diagrams inside the parenthesis have been squared, such that their product will lead to three disconnected diagrams that have a total of 6 sources/sinks, and the corresponding propagators that connect them to the corresponding pair of vertices. We note that this simplification could have also been used for the last three disconnected diagrams of Eq.~\eqref{eq:chi_squeezed}, which can also be grouped as the square of the sum of the two first ones.

Similar diagrammatic expansions can be derived for $n\geq 4$, which will include more types of connected diagrams as the vertices have a larger number of legs to be combined. The functional techniques hereby presented yield directly the symmetry factors that count all possible ways in which these legs can be connected to yield the final Feynman diagrams. In the following section, we will perform the explicit time integrals which, together with the symmetry factors, give the precise contribution to the characteristic functional for resonant sources at a given order of the sources and vertices.

\subsubsection{ Standard squeezing persistence amplitude }
\label{eq:gaussian_integrals}

In this section, we focus on the characteristic function for the $n=2$ squeezed state~\eqref{eq:squeezed}, and derive the explicit time-dependence underlying the Feynman diagrams when considering the harmonic sources of frequency~\eqref{eq:rwa_sources}. We will present a generic expression that is valid out of resonance $\Delta\neq0$, and also specialize to the resonant limit $\Delta=0$ to connect to the regime of the recent experiments~\cite{Fl_hmann_2020,matsos2023robust,bazavan2024squeezing} in which the phase-space coordinates grow linearly in time~\eqref{eq_resonant_source}.

The lowest-order contribution is provided by the first two diagrams in Eq.~\eqref{eq:chi_squeezed}, in which a pair of excitations created (annihilated) at the initial (final) time are absorbed (emitted) by a couple of sinks (sources) at a later (previous) instant of time. Translating these diagrams into time integrals such as Eq.~\eqref{eq:feyn_diag_example}, and performing the corresponding integrals, we find
 \beq
 \label{eq:fd_1}
 \setlength{\unitlength}{1cm}
\thicklines
\begin{picture}(19,0)
\put(0.6,0.01){$\frac{\ii}{2}$}
\put(1.05,.05){\line(2,1){0.5}}
\put(1.05,0.05){\vector(2,1){0.4}}
\put(1.05,0.05){\vector(2,-1){0.4}}
\put(1.05,.05){\line(2,-1){0.5}}
\put(0.8,-0.05){${\color{amber}\boldsymbol{\circledcirc}_{\!\!_{_{0}}}}$}
\put(1.5,.27){${\color{forestgreen}\boldsymbol{\ominus}}$}
\put(1.5,-.35){${\color{forestgreen}\boldsymbol{\ominus}}$}
\put(2.0,0.01){$\!\!\!\!\!=-\frac{\ii}{2}\lambda_0\ee^{+2\ii\delta t_0}J_0^{*2}\,C^2_{\delta}(t_{\rm f}-t_0)\approx-\frac{r}{2}\xi^{*2}\ee^{\ii\theta}$,}
\end{picture}
 \eeq
 \beq
 \label{eq:fd_2}
 \nonumber
 \setlength{\unitlength}{1cm}
\thicklines
\begin{picture}(19,0)
\put(0.65,0.01){$\frac{\ii}{2}$}
\put(0.95,0.27){${\color{forestgreen}\boldsymbol{\oplus}}$}
\put(0.95,-0.35){${\color{forestgreen}\boldsymbol{\oplus}}$}
\put(1.2,.29){\line(2,-1){0.4}}
\put(1.2,-.19){\line(2,1){0.4}}
\put(1.2,0.29){\vector(2,-1){0.3}}
\put(1.2,-0.18){\vector(2,1){0.3}}
\put(1.55,-0.05){${\color{amber}\boldsymbol{\circledast}_{\!\!_{_{\rm f}}}}$}
\put(2.02,0.01){$\!\!\!\!\!=+\frac{\ii}{2}\lambda_0^*\ee^{-2\ii\delta t_{\rm 0}}J_0^{2}\,C^{*2}_{\delta}(t_{\rm f}-t_0)\approx-\frac{r}{2}\xi^2\ee^{-\ii\theta}$,}
\end{picture}
 \eeq
 In these expressions, we have introduced the circle function
\beq
C_{\Delta}(t)=\frac{1}{\Delta}\left(\ee^{\ii\Delta t}-1\right).
\eeq
and taken the limit of $\Delta\to 0$, which amounts to the resonant regime in which one can identify a characteristic function parameter that grows linearly with the time
\beq
\label{eq:xi_dispalcement}
\xi\approx\ii J_0(t_{\rm f}-t_0)=\Omega\eta(t_{\rm f}-t_0)\ee^{\ii\Delta\varphi}.
\eeq
 As discussed in Appendix~\ref{app:exact}, in this resonant regime, one can forget about the time-dependence, and simply use the exact expressions for the characteristic function of Gaussian states as a function of $\xi$, finding a perfect agreement of these two diagrams with the corresponding order in Eq.~\eqref{eq:char_squeezed_pert_exact}. Going to second order in the vertices, we find
\beq
 \label{eq:fd_3}
 \setlength{\unitlength}{1cm}
\thicklines
\begin{picture}(19,0)
\put(0.6,.05){\line(2,1){0.5}}
\put(0.6,0.05){\vector(2,1){0.4}}
\put(1.45,-0.27){\vector(1,0){0.2}}
\qbezier(0.6,-.075)(1.35,-0.5)(2.3,-0.075)
\put(0.32,-0.1){${\color{amber}\boldsymbol{\circledcirc}_{\!\!_{_{0}}}}$}
\put(1.05,.27){${\color{forestgreen}\boldsymbol{\ominus}}$}
\put(1.6,.27){${\color{forestgreen}\boldsymbol{\oplus}}$}
\put(1.85,.3){\line(2,-1){0.44}}
\put(1.85,0.3){\vector(2,-1){0.3}}
\put(2.25,-0.1){${\color{amber}\boldsymbol{\circledast}_{\!\!_{_{\rm f}}}}$}\put(2.75,0.01){$\!\!\!\!\!=-|\lambda_0J_0|^{2}\,|C_{\delta}(t_{\rm f}-t_0)|^2\approx-r^2|\xi|^2$,}%
\end{picture}
 \eeq
 which accounts for situations in which one of the created excitations travels all the way from the initial to the final time, and agrees with the second contribution in Eq.~\eqref{eq:char_squeezed_pert_exact}.
The remaining diagrams in Eq.~\eqref{eq:chi_squeezed} are disconnected and can be obtained from the square of the first ones, such that 
\beq
\label{eq:fd_4}
 \setlength{\unitlength}{1cm}
\thicklines
\begin{picture}(19,0)
\put(0.0,0.0){$\phantom{-}\half\bigg(\!\!\frac{\ii}{2}$}
\put(1.2,.05){\line(2,1){0.5}}
\put(1.2,0.05){\vector(2,1){0.4}}
\put(1.2,0.05){\vector(2,-1){0.4}}
\put(1.2,.05){\line(2,-1){0.5}}
\put(0.94,-0.05){${\color{amber}\boldsymbol{\circledcirc}_{\!\!_{_{0}}}}$}
\put(1.65,.27){${\color{forestgreen}\boldsymbol{\ominus}}$}
\put(1.65,-.35){${\color{forestgreen}\boldsymbol{\ominus}}$}
\put(1.9,0.){$+\,\,\frac{\ii}{2}$}
\put(2.55,0.27){${\color{forestgreen}\boldsymbol{\oplus}}$}
\put(2.55,-0.35){${\color{forestgreen}\boldsymbol{\oplus}}$}
\put(2.8,.26){\line(2,-1){0.4}}
\put(2.8,-.19){\line(2,1){0.4}}
\put(2.8,0.25){\vector(2,-1){0.3}}
\put(2.8,-0.18){\vector(2,1){0.3}}
\put(3.18,-0.05){${\color{amber}\boldsymbol{\circledast}_{\!\!_{_{\rm f}}}}\!\!\bigg)^2=\frac{1}{2}\big( \frac{r}{2}\xi^{*2}\ee^{\ii\theta}+ \frac{r}{2}\xi^2\ee^{-\ii\theta}\big)^{\!2}$,}
\end{picture}
\eeq
which coincides with the last contribution to Eq.~\eqref{eq:char_squeezed_pert_exact}. 

Adding all of the Feynman diagrams, we find the following expression for the resonant characteristic functional
\beq
\label{eq:Feynman_n_2}
\frac{\chi}{\chi_0}\approx1-r\,{\rm Re}\{\xi^2\ee^{-\ii\theta}\}-{r^2}|\xi|^2+\frac{r^2}{2}\big({\rm Re}\{\xi^2\ee^{-\ii\theta}\}\big)^{\!\!2}\!,
\eeq
where $\chi_0={\rm exp}\{-\xi^*\xi/2\}$ is the free Gaussian part. One can see that the coefficients of the different terms are the combination of  the symmetry factors in Eq.~\eqref{eq:chi_squeezed} with additional factors that arising from the explicit time integrals of the propagators.  Altogether, the  expansion agrees with the exact Weyl characteristic function of the squeezed state~\eqref{eq:char_squeezed_pert_exact}, considering a  precise time-dependence that results from the combined integrals of the boson propagator and the source/sink functions.

\subsubsection{Non-Gaussian persistence amplitude }

After benchmarking our path-integral approach with the exact resonant expressions for the $n=2$ squeezed state, we can now move to a non-Gaussian state, such as the tri-squeezed state $n=3$ in Eq.~\eqref{eq:squeezed}. In this case, there is no closed analytical solution, and we will postpone a benchmark of the path-integral predictions to the following section.

We have already presented the Feynman diagram expansion for the tri-squeezed state in Eq.~\eqref{eq:chi_squeezed_3}, and it is then a matter of translating the expressions to specific time integrals, and evaluating those integrals in the case of the harmonic sources and sinks~\eqref{eq:rwa_sources}. The first two diagrams yield

 \beq
 \label{eq:fd_1_non_gaussian_psa}
 \setlength{\unitlength}{1cm}
\thicklines
\begin{picture}(19,0)
\put(0.3,0.01){$-\frac{1}{6}$}
\put(1.05,.05){\line(2,1){0.5}}
\put(1.05,0.05){\vector(2,1){0.4}}
\put(1.05,0.05){\vector(2,-1){0.4}}
\put(1.05,.05){\line(2,0){0.6}}
\put(1.1,0.05){\vector(2,0){0.5}}
\put(1.05,.05){\line(2,-1){0.5}}
\put(0.8,-0.05){${\color{amber}\boldsymbol{\circledcirc}_{\!\!_{_{0}}}}$}
\put(1.5,.27){${\color{forestgreen}\boldsymbol{\ominus}}$}
\put(1.575,-.05){${\color{forestgreen}\boldsymbol{\ominus}}$}
\put(1.5,-.35){${\color{forestgreen}\boldsymbol{\ominus}}$}
\put(2.1,0.01){$\!\!\!\!\!=-\frac{\ii}{6}\lambda_0\ee^{+3\ii\delta t_0}J_0^{*3}\,C^3_{\delta}(t_{\rm f}-t_0)\approx+\frac{r}{6}\xi^{*3}\ee^{\ii\theta}$,}
\end{picture}
 \eeq
 \beq
 \label{eq:fd_2_}
 \nonumber
 \setlength{\unitlength}{1cm}
\thicklines
\begin{picture}(19,0)
\put(0.3,0.01){$-\frac{1}{6}$}
\put(0.95,0.27){${\color{forestgreen}\boldsymbol{\oplus}}$}
\put(0.95,-0.35){${\color{forestgreen}\boldsymbol{\oplus}}$}
\put(1.2,.29){\line(2,-1){0.4}}
\put(1.2,-.19){\line(2,1){0.4}}
\put(1.2,0.29){\vector(2,-1){0.3}}
\put(1.2,-0.18){\vector(2,1){0.3}}
\put(1.15,.05){\line(2,0){0.45}}
\put(1.1,0.05){\vector(2,0){0.3}}
\put(0.85,-0.05){${\color{forestgreen}\boldsymbol{\oplus}}$}
\put(1.55,-0.05){${\color{amber}\boldsymbol{\circledast}_{\!\!_{_{\rm f}}}}$}
\put(2.1,0.01){$\!\!\!\!\!=+\frac{\ii}{6}\lambda_0^*\ee^{-3\ii\delta t_{\rm 0}}J_0^{3}\,C^{*3}_{\delta}(t_{\rm f}-t_0)\approx-\frac{r}{6}\xi^3\ee^{-\ii\theta}$,}
\end{picture}
 \eeq
where the approximations require setting $\delta\to 0$, and making use of the resonant characteristic function parameter in Eq.~\eqref{eq:xi_dispalcement}. It is interesting to note that these contributions are odd with respect to $J_0\to-J_0$, which shows that the full characteristic function cannot be a Gaussian, which is consistent with the fact that the tri-squeezed state does not belong to the family of Gaussian states~\cite{WANG20071,PRXQuantum.2.030204}. Moving to the connected Feynman diagrams, we now find two contributions

\beq
\setlength{\unitlength}{1cm}
\thicklines
\begin{picture}(19,0)
\put(0.25,0.){$\hspace{1ex}\half $}
\put(1.0,.05){\line(4,3){0.4}}
\put(1.,0.04){\vector(4,3){0.3}}
\put(1.,-.02){\line(1,0){1.65}}
\put(1.25,-0.02){\vector(1,0){0.68}}
\put(1.85,-0.31){\vector(1,0){0.1}}
\qbezier(1.,-.1)(1.75,-0.55)(2.65,-0.07)
\put(0.7,-0.1){${\color{amber}\boldsymbol{\circledcirc}_{\!\!_{_{0}}}}$}
\put(1.36,.3){${\color{forestgreen}\boldsymbol{\ominus}}$}
\put(2.,.3){${\color{forestgreen}\boldsymbol{\oplus}}$}
\put(2.27,.35){\line(4,-3){0.4}}
\put(2.27,0.35){\vector(4,-3){0.3}}
\put(2.65,-0.1){${\color{amber}\boldsymbol{\circledast}_{\!\!_{_{\rm f}}}}$}
\put(3.15,0.01){$\!\!\!\!\!=-\frac{1}{2}|\lambda_0J_0|^{2}\,|C_{\delta}(t_{\rm f}-t_0)|^2\!\approx\!-\frac{r^2}{2}|\xi|^2$\!,}
\end{picture}
\eeq
 \beq
 \nonumber
  \setlength{\unitlength}{1cm}
\thicklines
\begin{picture}(19,0)
\put(0.2,0.0){$-\frac{1}{6}$}
\put(1.0,.05){\line(4,3){0.4}}
\put(1.0,0.04){\vector(4,3){0.3}}
\put(1.0,-.02){\line(1,0){0.4}}
\put(1.0,-0.02){\vector(1,0){0.3}}
\put(1.85,-0.31){\vector(1,0){0.1}}
\qbezier(1.,-.1)(1.75,-0.55)(2.65,-0.07)
\put(0.76,-0.1){${\color{amber}\boldsymbol{\circledcirc}_{\!\!_{_{0}}}}$}
\put(1.36,.3){${\color{forestgreen}\boldsymbol{\ominus}}$}
\put(1.36,-.1){${\color{forestgreen}\boldsymbol{\ominus}}$}
\put(2.0,.3){${\color{forestgreen}\boldsymbol{\oplus}}$}
\put(2.0,-.1){${\color{forestgreen}\boldsymbol{\oplus}}$}
\put(2.25,.3){\line(4,-3){0.4}}
\put(2.25,0.3){\vector(4,-3){0.3}}
\put(2.25,-.025){\line(1,0){0.4}}
\put(2.25,-0.025){\vector(1,0){0.3}}
\put(2.625,-0.1){${\color{amber}\boldsymbol{\circledast}_{\!\!_{_{\rm f}}}}$}
\put(3.15,0.01){$\!\!\!\!\!=+\frac{1}{6}|\lambda_0J_0^2|^{2}\,|C_{\delta}(t_{\rm f}-t_0)|^4\approx\frac{r^2}{6}|\xi|^4$,}
\end{picture}
 \eeq
 which scale differently with the source and sink functions. Finally, the last disconnected diagrams in Eq.~\eqref{eq:chi_squeezed_3} read
\beq
 \setlength{\unitlength}{1cm}
\thicklines
\begin{picture}(19,0)
\put(0.1,0.0){$+\half\bigg(\frac{1}{6}$}
\put(1.35,.05){\line(2,1){0.5}}
\put(1.35,0.05){\vector(2,1){0.4}}
\put(1.35,.05){\line(2,0){0.6}}
\put(1.35,0.05){\vector(2,0){0.55}}
\put(1.35,.05){\line(2,-1){0.5}}
\put(1.35,0.05){\vector(2,-1){0.4}}
\put(1.35,.05){\line(2,-1){0.5}}
\put(1.1,-0.05){${\color{amber}\boldsymbol{\circledcirc}_{\!\!_{_{0}}}}$}
\put(1.85,.27){${\color{forestgreen}\boldsymbol{\ominus}}$}
\put(1.95,-.04){${\color{forestgreen}\boldsymbol{\ominus}}$}
\put(1.85,-.35){${\color{forestgreen}\boldsymbol{\ominus}}$}
\put(2.3,0.){$+\,\frac{1}{6}$}
\put(3.,0.27){${\color{forestgreen}\boldsymbol{\oplus}}$}
\put(2.87,-0.04){${\color{forestgreen}\boldsymbol{\oplus}}$}
\put(3,-0.35){${\color{forestgreen}\boldsymbol{\oplus}}$}
\put(3.2,.26){\line(2,-1){0.4}}
\put(3.1,0.035){\line(2,0){0.5}}
\put(3.2,-.19){\line(2,1){0.4}}
\put(3.2,0.25){\vector(2,-1){0.3}}
\put(3.15,0.035){\vector(2,0){0.3}}
\put(3.2,-0.18){\vector(2,1){0.3}}
\put(3.62,-0.05){${\color{amber}\boldsymbol{\circledast}{\!_{_{\rm f}}}\boldsymbol{}}\!\!\bigg)^{\!\!\!2}=\frac{1}{2}\big( \frac{r}{6}\xi^{*3}\ee^{\ii\theta}- \frac{r}{6}\xi^3\ee^{-\ii\theta}\big)^{\!2}$}
\end{picture}
\eeq
Adding all of these diagrams, we can approximate the characteristic functional by truncating at this particular order of the source and vertex functions 
\beq
\label{eq:Feynman_n_3}
\frac{\chi}{\chi_0}\!\approx\!1-\ii\frac{r}{3}\,{\rm Im}\{\xi^3\ee^{-\ii\theta}\}-\frac{r^2}{2}|\xi|^2+\frac{r^2}{6}|\xi|^4-\frac{r^2}{18}\big({\rm Im}\{\xi^3\ee^{-\ii\theta}\}\big)^{\!\!2}\!\!.
\eeq
Once more, we remark that the terms in this expansion are not just the symmetry factors of the Feynman diagrams, but the result of performing the various time integrals. For instance, in this case, we see that the different orders can lead to both real and imaginary contributions of the characteristic, which is a result of the breakdown of the symmetry $J_0\to-J_0$ when considering the non-Gaussian tri-squeezed state. 

\subsection{\bf Quantum tomography of Feynman diagrams}\label{sec:statistical_inference}

In this section, we show how to use  methods of statistical inference for the estimation of Feynman diagrams via the probe qubit (see Fig.~\ref{fig:scheme}). We have shown in Eq.~\eqref{eq:sx_sy} that
the  complex-valued characteristic functional evaluated at the specific source functions~\eqref{eq:sd_force_sources}-\eqref{eq:rwa_sources} can be recovered from two simple Pauli observables on the qubit.
Let us emphasise that the qubit dynamics~\eqref{eq:density_matrix_qubit} will differ markedly from a Lindbladian Markovian dephasing~\cite{Lindblad1976,Gorini1976,Breuer2002}, in which the density matrix $
\rho_{\rm I}(t_{\rm f},t_0)=\half(\mathbb{I}_2+\ee^{-(t_{\rm f}-t_0)/T_2}\sigma_x)$ decays exponentially towards the maximally-mixed state with a certain dephasing time $T_2$. In fact, as the results of the previous sections show, for a vanishing squeezing parameter and for resonant sources, the qubit coherence evolves instead with a Gaussian decay that is controlled by the strength of the harmonic sources $
\rho_{\rm I}(t_{\rm f},t_0)=\half(\mathbb{I}_2+\ee^{-(t_{\rm f}-t_0)^2/2|J_0|^{-2}}\!\sigma_x)$. When switching on the squeezing, this decay will change, depending on the Feynman diagrams discussed in the previous section. The goal of this section is to discuss how, using methods of statistical inference~\cite{alma9924122381802466}, one can estimate these Feynman diagrams from a finite number of projective qubit measurements. 

In principle, one can reconstruct any time evolution of the qubit by maximum-likelihood estimation, which requires preparing and reading the qubit in a set of informationally-complete states and positive operator-valued measures~\cite{doi:10.1080/09500349708231894,PhysRevLett.78.390,PhysRevA.63.020101,PhysRevA.63.054104,PhysRevA.68.012305}. However, this procedure
requires a considerable amount of experimental shots, and must be repeated for all  times of interest  $t\in[t_0,t_{\rm f}]$.
As we now discuss, one can follow a similar philosophy for the maximum-likelihood estimation of Feynman diagrams, which reduces considerably the required resources by exploiting the fact that we are not interested in estimating any generic time evolution, but only in the information related to the characteristic functional~\eqref{eq:sx_sy}.

\subsubsection{ Maximum-likelihood Ramsey estimator}

We consider a single initial state for the qubit-oscillator system $\rho_{\rm I}(t_{0})=\ket{+}\!\bra{+}\otimes\ket{\psi_b}\!\bra{\psi_b}$, let it evolve $\rho_{\rm I}(t_{0})\to\rho_{\rm I}(t_{\rm f})$ under the qubit-oscillator coupling in Eq.~\eqref{eq:sd_force_sources}, and then measure projectively in the $s\in\{x,y\}$ Pauli bases with two possible outcomes $m_s\in\{-1,+1\}$. According to Born's rule, the probabilities are given by the following expression
\beq
\label{eq:Born_rule}
 p_{s}(m_s)=\mathrm{Tr}\left\{ M_{s,m_s} \rho_{\rm I}(t_{\rm f})\right\}, \hspace{1ex}M_{s,m_s}=\half(\mathbb{I}_2-m_b\sigma_b).
\eeq
These measurements contain the desired information of the real and imaginary parts of the GSPA. In particular,
$p_{x}(\pm1) = \frac{1}{2}\!\left(1\pm \mathrm{Re}\ \chi[J^*,J]\right)$ and $p_{y}(\pm1) = \frac{1}{2}\!\left(1\pm \mathrm{Im}\ \chi[J^*,J]\right)$, where the characteristic functional must be evaluated at the specific source and sink functions~\eqref{eq:rwa_sources}.
According to our previous discussion, the GSPA in the resonant case can be written as a series of Feynman diagrams, each of which scales with a different power of the sources, sinks, and vertices, and thus has a different time dependence
\beq
\label{eq:expansion}
\chi[J^*,J]\!=\!\sum_{\boldsymbol{n}}c_{\boldsymbol{n}}{\rm Re}\big\{\lambda^{\!\!*\,n_1}_0\lambda_0^{n_2}J^{*\,n_3}_0J_0^{n_4}\big\}(t_{\rm f}-t_0)^{n_3+n_4}\ee^{-\frac{|J_0|^2\!}{2}(t_{\rm f}-t_0)^2}\!\!\!,
\eeq
where $\boldsymbol{n}=(n_1,n_2,n_3,n_4)\in\mathbb{Z}_{+}^4$, and $c_{\boldsymbol{0}}=1$. Rather than repeating quantum process tomography at each time of interest, we can instead parametrize the GSPA via the Feynman diagrams and the specific constants $c_{\boldsymbol{n}}\in\mathbb{C}$. By truncating the above expansion~\eqref{eq:expansion} at some particular order $\boldsymbol{n}\in\mathbb{S}$, leading to $|\mathbb{S}|=N+1$ terms in the above series, we thus turn the problem of calculating the Feynman diagrams with our hybrid quantum processor into a problem of statistical inference: the $N$-parameter estimation of $\{c_{\boldsymbol{n}}\}$, which can be arranged in a single vector $\boldsymbol{\theta}\in{\Theta}=\mathbb{C}^{N}$~\cite{alma9924122381802466}. By solving this problem, we will obtain a certain estimation $\boldsymbol{\hat{\theta}}$ that approximates the exact values with which we can infer the Feynman diagrams for any real time. From now on, we will use carets for all quantities that depend on the estimation parameters.

The parameter estimation is performed by a maximum-likelihood approach. After truncating Eq.~\eqref{eq:expansion}, $\chi[J^*,J]\to \bar{\chi}_{{\boldsymbol{\theta}}}[J^*,J]$, the probabilities~\eqref{eq:Born_rule} are parameterized as
\beq 
\begin{split}
\label{eq:approx_likelihood_function}
\bar{p}_x(\pm1|{\boldsymbol{\theta}},\boldsymbol{g}) & = \half\left(1\pm {\rm Re} \bar{\chi}_{\boldsymbol{\theta}}[J^*,J]\right),\\
\bar{p}_y(\pm1|{\boldsymbol{\theta}},\boldsymbol{g}) & = \half\left(1\pm {\rm Im} \bar{\chi}_{\boldsymbol{\theta}}[J^*,J]\right),\\
\end{split}
\eeq
where we have made explicit that these probabilities will depend on the parameters to be estimated, and also on a vector of microscopic couplings $\boldsymbol{g}=(J^*_0,J_0,\lambda_0^{\!\!*},\lambda_0)$.
We aim to estimate $N$ parameters  by finding the probability distribution within this family that maximizes the likelihood of reproducing a set of projective measurements in  two different Pauli bases. We advance that this will require distributing the measurement shots considering a certain range of microscopic couplings, and thus assume that those can be controlled in an experiment, $\{\boldsymbol{g}_k\}$.  All of the readout results are then fed into a likelihood function that we now define.

To study this estimation, we simulate measurements under realistic experimental conditions (see Sec.~\ref{sec:bosonic_states_sim}) for the set $\{\boldsymbol{g}_k\}$, assuming that each measurement value can be independently calibrated. In such an experiment, the Pauli-basis measurements correspond to simple Bernoulli trials for the mutually exclusive $\pm 1$ outcomes, considering only a finite amount of measurement shots $N_{s,\boldsymbol{g}}$ for each value of the microscopic couplings $\boldsymbol{g}$ and measurement basis $s$. We can thus define a relative frequency approximation to the underlying probability distribution by counting the number of repeated outcomes for each binary measurement 
\beq
\tilde{f}_s(+1|{\boldsymbol{\theta}_{\!\star}},\boldsymbol{g})=\frac{N_{s,\boldsymbol{g},+1}}{N_{s,\boldsymbol{g}}}, \hspace{2ex} \tilde{f}_s(-1|{\boldsymbol{\theta}_{\!\star}},\boldsymbol{g})=1-\frac{N_{s,\boldsymbol{g},+1}}{N_{s,\boldsymbol{g}}}.
\eeq
Here, we use tildes to emphasize that the relative frequencies are stochastic and related to a binomial distribution, and we note that these depend on the real values of the parameters we want to estimate $\boldsymbol{\theta}_{\!\star}$. In the theoretical analysis presented below, these relative frequencies can be obtained by randomly sampling the untruncated probability distribution, which either follows the exact expression for the $n=2$ squeezing~\eqref{eq:char_squeezed} or, otherwise, requires a quasi-exact numerical calculation of the characteristic function for $n\geq 3$ as in Fig.~\ref{fig:char_functions}.

Once we have performed the samplings for several couplings $\{\boldsymbol{g}_k\}$, we can define a log-likelihood cost function
\beq
  \label{eq:ML_cost_function}
  \mathsf{C}_{\mathrm{ML}}(\boldsymbol{\theta}) = -\sum_{k,s} N_{s,\boldsymbol{g}_k}\sum_{m_s=\pm 1} \tilde{f}_s(m_s|\boldsymbol{\theta}_{\!\star},\boldsymbol{g}_k) \log \bar{p}_s
(m_s|\boldsymbol{\theta},\boldsymbol{g}_k),
\eeq
the minimization of which yields our estimate 
 \beq
 \label{eq:ml_estimate}
\boldsymbol{\hat{\theta}}_{\rm F}=\texttt{argmin}_{\boldsymbol{\theta}}\big\{ \mathsf{C}_{\mathrm{ML}}(\boldsymbol{\theta}): \, \boldsymbol{\theta}\in\Theta=\mathbb{R}^{N
}\big\}.
 \eeq
Let us note that, when the number of shots per point $N_{s,\boldsymbol{g}_k}$ is sufficiently large, $\tilde{f}_s(m_s|\boldsymbol{\theta},{\boldsymbol{g}_k})$ behaves like a normal random variable and we can equivalently use a weighted least-squares cost function 
\beq
\mathsf{C}_{\rm LS}(\boldsymbol{\theta}) = \sum_{k,s} \left(\frac{\bar{p}_s
  (+1|\boldsymbol{\theta},\boldsymbol{g}_k) - \tilde{f}_s(+1|\boldsymbol{\theta}_{\!\star},\boldsymbol{g}_k)}{\tilde{\sigma}_{s,\boldsymbol{g}_k}}\right)^{\!\!\!2},
  \eeq
where $\tilde{\sigma}_{s,\boldsymbol{g}_k}$ is the standard deviation of the measurements taken at the point ${\boldsymbol{g}_k}$ and basis $s$. For a large number of shots, this will take the expected value
\beq
\label{eq:bin_variance}
\mathbb{E}( \tilde{\sigma}^2_{s,\boldsymbol{g}_k} )=\frac{{p}_s
  (+1|\boldsymbol{\theta}_{\!\star},\boldsymbol{g}_k) \big(1-{p}_s
(+1|\boldsymbol{\theta}_{\!\star},\boldsymbol{g}_k) \big)}{N_{s,\boldsymbol{g}_k}},
\eeq
where $p_b(+1|\boldsymbol{\theta}_{\!\star},\boldsymbol{g}_k)$ is the exact expression for the probabilities of Eq.~\eqref{eq:Born_rule}, in contrast to Eq.~\eqref{eq:approx_likelihood_function} which is just an approximation due to the truncation of the Feynman-diagram expansion. More explicitly, $p_x(\pm1|\boldsymbol{\theta}_{\!\star},\boldsymbol{g}_k) = \frac{1}{2}(1\pm {\rm Re}\chi[J^*, J]) = \mathbb{E}(\tilde{f}_x(\pm 1|\boldsymbol{\theta}_{\!\star},\boldsymbol{g}_k))$, and $p_y(\pm1|\boldsymbol{\theta}_{\!\star},\boldsymbol{g}_k) = \frac{1}{2}(1\pm {\rm Im}\chi[J^*, J]) = \mathbb{E}(\tilde{f}_y(\pm 1|\boldsymbol{\theta}_{\!\star},\boldsymbol{g}_k))$. We will reduce this cost function using the trust-region reflective algorithm implemented in SciPy \cite{scipy}.

\subsubsection{Analysis of systematic and statistical errors }

Let us now discuss the errors in this statistical inference. To be more specific, we will focus in this section on the case of the Feynman diagrams for the GSPA with standard $n=2$ squeezing, where the leading order Feynman diagrams yielded Eq.~\eqref{eq:Feynman_n_2}.
The generic series~\eqref{eq:expansion} can then be truncated at second order in the vertices and fourth-order in the sources, leading to the following 
\beq
\label{eq:char_squeezed_pert_learning}
\frac{\bar{\chi}_{{\boldsymbol{\theta}}}}{\chi_0}\!=\!\bigg(\!\!1+c_1 r\,{\rm Re}\{\xi^2\ee^{-\ii\theta}\}+c_2 {r^2}|\xi|^2 
\left.+c_3 r^2 \big({\rm Re}\{\xi^2\ee^{-\ii\theta}\}\big)^{\!\!2}\!\!\bigg)\!\right|_0^1\!\!.
\eeq
Here, we have absorbed the time dependence of Eq.~\eqref{eq:expansion} inside the $\xi$ parameter~\eqref{eq:xi_dispalcement}. The notation $|_0^1$ at the end of the expression denotes that the truncated $\bar{\chi}_{{\boldsymbol{\theta}}}$ will be set to 1 if the result is bigger than 1, and to 0 if the result is lower than 0, so that we have a well-defined likelihood function with well-defined probabilities even after the truncation. This incorporates the previous knowledge that for Gaussian states, the exact characteristic function must indeed be constrained $\chi(\xi^*,\xi)\in[0,1]$, such that we also ascertain that the parameters $c_1,c_2,c_3\in\mathbb{R}$. Likewise, given that the characteristic function is real, we only need to measure in the $b=x$ basis, optimizing the resources by allocating all the measurement shots to the measurement of $\langle\sigma_x\rangle$.
In the following, we will thus only consider this basis, and simplify the notation by avoiding the $b$ sub-index and the corresponding summation.

\begin{figure*}
\subfloat[\label{fig:c_vs_N}]{%
 \includegraphics[width=0.31\linewidth]{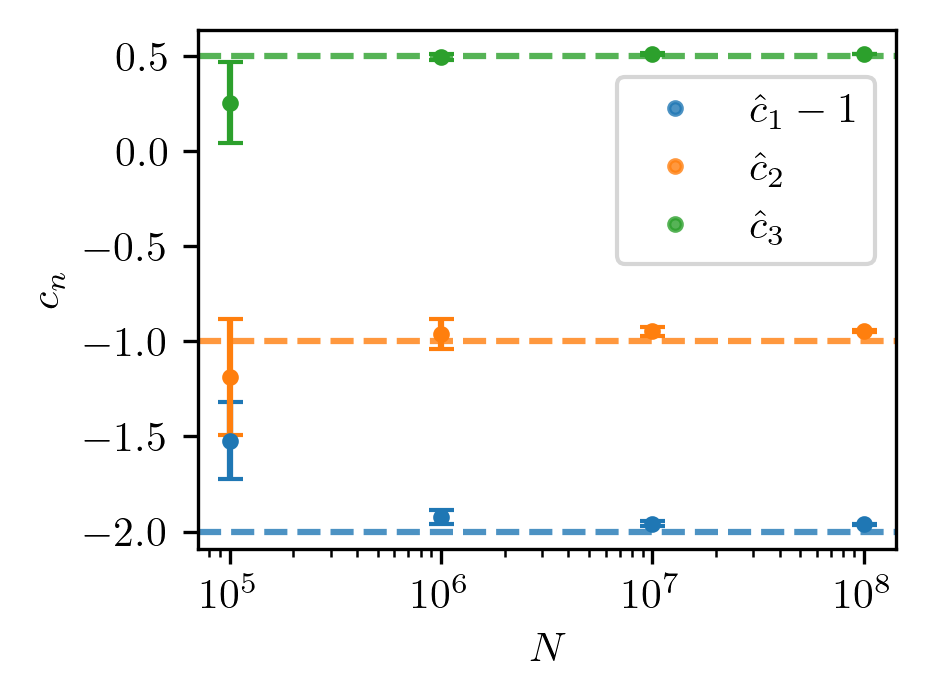}%
}\hfill
\subfloat[\label{fig:fd_time}]{%
 \includegraphics[width=0.31\linewidth]{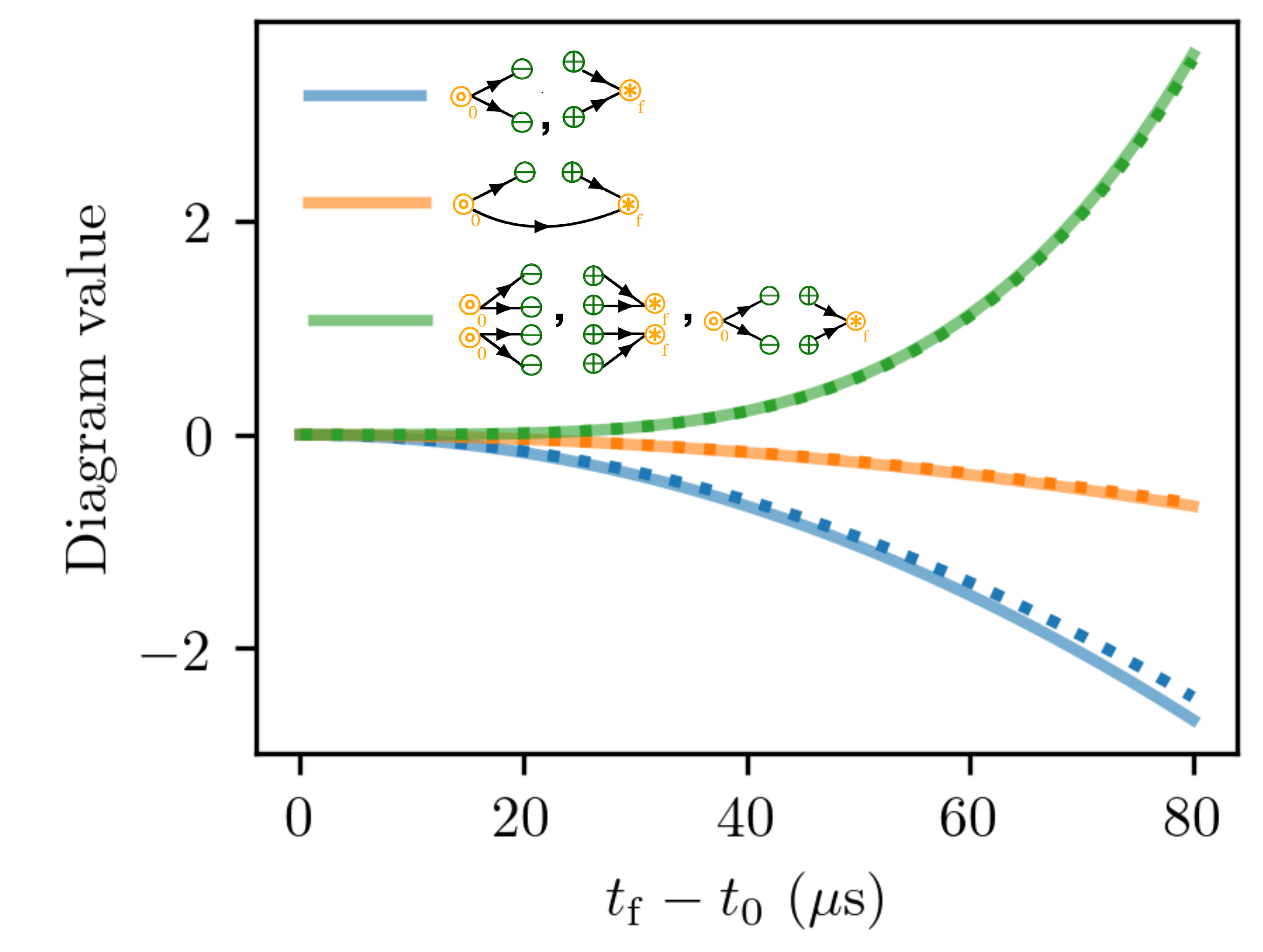}%
}\hfill
\subfloat[\label{fig:rmse}]{%
 \includegraphics[width=0.35\linewidth]{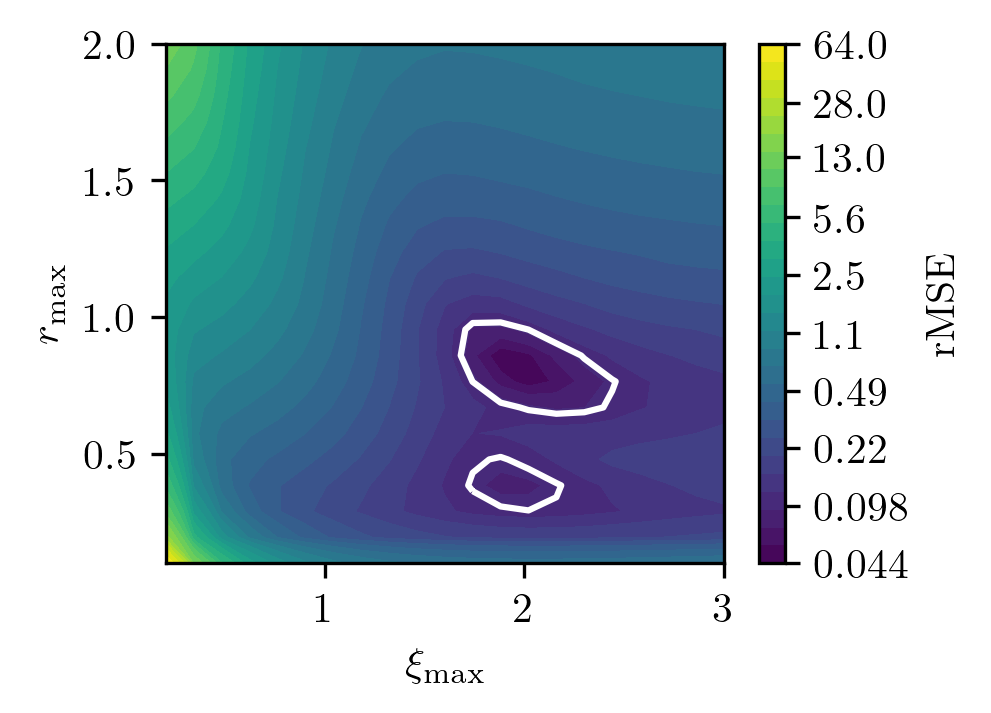}%
}
\caption{{\bf Feynman-diagram  estimates and real-time reconstruction for standard squeezing:} (a) Expected value of the estimates $\hat{\boldsymbol{\theta}}_{\rm F}$ for different total numbers of measurement shots $N$ in case of $n=2$ standard squeezing.
Error bars indicate two times the standard deviation of $\hat{\boldsymbol{\theta}}_{\rm F}$ and they are smaller than the marker when not visible. The horizontal dashed lines show the exact value of $c_n$. Note that for $N\gtrsim 10^6$ we are in the asymptotic regime where $\hat{\boldsymbol{\theta}}_{\rm F}$ behaves as Eq.~\eqref{eq:c_estimation_normal_dist}. Thus, in this regime the error bars scale as $O(1/\sqrt{N})$, while the expected value does not change and equals the exact value plus the systematic error. We note that $c_1$ is shifted one unit down for better visibility. The measurement points of the estimator $\{\boldsymbol{g}_k\}$ are located in a grid with $\xi\in[0.02, 2]$ and $r\in[0.02, 0.78]$ and spacing $\Delta \xi=0.02$, $\Delta r = 0.02$. (b)  Real-time dynamics of Feynman diagrams. Values taken by the first three Feynman diagrams for $n=2$ squeezing, i.e., the first three terms in the expansion \eqref{eq:char_squeezed_pert_learning}. Here, $\xi=e^{\ii \Delta \varphi} \Omega \eta (t_{\rm f}-t_0)$, with $\Delta \varphi = 0$, $r=0.25$ and $\theta=0$. The Rabi frequency and Lamb-Dicke parameter are set to typical values found in experimental settings, $\Omega \eta = 2\pi \times 4.7\, \rm kHz$. The dotted lines use the coefficients obtained from the expected value of the estimator presented in (a) for $N=10^6$ measurements, while the solid lines take the exact values. (c)  Optimal regime for the estimation. We represent the expected root mean squared error ${\rm rMSE}=\left(\frac{1}{3}\sum_n{\rm MSE}_n\right)^{\!\!1/2}$, where ${\rm MSE}_n= \mathbb{E} [(\hat{c}_n - c_{n\star})^2] \approx   \Delta \boldsymbol{\theta}^2_{\mathrm{sys},n} + [\Sigma_{\boldsymbol{\theta}}]_{nn} $, for Ramsey estimators with different $\xi_{\max}$ and $r_{\max}$ in case of $n=2$ squeezing. The total number of measurements is $N=1.6\times 10^6$ for all estimators, which is within the practical reach of experimental capabilities. The white contour line indicates the level with total error 0.1, and surrounds areas where the total error is low. Minimum is located at $\xi_{\max}\approx 2.0$, $r_{\max} \approx 0.78$.
}\label{fig:a}
\end{figure*}

The Feynman-diagram parameters $\boldsymbol{\theta}$ we aim at estimating take the exact values $\boldsymbol{\theta}_{\!\star}=(c_{1\star}, c_{2\star}, c_{3\star})=(-1, -1, 1/2)$ according to Eq.\ \eqref{eq:chi_squeezed} and the comparison with the exact solution discussed in Appendix~\ref{app:exact}. Therefore, this is an ideal situation to benchmark the statistical inference, and analyse in detail the different errors that can afflict it. Our estimate \eqref{eq:ml_estimate}, $\boldsymbol{\hat \theta}_{\rm F} = (\hat{c}_1, \hat{c}_2, \hat{c}_3)$, will be affected by two sources of error. First, the limited number of measurement shots per point $N_{\boldsymbol{g}_k}$  will cause $\tilde{f}(\pm1|\boldsymbol{\theta},\boldsymbol{g}_k)$ to behave as a normal random variable in the asymptotic limit of large $N_{\boldsymbol{g}_k}$, and will produce a statistical error in the final estimation of the  Feynman-diagram parameters $\hat{\boldsymbol{\theta}}_{\rm F}$. This is typically known as shot noise or quantum projection noise~\cite{PhysRevA.47.3554}.
Additionally, we need to take into account that we will be using a parameterized probability distribution that only considers a finite number of diagrams, which will introduce a systematic error in our estimation due to the truncation. 
According to the theory of maximum-likelihood estimators~\cite{alma9924122381802466}, in the asymptotic limit of a large total number of measurement shots $N=\sum_k N_{\boldsymbol{g}_k}$ our estimate $\hat{\boldsymbol{\theta}}_{\rm F}$ will be distributed as a normal random variable
\beq\label{eq:c_estimation_normal_dist}
\hat{\boldsymbol{\theta}}_{\rm F} \sim \mathcal{N}\!\left[\mu=\boldsymbol{\theta}_{\!\star}+\Delta \boldsymbol{\theta}_{\rm sys},\ \Sigma_{\boldsymbol{\theta}} = I^{-1}\right].
\eeq
Here, $\Delta\boldsymbol{\theta}_{\rm sys}$ is the systematic error, which can be obtained by applying a linear transformation to the difference between the ideal and truncated likelihood functions, $\Delta p_{k}|_{\mathrm{sys}} = p(+1 | \boldsymbol{\theta}_{\!\star}, \boldsymbol{g}_k) - \bar{p}(+1|{\boldsymbol{\theta}}_\star,\boldsymbol{g}_k) = ({\chi}[J^*,J]-\bar{\chi}_{\boldsymbol{\theta}_{\!\star}}[J^*,J])/2$. The covariance matrix is the inverse of $I = \sum_k N_{\boldsymbol{g}_k} I_{\boldsymbol{g}_k}$, the linear combination of the Fisher information matrices at points $\boldsymbol{g}_k$, where each matrix is weighted by the number of measurements taken at that point. This covariance gives rise to the statistical error (see Appendix~\ref{sec:error_analysis} for further details on the error analysis). Equation~\eqref{eq:c_estimation_normal_dist} shows that the estimation $\hat{\boldsymbol{\theta}}_{\rm F}$ will have the mean squared error ${\rm MSE}_n=\mathbb{E}[(\hat{c}_n - c_n)^2] = \Delta \boldsymbol{\theta}^2_{\mathrm{sys},n} + [\Sigma_{\boldsymbol{\theta}}]_{nn}$, i.e. the quadratic sum of the systematic error and the statistical error, or, equivalently, the bias and the variance. Note that the statistical error scales as $O(1/\sqrt{N})$, while the systematic error is independent of the number of measurement shots that is taken. 
This behaviour can be seen in Fig.~\ref{fig:c_vs_N}, where we present the results of our maximum-likelihood estimation as a function of the total number of shots. We can see how the estimates approach the exact values, and how the precision increases as one considers larger number of measurements. From the estimated coefficients, we can reconstruct the corresponding time dependence of the Feynman diagrams, which lead to the real-time dynamics shown in Fig. \ref{fig:fd_time}.

Building on these results, we now aim at minimizing the total error in the estimation by selecting a suitable set of measurement points $\boldsymbol{g}_k$, which we recall depend on the microscopic parameters of the displacement and squeezing operations. For simplicity, we will consider that $\theta=\varphi=0$, such that we only have a couple of microscopic couplings $\boldsymbol{g}=(\xi,r)\in\mathbb{R}^2$. Intuitively, to minimize the statistical error we should select points with high values of $\xi$ and $r$. If $\xi$ and $r$ are too small, the effect of higher-order Feynman diagrams on the GSPA will be rather small and it will be difficult to estimate their corresponding coefficients. Therefore, it is important to include these points in the estimator. On the other hand, from the perspective of the systematic error, the situation is exactly the opposite. Higher values of $\xi$ and $r$ take us into an area where $\Delta p_{k}|_{\mathrm{sys}}$ is higher, since the truncated expansion \eqref{eq:char_squeezed_pert_learning} needs small values of $\xi$ and $r$ to better approximate $\chi$. Thus, we need to find the explicit trade-off between these two trends.

\begin{figure*}
\subfloat[\label{fig:c_vs_N_tri_real}]{%
 \includegraphics[width=0.24\linewidth]{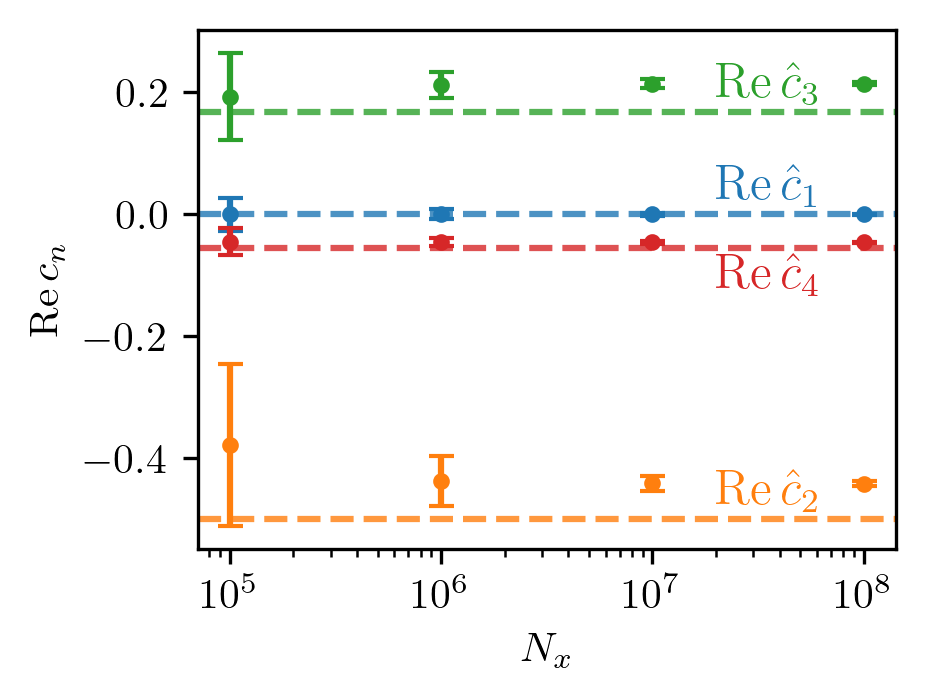}%
}\hfill
\subfloat[\label{fig:c_vs_N_tri_imag}]{%
 \includegraphics[width=0.23\linewidth]{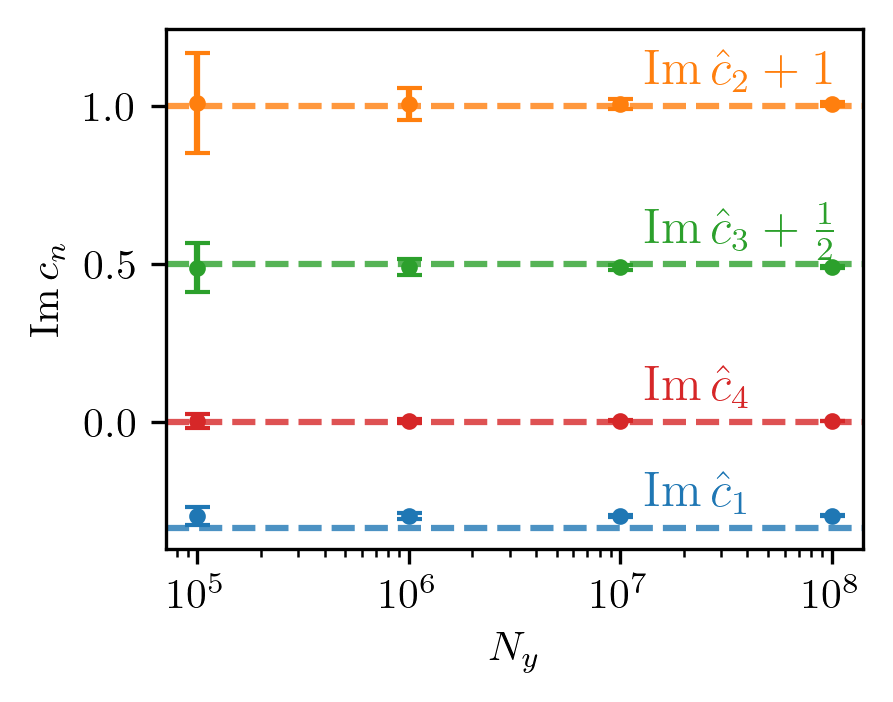}%
}
\hfill
\subfloat[\label{fig:fd_time_a}]{%
 \includegraphics[width=0.25\linewidth]{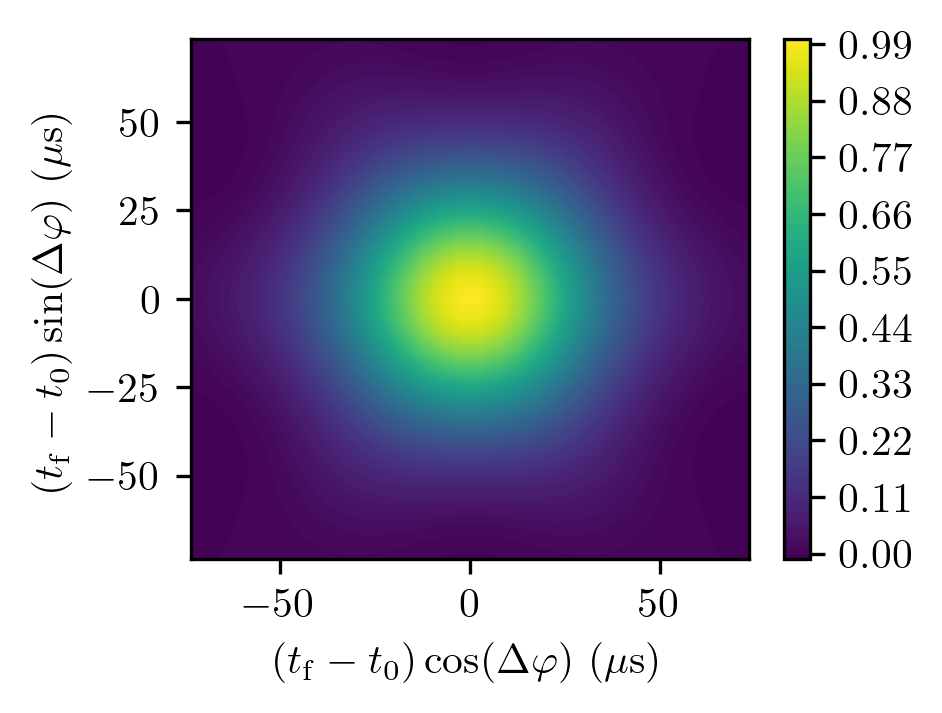}%
}
\hfill
\subfloat[\label{fig:fd_time_b}]{%
 \includegraphics[width=0.27\linewidth]{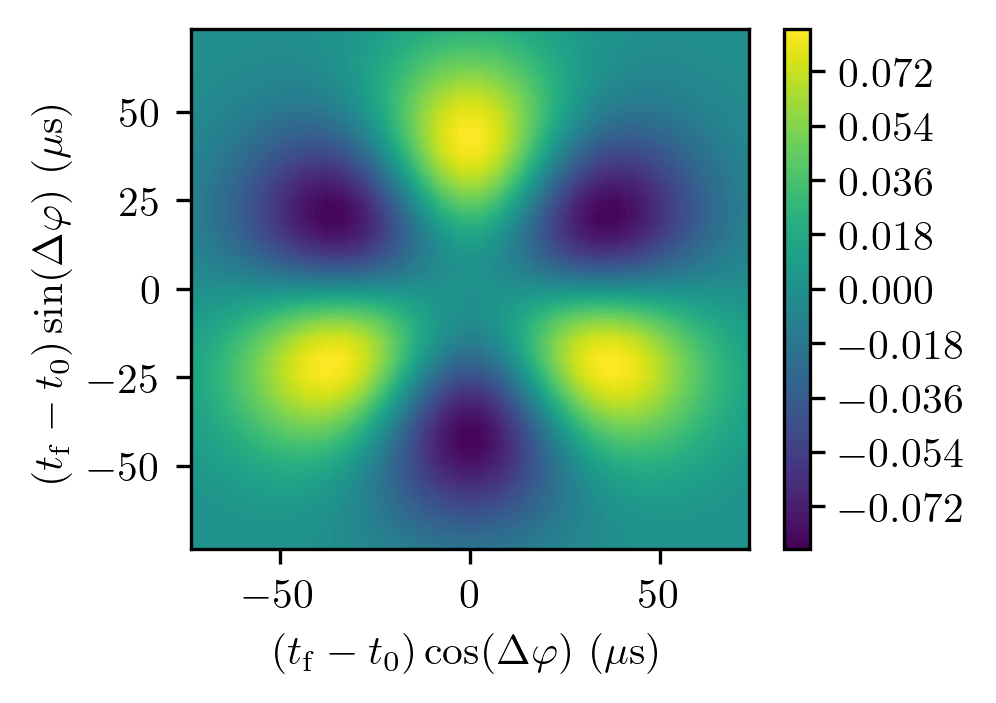}%
}

\caption{{\bf Feynman-diagram estimates and real-time reconstruction for generalised tri-squeezing:} (a,b) Real and imaginary part of expected value of the estimates $\hat{\boldsymbol{\theta}}_{\rm F}$ for different total numbers of measurement shots $N_s$ in case of $n=3$  squeezing. Here, $N_s$ refers to the number of shots taken at $s=x$ (a) or $s=y$ (b). $\mathrm{Im}\,\hat{c}_2$ and $\mathrm{Im}\,\hat{c}_3$ are shifted for better visibility. The measurement points $\{\boldsymbol{g}_k\}$ are located in a 3-dimensional grid with range $\mathrm{Re}\,\xi\in [0,1.7]$, $\mathrm{Im}\,\xi\in [0,1.7]$, $r\in [0,0.5]$, $\theta=0$ and a total of $10\times10\times3 = 300$ points. (c,d)  Real-time dynamics of characteristic functional: Time evolution of the characteristic function expansion of Eq.~\eqref{eq:char_trisqueezed_pert_learning} for different complex values of $\xi$, where $\xi=e^{\ii \Delta \varphi} \Omega \eta (t_{\rm f}-t_0)$, for $n=3$ tri-squeezing with magnitude $r= 0.25$. The Rabi frequency and Lamb-Dicke parameter are set to typical values found in experimental settings, $\Omega \eta = 2\pi \times 4.7\, \rm kHz$. Upper plot shows $\mathrm{Re}\,\bar{\chi}_{\boldsymbol{\theta}}$ and lower plot shows $\mathrm{Im}\,\bar{\chi}_{\boldsymbol{\theta}}$. The values of the coefficients $\boldsymbol{\theta}$ are the expected value of the estimator $\hat{\boldsymbol{\theta}}_{\rm F}$ for $N_x = N_y = 10^6$ shown in Figs.~\ref{fig:c_vs_N_tri_real} and \ref{fig:c_vs_N_tri_imag}.}
\label{fig:a_}
\end{figure*}

We can illustrate this by comparing different estimators, each of which selects a different set of points $\{\boldsymbol{g}_k\}$. The set of selected points will be the vertices of a square lattice in parameter space with spacings $\Delta r = 0.02$ and $\Delta \xi = 0.02$. This extends from $\Delta r$ to $r_{\max}$ in the $r$ direction, and from $\Delta \xi$ to $\xi_{\max}$ in the $\xi$ direction. We will compare estimators with different $\xi_{\max}$ and $r_{\max}$ values, while keeping the total number of shots $N=\sum_k N_{\boldsymbol{g}_k}$ fixed. As expected, the statistical error is lower for higher $\xi_{\max}$ and $r_{\max}$. When we take into account both the random and systematic errors, it is no longer optimal to make $\xi_{\max}$ and $r_{\max}$ as high as possible, as shown in Fig.\ \ref{fig:rmse}. The statistical error in this figure is obtained by computing the asymptotic covariance matrix $\Sigma_{\boldsymbol{\theta}}$. We estimate the systematic error by performing a least-squares fit to some simulated data where the number of shots $N$ is extremely high, so that the difference between the estimation and the exact values is very close to the systematic error, since the statistical error is vanishingly small. The optimal values minimizing the total error are $\xi_{\max}\approx 2.0$, $r_{\max} \approx 0.78$.

 Let us note that in a different regime, e.g. for non-Gaussian characteristic functions, it is still possible to estimate the systematic truncation ratio by comparing the Feynman-diagram calculations to the quasi-exact numerical calculation of the characteristic function. Since going to larger orders may lead to more complicated calculations, the complexity of which will increase when going beyond the single-mode case towards QFTs, one should find more general upper bound of the systematic error to estimate the best measurement points.

\subsubsection{Non-Gaussian Feynman diagrams}

We now focus on the $n=3$ squeezing case where the leading order Feynman diagrams yield Eq.~\eqref{eq:Feynman_n_3}. Truncating the expansion to third order in the vertices, we obtain
\begin{align}
\label{eq:char_trisqueezed_pert_learning}
  \frac{\bar{\chi}_{\boldsymbol{\theta}}}{\chi_0}\!=\! \big(1&+c_1 r\,{\rm Im}\{\xi^3\ee^{-\ii\theta}\}+c_2 r^2|\xi|^2+c_3 r^2|\xi|^4 \nonumber \\ 
  & \left. +c_4 r^2\big({\rm Im}\{\xi^3\ee^{-\ii\theta}\}\big)^{\!\!2} \big)\right|_{-1}^{1}.
\end{align}
Analogously to what we do in Eq.~\eqref{eq:char_squeezed_pert_learning} for the $n=2$ squeezing, we clip the output of the real and imaginary parts of this expression between -1 and 1, so that we have a well-defined likelihood function when learning the coefficients. These are now complex numbers, $c_1, c_2, c_3, c_4 \in \mathbb{C}$, in contrast to the $n=2$ squeezing, and the characteristic function also yields complex numbers. The exact values of the coefficients are $\boldsymbol{\theta}_\star = (c_{1\star}, c_{2\star}, c_{3\star}, c_{4\star}) = \left(-\frac{\ii}{3}, -\frac{1}{2}, \frac{1}{6}, -\frac{1}{18}\right)$. We can measure the $\sigma_x$ operator to obtain information about the real part of the characteristic function, while $\sigma_y$ will give us information about the imaginary part, as shown in Eq.~\eqref{eq:approx_likelihood_function}. These measurements are then input into the maximum-likelihood cost function \eqref{eq:ML_cost_function}, and by minimizing this function, we can determine the coefficients. Since $\sigma_x$ measurements only provide information about the real part of the characteristic function and, consequently, the real part of the coefficients, and $\sigma_y$ measurements do the same for the imaginary part, the cost function can be separated into two independent parts. Each part is associated with either the real or imaginary parts of the coefficients. Thus, we can minimize each part of the cost function independently to estimate the real and imaginary parts of the coefficients. The errors in this estimation have exactly the same characteristics as discussed in previous section and Eq.~\eqref{eq:c_estimation_normal_dist}, with a random error due to the limited number of shots and a systematic error due to the truncation of the characteristic function. This is illustrated in Figs.~\ref{fig:c_vs_N_tri_real} and \ref{fig:c_vs_N_tri_imag}. $\mathrm{Re}\,c_2$ and $\mathrm{Re}\,c_3$ have a high systematic error in comparison to the rest of the coefficients. We can reduce this systematic error by selecting another set of measurement points $\{\boldsymbol{g}_k\}$, but this will probably come at the cost of increasing the systematic error in other real coefficients or increasing the random error, as we discussed before for the $n=2$ squeezing. In contrast, the imaginary coefficients are very well behaved. The real-time dynamics of characteristic functional are shown in Figs.~\ref{fig:fd_time_a} and~\ref{fig:fd_time_b}.

\subsubsection{Zero thermal noise extrapolation}\label{sec:zero_thermal_noise_extrapolation}
In addition to the error caused by the truncation of the GSPA, there can be other sources of systematic error in an experiment. A particularly-relevant example is the error caused by a non-zero temperature in the initial state, which is a manifestation of the practical limits one may encounter when performing laser cooling on trapped ions~\cite{Wineland1998experimental,3981}. Due to this non-zero temperature,
we will be sampling from a binomial distribution with a slightly different probability,
which also leads to a systematic error and a bias in our estimates $\hat{\boldsymbol{\theta}}_{\rm F}$. Since the temperature can be increased by decreasing the time of laser cooling, one can thus think of obtaining several different estimates of the coefficients for different (small) values of the temperature, and then extrapolate to zero temperature to obtain an estimation of the $T=0$ Feynman diagrams even when no measurement is done at zero temperature. This procedure is analogous to the zero-noise extrapolation used in quantum error mitigation for quantum algorithms~\cite{PhysRevLett.119.180509,PRXQuantum.2.040326,vandenBerg2023,Cai_2023}. 

\begin{figure}
  \centering
  \includegraphics[width=.9\linewidth]{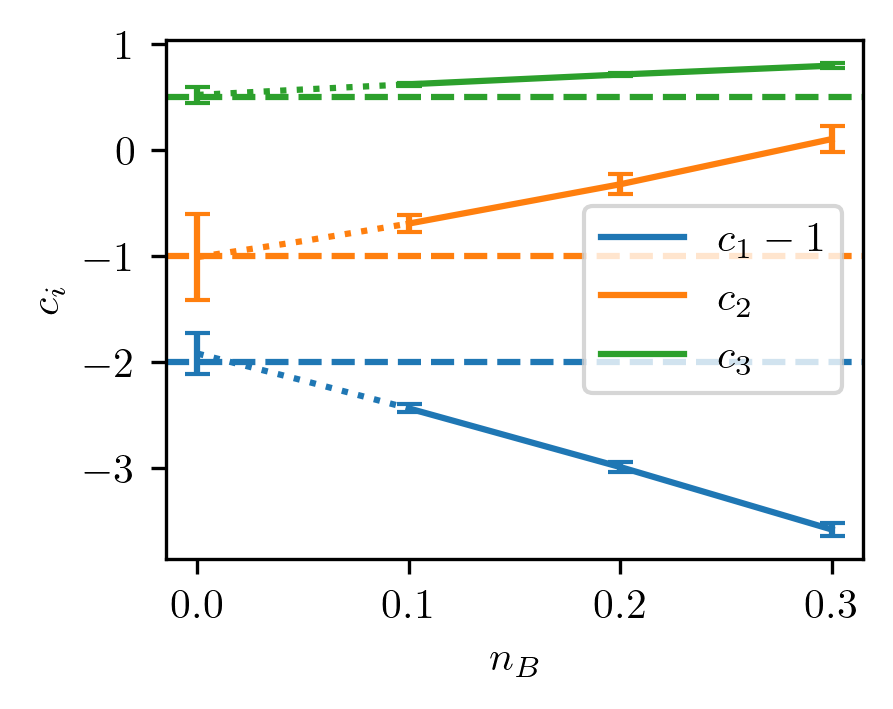}
  \caption{{\bf Zero-noise extrapolation of Feynman-diagram coefficients:} Expected value of the estimation $\hat{\boldsymbol{\theta}}_{\rm F}$ for different values of $n_B$ in the 2-squeezing case. The estimation at $n_B=0$ is obtained by fitting a second degree polynomial to the $n_B\in\{0.1, 0.2, 0.3\}$ estimates and extrapolating to 0 (dotted line). The horizontal dashed lines indicates the true value of the coefficients. Error bars indicate two times the standard deviation of $\hat{\boldsymbol{\theta}}_{\rm F}$, which in the asymptotic regime should equal $2[\Sigma_{\boldsymbol{\theta}}]_{ii}^{1/2}$ for $n_B\neq 0$ estimates. Plot produced simulating 1000 runs of the least squares algorithm for each $n_B\neq 0$ value to obtain the expected estimation and the standard deviation. Number of shots for each one of the three $n_B\neq 0$ estimations is $N=\frac{1.6}{3}\times 10^6$. The measurement points are taken from a square lattice with $\xi_{\max} = 2.0$, $r_{\max} = 0.78$. $c_1$ is shifted one unit down for better visibility.}
  \label{fig:zero_noise_extrapolation}
\end{figure}

We follow this strategy in Fig.~\ref{fig:zero_noise_extrapolation}, where we obtain three different estimates of the Feynman diagram parameters $\hat{\boldsymbol{\theta}}_{\rm F}=(\hat{c}_1,\hat{c}_2,\hat{c}_3)$ resulting from the minimization of Eq.~\eqref{eq:ml_estimate} for three different mean phonon numbers $n_B>0$ in the 2-squeezing case. Here, the relative frequencies are obtained from a finite sampling of the untruncated thermal probability distribution which, for the $n=2$ squeezing case, can be obtained from the exact expression in Eq.~\eqref{eq:char_squeezed_thermal}.  The zero-temperature estimation is obtained by fitting a polynomial to the $n_B>0$ estimates, and extrapolating to $n_B=0$. Note that this procedure allows us to eliminate the systematic error and bias due to non-zero temperature, as depicted in Fig.~\ref{fig:zero_noise_extrapolation}. On the other hand, as customary of quantum error mitigation techniques, this comes at the cost of amplifying the statistical errors due to the propagation of uncertainty from estimates with $n_B>0$ to the extrapolation at $n_B=0$. This is thus an inevitable consequence of the extrapolation, and would require increasing the number of shots to maintain the same level of the overall mean squared error. This can be seen in the larger error bars of the $n_B=0$ estimation, which show that the increase in accuracy (i.e. smaller bias) is associated to a decrease in precision (i.e. increased variance). Once more, we remark that we could reconstruct the dynamical estimates of the Feynman diagrams from these coefficients, leading to similar real-time plots of Fig.~\ref{fig:fd_time}.

 \subsection{\bf  Keldysh contours and finite temperatures }
\label{sec:finte_temp}
 As we have seen in the previous section, the accuracy and precision in the estimation of the Feynman diagrams can be affected by the non-zero temperature of the initial state.
 Actually, this need not be a limitation if one abandons the $T=0$ formalism for the characteristic functional, seeing  our $D=0+1$ dimensional problem from the perspective of real-time thermal QFTs~\cite{Landsman1987RealAI,LeBellac}. Indeed, novel effects in QFTs such as symmetry restoration~\cite{Dolan:1973qd,PhysRevD.9.3357} or deconfinement and the quark-gluon plasma~\cite{PhysRevLett.34.1353,Cabibbo:1975ig}, can appear when allowing for finite temperatures in fully-fledged QFTs. Here, we aim once more to a first step in the incorporation of thermal effects, which generalizes the above estimation of Feynman diagrams to non-zero temperatures.

 For non-zero temperatures, our starting point is no longer Eq.~\eqref{eq:charac_time} but, instead,  the thermal characteristic functional
 \begin{equation}
\label{eq:charac_time_thermal}  \chi[J^*,J]={\rm Tr}\Bigl\{S_n(\zeta)\rho_\beta S_n(-\zeta)U_J(t_{\rm f},t_0)\Bigr\},
\end{equation}
where $\rho_\beta=\ee^{-\beta\omega_ba^{\dagger}a}/{Z_\beta}$ is the Gibbs state of the oscillator for an inverse temperature $\beta$, and ${Z_\beta}$ the partition function.
In analogy to the zero temperature case, one can rewrite the thermal GSPA underlying this characteristic functional as a path-integral
for dynamical fields,
but the time must be extended to the complex plane.
 This follows from the observation that $\rho_\beta$ can be interpreted in terms of the time-evolution operation in imaginary time~\cite{Matsubara1955ANA}. Letting
 $t\mapsto{z}\in\mathcal{C}\subset\mathbb{C}$, 
we can recover Matsubara's imaginary-time formalism for equilibrium finite-$T$ problems~\cite{Landsman1987RealAI,LeBellac,Matsubara1955ANA,Dolan:1973qd} when the path $\mathcal{C}$ lies along the purely-imaginary axis $z(s)=t_0-\ii\beta s$ for $s\in[0,1]$. On the other hand, choosing a different contour $\mathcal{C}=\cup_a\mathcal{C}_a$ allows one to consider also out of equilibrium phenomena at finite temperatures~\cite{Schwinger:1960qe,Keldysh:1964ud,osti_5015361}.
This so-called Keldysh contour starts by first extending along the real-time axis $\mathcal{C}_1:z(s)=t_0+(t_{\rm f}-t_0)\beta s$, then proceeds into the imaginary direction $\mathcal{C}_3:z(s)=t_{\rm f}-\ii{\beta} s/2$, to then come back to the origin $\mathcal{C}_2:z(s)=t_{\rm f}-\ii\beta/2+(t_0-t_{\rm f})s$, and finally to the purely imaginary axis corresponding to the thermal state $\mathcal{C}_4:z(s)=t_{\rm 0}-\ii\beta(1+s)/2$~\cite{osti_5015361}.
We note that the time ordering along a contour proceeds in terms of the underlying monotonous parameter $s\in[0,1]$. Hence, on the backward branch $C_2$, it orders the operators anti-chronologically. There are a variety of choices to select the contour along which the dynamics takes place, which should yield equivalent outcomes.

It can be shown that, in the limit of sufficiently distant evolution times~\cite{osti_5015361}, the path integral can be expressed in terms of only the forward and backward branches $\mathcal{C}_1\cup\mathcal{C}_2$, leading to a closed-time contour, provided one allows for a doubling of the degrees of freedom. In our case, we need to define the sources  $J_1(t)=J(t),J_2(t)=J(t-\ii\beta/2)$ and sinks $J^*_1\!(t)=J^*\!(t),J^*_2\!(t)=J^*\!(t-\ii \beta/2)$ for both the forward and backward branches, and a matrix propagator that incorporates thermal effects and possible connections between these
branches. As with the choice of the contour, there are various options for defining the sources. These include taking linear combinations of $J_1(t)$ and $J_2(t)$, which might simplify certain calculations but must ultimately yield the same final results.

Including the previous delta-type switching of the interaction vertices, one can rewrite the thermal characteristic functional paralleling the zero-temperature case~\eqref{eq:char_functional} as 
\beq
\label{eq:char_functional_thermal}
\chi[{J}_1^{*},{J}_1]=\frac{\ee^{-\ii\!\bigintssss_{t\!}\!\tilde{V}_{\lambda(t)}\!\!\big(\!-\ii\delta_ {\boldsymbol{J}(t)},-\ii\delta_ {\boldsymbol{J}^{\dagger}\!(t)}\big)}\!\ee^{-\!\bigintssss_{t_1}\!\!\bigintssss_{t_2}\!\! \boldsymbol{J}^{\dagger}\!(t_1)\mathbb{G}_0(t_1-t_2)\boldsymbol{J}(t_2)}{\big|_{{J_2}={0}}}}
{\ee^{-\ii\!\bigintssss_{t\!}\!\tilde{V}_{\lambda(t)}\!\!\big(\!-\ii\delta_ {\boldsymbol{J}(t)},-\ii\delta_ {\boldsymbol{J}^{\dagger}\!(t)}\big)}\!\ee^{-\!\!\bigintssss_{t_1}\!\!\bigintssss_{t_2}\! \!\boldsymbol{J}^{\dagger}\!(t_1)\mathbb{G}_0(t_1-t_2)\boldsymbol{J}(t_2)}\big|_{\boldsymbol{J}=\boldsymbol{0}}}.
\eeq
Here, we have introduced a vectorial notation for the source and sink functions on both the forward an backward branches $\boldsymbol{J}=(J_1,J_2)^{\rm T}$, $\boldsymbol{J}^{\dagger}=(J^*_1,J^*_2)$, and the new interaction potential
\beq
\tilde{V}_{\lambda(t)}={V}_{\lambda(t)}\!\big(-\ii\delta_ {{J}_1(t)},-\ii\delta_ {{J}_1^{*}\!(t)}\big)-{V}_{\lambda(t)}\!\big(-\ii\delta_ {{J}_2(t)},-\ii\delta_ {{J}_2^{*}\!(t)}\big). 
\eeq

The problem has again reduced to the calculation of functional derivatives on the free thermal functional
\beq
\label{eq:free_charact_finite_T}
\chi_0[\boldsymbol{J}^\dagger,\boldsymbol{J}]={\rm exp}\left\{- \int_{t_1}\!\int_{t_2}\, \boldsymbol{J}^\dagger\!(t_1)\mathbb{G}_0(t_1-t_2)\boldsymbol{J}(t_2)\right\},
\eeq
where the thermal propagator has now become a matrix 
\beq
\label{eq:thermal_propagator}
\begin{split}
[\mathbb{G}_0(t)]_{11}\!&=(1+n_B)\,\ee^{-\ii\omega_bt}\theta(t)+n_B\,\ee^{-\ii\omega_bt}\theta(-t)\,\\
[\mathbb{G}_0(t)]_{22}\!&=n_B\,\ee^{-\ii\omega_bt}\theta(t)\,+(1+n_B)\,\ee^{-\ii\omega_bt}\theta(-t),\\
[\mathbb{G}_0(t)]_{12}\!&=\sqrt{n_B(1+n_B)}\ee^{-\ii\omega_bt}=[\mathbb{G}_0(t)]_{21},
\end{split}
\eeq
 depending on the Bose-Einstein distribution 
\beq
n_B=\frac{1}{\ee^{\beta\omega_b}-1}.
\eeq
Here we can see how, as a result of the thermal background, we can have both excitations and holes that propagate as time evolves along the forward branch $[\mathbb{G}_0(t)]_{11}$, each of which has a different bosonic enhancement factor. These will lead to additional contributions with respect to the Feynman-diagram expansion in Eq.~\eqref{eq:chi_squeezed},
and will be represented by two different oriented lines $\boldsymbol{\rightarrow}\hspace{-1.2ex}\boldsymbol{-}\,=\theta(t_i-t_j)(1+n_B)\ee^{-\ii\omega_{b}(t_i-t_j)}$ for particles, and
$\boldsymbol{-}\hspace{-1.2ex}\boldsymbol{\leftarrow}\,=\theta(t_j-t_i)\,n_B\,\ee^{-\ii\omega_{b}(t_i-t_j)}$ for holes. From this perspective, the source (sink) function
$ {\color{forestgreen}\boldsymbol{\oplus}}=J_1(t_j)$, (${\color{forestgreen}\boldsymbol{\ominus}}=J_1^*\!(t_i)$) at finite temperatures can also act by absorbing (creating) holes, which can also connect to new vertices acting on the boundaries of the path integral $ {\color{amber}\boldsymbol{\circledcirc}_{_{\rm f}}}=\lambda_-(t_{\rm f})$ (${\color{amber}\boldsymbol{\circledast}_{_{\rm 0}}}=\lambda_+^*\!(t_{\rm 0})$). These vertices, which correspond to the complex conjugate terms in Eq.~\eqref{eq:n_squeezing_v}, can now lead to new contributions as they no longer annihilate the bosonic state as occurred for the $T=0$ case. 

In addition, we have the propagator along the backwards branch of the contour $[\mathbb{G}_0(t)]_{22}$, which presents an inverted role of the particles and holes, and will be represented by directed dashed lines $ \tikz[baseline=1ex]{\draw[dashed,very thick, ->] (-0.2,0.25) -- (0.15,0.25)}\tikz[baseline=1ex]{\draw[dashed,very thick,-] (-0.1,0.25) -- (0.2,0.25)}=\theta(t_i-t_j)\,n_B\,\ee^{-\ii\omega_{b}(t_i-t_j)}$ for particles, and
$ \tikz[baseline=1ex]{\draw[dashed,very thick, -] (-0.2,0.25) -- (0.1,0.25)}\tikz[baseline=1ex]{\draw[dashed,very thick,<-] (-0.3,0.25) -- (0.1,0.25)}=\theta(t_i-t_j)(1+n_B)\ee^{-\ii\omega_{b}(t_i-t_j)}$ for holes. Although these lines can in principle be connected to the new sources and sinks $J_2,J_2^*$, the final characteristic functional~\eqref{eq:char_functional_thermal} cannot depend on those. Hence, these lines will only appear in the Feynman diagrams as internal lines connecting interaction vertices. We also note that the off-diagonal propagators $[\mathbb{G}_0(t)]_{12}=[\mathbb{G}_0(t)]_{21}$ connect the forward and backward branches, and can appear in both internal and external lines. These propagators do not make any difference regarding particles and holes, and will be represented by mixed directed lines that are independent on the time ordering $ \tikz[baseline=1ex]{\draw[dashed,very thick, ->] (-0.2,0.25) -- (0.15,0.25)}\tikz[baseline=1ex]{\draw[very thick,-] (-0.1,0.25) -- (0.2,0.25)}=\tikz[baseline=1ex]{\draw[very thick, ->] (-0.2,0.25) -- (0.15,0.25)}\tikz[baseline=1ex]{\draw[dashed,very thick,-] (-0.1,0.25) -- (0.2,0.25)}=\,\sqrt{n_B(1+n_B)}\,\ee^{-\ii\omega_{b}(t_i-t_j)}=\tikz[baseline=1ex]{\draw[dashed,very thick, -] (-0.2,0.25) -- (0.1,0.25)}\tikz[baseline=1ex]{\draw[very thick,<-] (-0.3,0.25) -- (0.1,0.25)}=\tikz[baseline=1ex]{\draw[very thick, -] (-0.2,0.25) -- (0.1,0.25)}\tikz[baseline=1ex]{\draw[dashed,very thick,<-] (-0.25,0.25) -- (0.1,0.25)}$ . 

\subsubsection{Thermal Feynman diagrams}

In the higher-dimensional QFTs, all of these additional propagators lead to further Feynman diagrams that can complicate the analysis and, often, one works instead with advanced, retarded, and statistical propagators ~\cite{Aarts:1997kp}. In our case, fortunately, one finds several simplifications when focusing on the thermal GSPA~\eqref{eq:char_functional_thermal}. Even if the vertices can in principle couple to the off-diagonal propagators, we find that their contribution cancels out for all diagrams in which the initial and final times are disconnected.
This occurs for instance in the lowest-order contributions to the $n=2$ squeezing, where all decorations with off-diagonal propagators cancel, and we find
\begin{widetext}
\beq
\label{eq:fd_1_thermal}
 \setlength{\unitlength}{1cm}
\thicklines
\begin{picture}(19,0)
\put(0.4,0.01){$\!\!\frac{\ii}{2}$}
\put(0.8,.05){\line(2,1){0.5}}
\put(0.8,0.05){\vector(2,1){0.4}}
\put(0.8,0.05){\vector(2,-1){0.4}}
\put(0.8,.05){\line(2,-1){0.5}}
\put(0.55,-0.05){${\color{amber}\boldsymbol{\circledcirc}_{\!\!_{_{0}}}}$}
\put(1.25,.27){${\color{forestgreen}\boldsymbol{\ominus}}$}
\put(1.25,-.35){${\color{forestgreen}\boldsymbol{\ominus}}$}
\put(1.75,0.01){$\!\!\!+$}
\put(1.90,0.01){$\frac{\ii}{2}$}
\put(2.20,0.27){${\color{forestgreen}\boldsymbol{\ominus}}$}
\put(2.20,-0.35){${\color{forestgreen}\boldsymbol{\ominus}}$}
\put(2.45,.26){\line(2,-1){0.4}}
\put(2.45,-.19){\line(2,1){0.4}}
\put(2.85,0.07){\vector(-2,1){0.3}}
\put(2.85,-0.0){\vector(-2,-1){0.3}}
\put(2.83,-0.05){${\color{amber}\boldsymbol{\circledcirc}_{\!\!_{_{\rm f}}}}$}
\put(3.45,0.01){$\!\!\!\!\!=-\frac{\ii}{2}\lambda_0\ee^{+2\ii\delta t_{\rm 0}}J_0^{*2}\,C^{\phantom{*}\!2}_{\delta}(t_{\rm f}-t_0)(1+n_B)^2+\frac{\ii}{2}\lambda_0\ee^{+2\ii\delta t_{\rm 0}}J_0^{*2}\,C^{\phantom{*
}\!2}_{\delta}(t_{\rm f}-t_0)n_B^2\approx -\frac{r}{2}\xi^{*2}\ee^{+\ii\theta}(1+2n_B)$,}
\end{picture}
 \eeq
 \beq
\nonumber
 \setlength{\unitlength}{1cm}
\thicklines
\begin{picture}(19,0)
\put(0.4,0.01){$\frac{\ii}{2}$}
\put(0.7,0.27){${\color{forestgreen}\boldsymbol{\oplus}}$}
\put(0.7,-0.35){${\color{forestgreen}\boldsymbol{\oplus}}$}
\put(0.95,.26){\line(2,-1){0.4}}
\put(0.95,-.19){\line(2,1){0.4}}
\put(0.95,0.25){\vector(2,-1){0.3}}
\put(0.95,-0.18){\vector(2,1){0.3}}
\put(1.31,-0.05){${\color{amber}\boldsymbol{\circledast}_{\!\!_{_{\rm f}}}}$}
\put(1.75,0.01){$\!\!\!+\frac{\ii}{2}$}
\put(2.75,0.27){${\color{forestgreen}\boldsymbol{\oplus}}$}
\put(2.75,-0.35){${\color{forestgreen}\boldsymbol{\oplus}}$}
\put(2.4,.1){\line(2,1){0.4}}
\put(2.4,-.0){\line(2,-1){0.4}}
\put(2.8,0.3){\vector(-2,-1){0.3}}
\put(2.8,-0.2){\vector(-2,1){0.3}}
\put(2.15,-0.05){${\color{amber}\boldsymbol{\circledast}_{\!\!_{_0}}}$}
\put(3.45,0.01){$\!\!\!\!\!=+\frac{\ii}{2}\lambda_0^*\ee^{-2\ii\delta t_{\rm 0}}J_0^{\phantom{*}\!2}\,C^{*2}_{\delta}(t_{\rm f}-t_0)(1+n_B)^2+\frac{\ii}{2}\lambda_0^*\ee^{-2\ii\delta t_{\rm 0}}J_0^{\phantom{*}\!2}\,C^{*2}_{\delta}(t_{\rm f}-t_0)n_B^2\approx -\frac{r}{2}\xi^{\phantom{*}\!2}\ee^{-\ii\theta}(1+2n_B)$.}
\end{picture}
 \eeq
\end{widetext}
Here, we note that the last approximations amount to taking the resonant limit $\Delta\to 0$, where the characteristic function parameter grows linearly with time
according to Eq.~\eqref{eq:xi_dispalcement}. For $n=2$, the squeezed thermal state falls under the family of Gaussian states and, as discussed in Appendix~\ref{app:exact}, there is an exact expression of the thermal GSPA~\eqref{eq:char_squeezed_thermal}. Performing a Taylor series of this expression, we see that there is a perfect agreement of these two sets of diagrams~\eqref{eq:fd_1_thermal} with the first non-trivial term in the expansion~\eqref{eq:char_squeezed_pert}. It is interesting to note that the overall bosonic enhancement is linear in the thermal boson number $(1+2n_B)$, which can be seen as a result of the interference of the propagators of particles and holes, each of which has a quadratic component scaling with $n_B^2$.

Let us now move to the next set of thermal Feynman diagrams, which include internal lines connecting the initial and final vertices, and no longer lead to the cancellation of the contributions of the off-diagonal and backwards propagators. In particular, we find the following contributions for the propagation of particles and holes
\begin{widetext}
\beq
 \setlength{\unitlength}{1cm}
\thicklines
\begin{picture}(19,0)
\put(0.6,.05){\line(2,1){0.5}}
\put(0.65,0.075){\vector(2,1){0.3}}
\put(1.35,-0.275){\vector(1,0){0.2}}
\qbezier(0.6,-.075)(1.35,-0.5)(2.15,-0.075)
\put(0.32,-0.1){${\color{amber}\boldsymbol{\circledcirc}_{\!\!_{_{0}}}}$}
\put(1.07,.25){${\color{forestgreen}\boldsymbol{\ominus}}$}
\put(1.4,.25){${\color{forestgreen}\boldsymbol{\oplus}}$}
\put(1.65,.3){\line(2,-1){0.5}}
\put(1.73,0.25){\vector(2,-1){0.3}}
\put(2.1,-0.1){${\color{amber}\boldsymbol{\circledast}_{\!\!_{_{\rm f}}}}$}
\put(2.5,-0,1){$-$}
\put(3.1,.05){\line(2,1){0.5}}
\put(3.2,0.1){\vector(2,1){0.3}}
\put(3.85,-0.275){\vector(1,0){0.2}}
\qbezier(3.1,-.075)(3.85,-0.5)(4.65,-0.075)
\put(2.82,-0.1){${\color{amber}\boldsymbol{\circledcirc}_{\!\!_{_{0}}}}$}
\put(3.57,.25){${\color{forestgreen}\boldsymbol{\ominus}}$}
\put(3.9,.25){${\color{forestgreen}\boldsymbol{\oplus}}$}
\put(4.15,.3){\line(2,-1){0.5}}
\put(4.23,0.25){\vector(2,-1){0.3}}
\put(4.6,-0.1){${\color{amber}\boldsymbol{\circledast}_{\!\!_{_{\rm f}}}}$}
\put(3.1,-.08){\tikz[baseline=1ex]{\draw[white,dashed, very thick, -] (0.1,0.3) -- (0.2,0.35)}}
\put(3.1,-.08){\tikz[baseline=1ex]{\draw[white,dashed, very thick, -] (3.1,0.15) .. controls (3.5,-0.03) .. (3.85,-0.05)}}
\put(5.0,-0,1){$-$}
\put(5.6,.05){\line(2,1){0.5}}
\put(5.65,0.075){\vector(2,1){0.3}}
\put(6.35,-0.275){\vector(1,0){0.2}}
\qbezier(5.6,-.075)(6.35,-0.5)(7.15,-0.075)
\put(5.32,-0.1){${\color{amber}\boldsymbol{\circledcirc}_{\!\!_{_{0}}}}$}
\put(6.07,.25){${\color{forestgreen}\boldsymbol{\ominus}}$}
\put(6.4,.25){${\color{forestgreen}\boldsymbol{\oplus}}$}
\put(6.65,.3){\line(2,-1){0.5}}
\put(6.65,0.30){\vector(2,-1){0.3}}
\put(7.1,-0.1){${\color{amber}\boldsymbol{\circledast}_{\!\!_{_{\rm f}}}}$}
\put(6.95,.0){\tikz[baseline=1ex]{\draw[white,dashed, very thick, -] (0.1,0.3) -- (0.2,0.25)}}
\put(6.6,-.27){\tikz[baseline=1ex]{\draw[white,dashed, very thick, -] (5.6,0.16) .. controls (5.9,0.24) .. (6.2,0.38)}}
\put(7.5,-0,1){$+$}
\put(8.1,.05){\line(2,1){0.5}}
\put(8.2,0.1){\vector(2,1){0.3}}
\put(8.85,-0.275){\vector(1,0){0.2}}
\qbezier(8.1,-.075)(8.85,-0.5)(9.65,-0.075)
\put(7.82,-0.1){${\color{amber}\boldsymbol{\circledcirc}_{\!\!_{_{0}}}}$}
\put(8.57,.25){${\color{forestgreen}\boldsymbol{\ominus}}$}
\put(8.9,.25){${\color{forestgreen}\boldsymbol{\oplus}}$}
\put(9.15,.3){\line(2,-1){0.5}}
\put(9.15,0.30){\vector(2,-1){0.3}}
\put(9.6,-0.1){${\color{amber}\boldsymbol{\circledast}_{\!\!_{_{\rm f}}}}$}
\put(9.45,.0){\tikz[baseline=1ex]{\draw[white,dashed, very thick, -] (0.1,0.3) -- (0.2,0.25)}}
\put(9.1,-.27){\tikz[baseline=1ex]{\draw[white,dashed, very thick, -] (8.1,0.16) .. controls (8.4,0.24) .. (8.7,0.38)}}
\put(8.1,-.08){\tikz[baseline=1ex]{\draw[white,dashed, very thick, -] (0.1,0.3) -- (0.2,0.35)}}
\put(8.1,-.08){\tikz[baseline=1ex]{\draw[white,dashed, very thick, -] (8.1,0.15) .. controls (8.5,-0.03) .. (8.85,-0.05)}}
\put(10.15,0.01){$\!\!\!\!\!=-|\lambda_0|^2|J_0|^{2}\,|C_{\delta}(t_{\rm f}-t_0)|^2(1+n_B)\approx\!-r^2|\xi|^2(1+n_B)$,}%
\end{picture}
 \eeq
 \beq
 \nonumber
 \setlength{\unitlength}{1cm}
\thicklines
\begin{picture}(19,0)
\put(0.6,.05){\line(2,1){0.5}}
\put(1.08,0.29){\vector(-2,-1){0.3}}
\put(1.38,-0.275){\vector(-1,0){0.2}}
\put(2.05,0.1){\vector(-2,1){0.3}}
\qbezier(0.62,-.075)(1.35,-0.5)(2.15,-0.075)
\put(0.32,-0.1){${\color{amber}\boldsymbol{\circledast}_{\!\!_{_{0}}}}$}
\put(1.07,.25){${\color{forestgreen}\boldsymbol{\oplus}}$}
\put(1.4,.25){${\color{forestgreen}\boldsymbol{\ominus}}$}
\put(1.65,.3){\line(2,-1){0.5}}
\put(2.1,-0.1){${\color{amber}\boldsymbol{\circledcirc}_{\!\!_{_{\rm f}}}}$}
\put(2.5,-0,1){$-$}
\put(3.1,.05){\line(2,1){0.5}}
\put(3.58,0.29){\vector(-2,-1){0.3}}
\put(4.02,-0.275){\vector(-1,0){0.2}}
\put(4.55,0.1){\vector(-2,1){0.3}}
\qbezier(3.1,-.075)(3.85,-0.5)(4.65,-0.075)
\put(2.82,-0.1){${\color{amber}\boldsymbol{\circledast}_{\!\!_{_{0}}}}$}
\put(3.57,.25){${\color{forestgreen}\boldsymbol{\oplus}}$}
\put(3.9,.25){${\color{forestgreen}\boldsymbol{\ominus}}$}
\put(4.15,.3){\line(2,-1){0.5}}
\put(4.6,-0.1){${\color{amber}\boldsymbol{\circledcirc}_{\!\!_{_{\rm f}}}}$}
\put(3.1,-.08){\tikz[baseline=1ex]{\draw[white,dashed, very thick, -] (0.1,0.3) -- (0.2,0.35)}}
\put(3.1,-.08){\tikz[baseline=1ex]{\draw[white,dashed, very thick, -] (3.1,0.15) .. controls (3.5,-0.03) .. (3.85,-0.05)}}
\put(5.0,-0,1){$-$}
\put(5.6,.05){\line(2,1){0.5}}
\put(5.6,.05){\line(2,1){0.5}}
\put(6.08,0.29){\vector(-2,-1){0.3}}
\put(6.45,-0.285){\vector(-1,0){0.2}}
\put(7.05,0.1){\vector(-2,1){0.3}}
\qbezier(5.6,-.075)(6.35,-0.5)(7.15,-0.075)
\put(5.32,-0.1){${\color{amber}\boldsymbol{\circledast}_{\!\!_{_{0}}}}$}
\put(6.07,.25){${\color{forestgreen}\boldsymbol{\oplus}}$}
\put(6.4,.25){${\color{forestgreen}\boldsymbol{\ominus}}$}
\put(6.65,.3){\line(2,-1){0.5}}
\put(7.1,-0.1){${\color{amber}\boldsymbol{\circledcirc}_{\!\!_{_{\rm f}}}}$}
\put(6.95,.0){\tikz[baseline=1ex]{\draw[white,dashed, very thick, -] (0.1,0.3) -- (0.2,0.25)}}
\put(6.6,-.27){\tikz[baseline=1ex]{\draw[white,dashed, very thick, -] (5.6,0.16) .. controls (5.9,0.24) .. (6.2,0.38)}}
\put(7.5,-0,1){$+$}
\put(8.1,.05){\line(2,1){0.5}}
\put(8.58,0.29){\vector(-2,-1){0.3}}
\put(9.05,-0.285){\vector(-1,0){0.2}}
\put(9.55,0.1){\vector(-2,1){0.3}}
\qbezier(8.1,-.075)(8.85,-0.5)(9.65,-0.075)
\put(7.82,-0.1){${\color{amber}\boldsymbol{\circledast}_{\!\!_{_{0}}}}$}
\put(8.57,.25){${\color{forestgreen}\boldsymbol{\oplus}}$}
\put(8.9,.25){${\color{forestgreen}\boldsymbol{\ominus}}$}
\put(9.15,.3){\line(2,-1){0.5}}
\put(9.6,-0.1){${\color{amber}\boldsymbol{\circledcirc}_{\!\!_{_{\rm f}}}}$}
\put(9.45,.0){\tikz[baseline=1ex]{\draw[white,dashed, very thick, -] (0.1,0.3) -- (0.2,0.25)}}
\put(9.1,-.27){\tikz[baseline=1ex]{\draw[white,dashed, very thick, -] (8.1,0.16) .. controls (8.4,0.24) .. (8.7,0.38)}}
\put(8.1,-.08){\tikz[baseline=1ex]{\draw[white,dashed, very thick, -] (0.1,0.3) -- (0.2,0.35)}}
\put(8.1,-.08){\tikz[baseline=1ex]{\draw[white,dashed, very thick, -] (8.1,0.15) .. controls (8.5,-0.03) .. (8.85,-0.05)}}
\put(10.15,0.01){$\!\!\!\!\!=-|\lambda_0|^2|J_0|^{2}\,|C_{\delta}(t_{\rm f}-t_0)|^2\,n_B\approx\!-r^2|\xi|^2\,n_B$,}%
\end{picture}
 \eeq
\end{widetext}
which, upon addition, lead to the same statistical enhancement $(1+2n_B)$, and agree with the corresponding term in the power expansion of the exact expression~\eqref{eq:char_squeezed_pert}.

The last term in this expansion~\eqref{eq:char_squeezed_pert} can be easily obtained by rewriting the corresponding disconnected diagrams as the square of the sum of the first contributions, paralleling our discussion of the $T=0$ case~\eqref{eq:fd_4}. As these diagrams, which contain 4 sources or sinks and two vertices, do not have any line connecting the initial and final vertices, the contributions of the off-diagonal propagators cancel once more. We thus obtain 
\begin{widetext}
\beq
\label{eq:fd_4_thermal}
 \setlength{\unitlength}{1cm}
\thicklines
\begin{picture}(19,0)
\put(0.15,0.01){$\frac{1}{2}\bigg($}
\put(0.55,0.01){$\!\frac{\ii}{2}$}
\put(0.95,.05){\line(2,1){0.5}}
\put(0.95,0.05){\vector(2,1){0.4}}
\put(0.95,0.05){\vector(2,-1){0.4}}
\put(0.95,.05){\line(2,-1){0.5}}
\put(0.72,-0.05){${\color{amber}\boldsymbol{\circledcirc}_{\!\!_{_{0}}}}$}
\put(1.4,.27){${\color{forestgreen}\boldsymbol{\ominus}}$}
\put(1.4,-.35){${\color{forestgreen}\boldsymbol{\ominus}}$}
\put(1.9,0.01){$\!\!\!+$}
\put(2.05,0.01){$\frac{\ii}{2}$}
\put(2.35,0.27){${\color{forestgreen}\boldsymbol{\ominus}}$}
\put(2.35,-0.35){${\color{forestgreen}\boldsymbol{\ominus}}$}
\put(2.6,.26){\line(2,-1){0.4}}
\put(2.6,-.19){\line(2,1){0.4}}
\put(3.0,0.07){\vector(-2,1){0.3}}
\put(3.0,-0.0){\vector(-2,-1){0.3}}
\put(3.,-0.05){${\color{amber}\boldsymbol{\circledcirc}_{\!\!_{_{\rm f}}}}$}
\put(3.4,0.01){$+\frac{\ii}{2}$}
\put(3.9,0.27){${\color{forestgreen}\boldsymbol{\oplus}}$}
\put(3.9,-0.35){${\color{forestgreen}\boldsymbol{\oplus}}$}
\put(4.1,.26){\line(2,-1){0.4}}
\put(4.1,-.19){\line(2,1){0.4}}
\put(4.1,0.25){\vector(2,-1){0.3}}
\put(4.1,-0.18){\vector(2,1){0.3}}
\put(4.52,-0.05){${\color{amber}\boldsymbol{\circledast}_{\!\!_{_{\rm f}}}}$}
\put(5.15,0.01){$\!\!\!+\frac{\ii}{2}$}
\put(6.15,0.27){${\color{forestgreen}\boldsymbol{\oplus}}$}
\put(6.15,-0.35){${\color{forestgreen}\boldsymbol{\oplus}}$}
\put(5.8,.1){\line(2,1){0.4}}
\put(5.8,-.0){\line(2,-1){0.4}}
\put(6.2,0.3){\vector(-2,-1){0.3}}
\put(6.2,-0.2){\vector(-2,1){0.3}}
\put(5.55,-0.05){${\color{amber}\boldsymbol{\circledast}_{\!\!_{_0}}}$}
\put(6.55,0.01){$\bigg)^2$}
\put(7.1,0.01){$\!\!\!\!\!={\rm Re}\big\{-\ii\frac{\lambda_0}{2}\ee^{2\ii\delta t_{\rm 0}}J_0^{*2}\,C^{2}_{\delta}(t_{\rm f}-t_0)\big\}^2(1+2n_B)^2\approx \frac{r^2}{2}\big({\rm Re}\{\xi^2\ee^{-\ii\theta}\}\big)^{\!\!2}(1+2n_B)^2$,}
\end{picture}
 \eeq
\end{widetext}
where the last approximation stems from setting the detuning $\delta\to 0$. Once again, we obtain a perfect agreement with the third contribution in Eq.~\eqref{eq:char_squeezed_pert}, which now scales with the square of the same statistical factor $(1+2n_B)^2$.

Altogether, the Schwinger-Keldysh formalism predicts  \beq
\begin{split}
\frac{\chi}{\chi^{n_B}_0}&\approx1-(1+2n_B)\,r\,{\rm Re}\{\xi^2\ee^{-\ii\theta}\}-(1+2n_B)\,{r^2}|\xi|^2\\
&+\frac{(1+2n_B)^2}{2}{r^2}\big({\rm Re}\{\xi^2\ee^{-\ii\theta}\}\big)^{\!\!2}\!,
\end{split}
\eeq
where $\chi^{n_B}_0={\rm exp}\{-(1+2n_B)\xi^*\xi/2\}$ is the thermal free Gaussian part. In this expression, we see how the different contributions have a very specific scaling with the temperature, which was not at all considered in our previous zero-noise extrapolation.

\subsubsection{Thermal Ramsey estimators}

\begin{figure}
  \centering
  \includegraphics[width=0.9\linewidth]{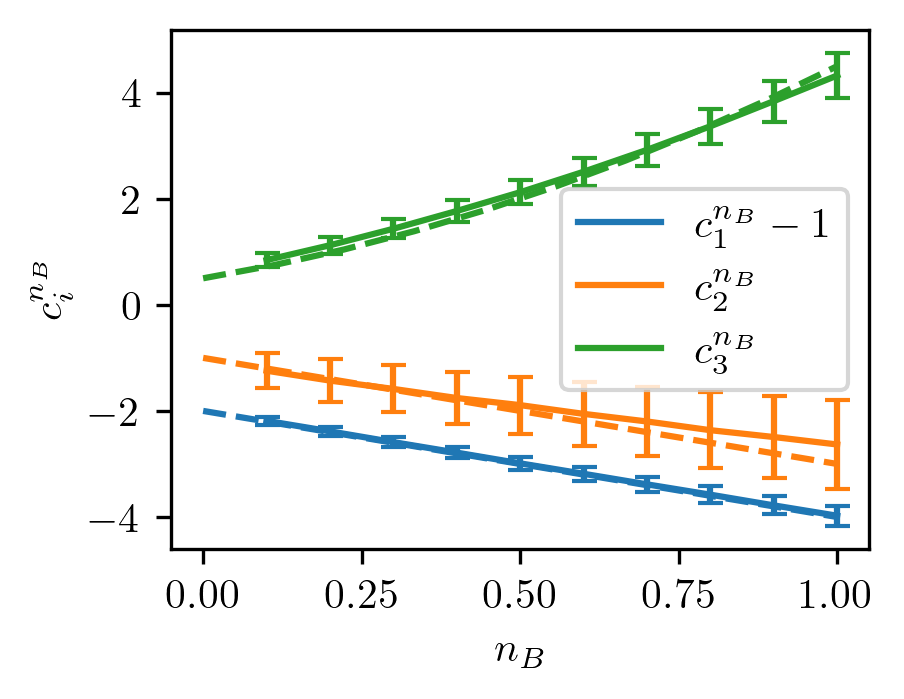}
  \caption{{\bf Schwinger-Keldysh Feynman diagram estimates:} Expected value of the estimation $\hat{\boldsymbol{\theta}}_{\rm F}(n_B)$ for different values of $n_B$. The dashed lines indicates the exact value of the coefficients $c_i^{n_B}$, which in this case are $n_B$ dependent. Error bars indicate two times the standard deviation of $\hat{\boldsymbol{\theta}}_{\rm F}$. Plot produced simulating 1000 runs of the least squares algorithm for each $n_B$ value to obtain the expected estimation and the standard deviation. Number of shots for each one of the estimations is $N=1.6\times 10^6$. The measurement points are taken from a square lattice with $\xi_{\max} = 1.46$, $r_{\max} = 0.3$. $c^{n_B}_1$ is shifted one unit down for better visibility.}
  \label{fig:non_zero_temperature_estimations}
\end{figure}

Similarly to what is done in Sec.~\ref{sec:statistical_inference} to estimate the value of the Feynman diagrams at zero temperature, we can now apply the same approach to estimate the coefficients at non-zero temperatures. These coefficients depend now on $n_B$ as we have seen in Eqs.~\eqref{eq:fd_1_thermal}-\eqref{eq:fd_4_thermal} for the $n=2$ squeezing case, which correspond to the first terms of the expansion \eqref{eq:char_squeezed_pert}. We need to experimentally measure the system at different points $\boldsymbol{g}=(\xi, r)$ for a specific value of $n_B$ to determine the coefficients at that temperature. We illustrate this here for the $n=2$ squeezing case, truncating the expansion and keeping only terms up to order $r^2$, 
\begin{equation}
\label{eq:char_squeezed_pert_learning_thermal}
\frac{\bar{\chi}_{{\boldsymbol{\theta}}}}{\chi_0^{n_B}}\!=\!\bigg(\!\!1+c^{n_B}_1 r\,{\rm Re}\{\xi^2\ee^{-\ii\theta}\}+c_2^{n_B} {r^2}|\xi|^2 
\left.+c_3^{n_B} r^2 \big({\rm Re}\{\xi^2\ee^{-\ii\theta}\}\big)^{\!\!2}\!\!\bigg)\!\right|_0^1\!\!.
\end{equation}
 In contrast to Eq.~\eqref{eq:char_squeezed_pert_learning} corresponding to the zero temperature case, now the coefficients we want to estimate depend on $n_B$ and take the exact values $c^{n_B}_{1\star}=c^{n_B}_{2\star}=-(1+2n_B)$ and $c^{n_B}_{3\star} = (1+2n_B)^2/2$, while the free part of the characteristic function has a $(1+2n_B)$ exponent. Once a $n_B$ value is chosen we can again estimate the coefficients by taking measurements at different points $\boldsymbol{g}_k$ and using the maximum likelihood approach. The estimation will be similarly affected by a random error as a consequence of the finite number of shots and by the systematic error due to the truncation of the expansion. This can be seen in Fig.~\ref{fig:non_zero_temperature_estimations}. It is important here to take into account again the systematic error. Since now coefficients equal the zero-temperature coefficients times a factor $(1+2n_B)^m$, with increasing exponent $m$ for higher order terms, higher $n_B$ values will increase the weight of higher order coefficients. This causes the truncated expansion \eqref{eq:char_squeezed_pert_learning_thermal} to be a worse approximation to the characteristic function. Ultimately, this increases the systematic error of our estimation. Thus, in order to estimate coefficients with $n_B\neq 0$ we reduce the $\xi_{\max}$ and $r_{\max}$ of the lattice of points $\boldsymbol{g}_k$ where we take measurements to minimize this effect. This comes at the cost of increasing the statistical error.

\subsubsection{Trapped-ion estimation and motional heating}\label{sec:motional_heating}
In the experimental setup we might have not only a non-zero temperature of the initial state but also heating during the experiment. This affects the characteristic function measured and is a new systematic error source if the heating effects are not taken into account by the model during the maximum-likelihood estimation. The effects of heating can be incorporated into the model, as shown in Ref.~\cite{PhysRevA.83.062120}. Assuming that the heating rate is small $\kappa t \ll 1$, we have that the main effect of heating modifies Eq.~\eqref{eq_resonant_source} as
\begin{equation}
    \xi'(t) = -\ii \int_0^t \!{\rm d}t'J(t')G_0(t') e^{-\kappa t'/2}\approx \frac{2\Omega\eta\ee^{\ii\Delta\varphi}}{\kappa} (e^{-\kappa t/2}-1),
\end{equation}
where we assume we are in the resonant regime $\Delta \approx 0$ and $\xi_0=0$. For small $\kappa t$ we obtain
\begin{equation}
    \xi'(t) \approx \Omega \eta \ee^{\ii\Delta\varphi}\left(t+\frac{\kappa}{4}t^2\right)= \xi(t) + c_{\rm h} \xi^2(t),
\end{equation}
with $c_{\rm h}$ a small heating parameter. In order to take into account this heating in the model, we can simply make the substitution $\xi \rightarrow \xi+c_{\rm h}\xi^2$ in Eq.~\eqref{eq:char_squeezed_pert_learning_thermal}. This new model has now four learning parameters: the ones corresponding to the three Feynman diagrams and the heating parameter. As shown in Tab.~\ref{tab:heating_abs_error}, this correction allows us to better learn the coefficients in the presence of heating.
\begin{table}[]
\begin{tabular}{lcccc}
\hline
\hline
                   & Heating & $c^{n_B}_1$ & $c^{n_B}_2$ & $c^{n_B}_3$ \\ 
                   \hline
\multirow{2}{*}{\hspace{2ex}Model of Eq.~\eqref{eq:char_squeezed_pert_learning_thermal}}\hspace{2ex} & 0       &\hspace{2ex} 0.23\hspace{2ex}  &\hspace{2ex} 0.13\hspace{2ex}  & \hspace{2ex}0.16\hspace{2ex}  \\
                   & 300 q/s & 0.22  & 0.53  & 0.08  \\  \hline
\multirow{2}{*}{\hspace{2ex}Model $\xi+c_{\rm h}\xi^2$\hspace{2ex}} & 0       & 0.30  & 0.04 & 0.15  \\
                   & \hspace{2ex}300 q/s\hspace{2ex} & 0.15  & 0.06  & 0.01  \\ \hline
                   \hline
\end{tabular}
\caption{Expected value of the absolute error in the estimation of the coefficients, $\mathbb{E}\left[|\hat{c}^{n_B}_i - c^{n_B}_{i\star} | \right]$, for $n=2$ squeezing and $n_B=0.1$ with and without heating. The actual value of the parameters is $c^{n_B}_{1\star} = c^{n_B}_{2\star} = -1.2$, $c^{n_B}_{3\star}=0.72$. For the heating model $c_{\rm h}$ takes the expected value 0.016 with no heating and 0.088 with heating. The measurement data includes here imperfections coming from the imperfect experimental preparation of the squeezing state. The measurement points are taken from a square lattice with $\xi_{\max} = 2.0$, $r_{\max} = 0.78$, $\Delta \xi=0.02$ and $\Delta r=0.02$, $N=1.6\times 10^6$.}
\label{tab:heating_abs_error}
\end{table}

\subsection{\bf Simulations and further experimental considerations}\label{sec:bosonic_states_sim}
In the analysis in Sec.~\ref{sec:statistical_inference}, we use simulated measurements of the characteristic function $\chi(\xi^*,\xi)$ at different values of $\xi$ for both squeezed and tri-squeezed states with various squeezing amplitudes $r$. We perform the numerical simulations of the system using {\it QuantumOptics.jl}~\cite{kramer2018quantumoptics} and model the harmonic oscillator with a Fock state cutoff of $100$. The harmonic oscillator is either initialised in Fock state $\ket{0}$ or a thermal state with mean phonon number $n_B$, with the latter case investigated in Sec.~\ref{sec:zero_thermal_noise_extrapolation}. In the experiment, different mean phonon number values can be achieved by cooling the ion close to the ground state and then allowing the motion to weakly heat, or by altering the cooling parameters~\cite{johnson2015sensing}.

The squeezed and trisqueezed states are simulated by employing a method proposed in Ref.~\cite{PhysRevA.104.032609} and recently demonstrated in experiments in Ref.~\cite{bazavan2024squeezing}. This involves considering the evolution of two simultaneously applied spin-dependent forces detuned from the motional mode by $\Delta$ and $-(n-1)\Delta$, respectively, where $n$ denotes squeezing order. The spin-conditioning of the two forces is set such that they are non-commuting. When adiabatically switching on and off this interaction (ramp duration long compared to $\Delta$), the dynamics can be  effectively described by 
\begin{equation}
{H}_\textrm{eff} =  \Omega_n {\sigma}_z ({a}^n + ({a}^\dagger)^n),
\end{equation} 
where $\Omega_n$ is an effective coupling strength that dirves the $n$-squeezing. After the generalised squeezed state is created, we consider the evolution of the  qubit-oscillator system under a spin-dependent force as described in Sec.~\ref{eq:squeezing_rwa}, and apply the corresponding qubit rotation to obtain either $\rm{Re}\chi$ or $\rm{Im}\chi$.

For synthesising the squeezed and tri-squeezed states, we use experimental parameters as described in Ref.~\cite{bazavan2024squeezing}, where a trapped $^{88}{\rm Sr}^+$ ion is used. The qubit is encoded in the electronic quadrupole transition $\ket{0}=\ket{5S_{1/2}, m_j=-1/2} \leftrightarrow \ket{1}=\ket{4D_{5/2}, m_j=-3/2}$ and the harmonic oscillator is represented by the axial in-phase motional mode with $\omega_z/2\pi= \SI{1.2}{\MHz}$. The magnitude of the spin-dependent force used to compute the characteristic function is given by $\Omega\eta = 2\pi \times \SI{4.7}{\kilo\hertz}$, determined by the Lamb-Dicke factor $\eta = 0.05$ and the qubit Rabi frequency $\Omega = 2\pi \times \SI{94}{\kHz}$.

Heating of the motional mode during the state generation and the measurement of the characteristic function, discussed in Sec.~\ref{sec:motional_heating}, is introduced in the simulation by solving the master equation with jump operators $\hat{L}_m = {\hat{a}, \hat{a}^\dagger}$ and rates of $\dot{\bar{n}} = 300\ \unit{quanta/s}$ ($\dot{\bar{n}} = 300(20)\ \unit{quanta/s}$ measured in the experiment). Other slow-acting decoherence channels can be introduced in a similar manner. One example is qubit decoherence; in the experimental system considered, the qubit coherence time is on the order of \SI{3}{\milli\second}. We have neglected its effect here because it primarily impacts the characteristic function measurement step, which lasts approximately \SI{500}{\micro\second}. We note, however, that these values are specific to the system considered here and are not fundamental limitations of trapped-ion platforms. In more optimised setups, motional heating rates below one quantum per second \cite{brownnutt2015ion} and motional coherence times exceeding \SI{100}{\milli\second} have been reported (see, e.g.,~\cite{lucas2007long}. Similarly, trapped-ion qubit coherence times ranging from seconds to several hours have been demonstrated \cite{pi2026beyond}. Therefore, in a more optimised system, these decoherence channels are not expected to be a limiting factor.

We can estimate the mean values of the experimental parameters, such as the squeezing amplitude \(r\) and the parameter \(\xi\), to better than \(1\%\) precision. However, it is worth noting that we will observe a larger variation in these parameters over time. To improve on our estimates of future experiments,%
one should also account for such fluctuations, and evaluate how they affect the estimation accuracy. In order to illustrate the relative variation in squeezing strength, consider that the optical power is stabilized to approximately \(1\%\). This stabilization results in a fluctuation in the Rabi frequency, \(\Delta \Omega / \Omega = 1 - \sqrt{1.01}\), which propagates to fluctuations in \(\Delta \xi / \xi = 1 - \sqrt{1.01}\) and in the squeezing parameter \(\Delta r / r = 1 - (1.01)^{n/2}\), where \(n\) denotes the order of the non-linear interaction. The timing, controlled to a resolution of \SI{1}{\nano\second}, does not introduce significant uncertainty.

\section{\bf A glimpse beyond characteristic functionals}
\label{sec:glimpse}

We have shown that an interferometric scheme to measure the characteristic function of a generalized squeezed state can be understood from the perspective of functional techniques and Feynman diagrams. We have applied maximum-likelihood techniques to show that one can accurately estimate Feynman diagrams at different orders of the source function and interaction vertices. Using parameter estimation techniques, we have identified the parameter regimes in which the combined systematic and statistical errors in the estimation can be minimized. Moreover, applying Keldysh contour techniques similar to those employed in thermal QFTs, we have show that the estimation can embrace thermal fluctuations, going beyond a zero-noise extrapolation and avoiding an increase of the estimation error. All of these techniques have been applied to a realistic modeling of recent trapped-ion experiments, considering various possible imperfections that include thermal populations and motional heating. In spite of these, we have shown that  Feynman diagrams can still be estimated with a considerable accuracy, uncovering an interesting direction for future trapped-ion experiments.

Let us close this manuscript by discussing possible future directions for the quantum computation of Feynman diagrams in hybrid quantum devices. First of all, in order to use the experimental trapped-ion data of~\cite{bazavan2024squeezing}, we would need to incorporate a heating mechanism in the real-time evolution, which would require generalizing our previous Schwinger-Keldysh formalism to account for finite heating terms typically accounted for in a Lindbladian. This would also likely permit introducing other forms of dephasing, including that of the probe qubit, within the Feynman diagrammatics.

Secondly, it would be interesting to consider the application of a pair of $n$-order squeezing operations at two different times $t_1,t_2\in(t_0,t_{\rm f})$. In this way, one can already see how a familiar concept in QFTs, that of Feynman loop integrals, would also appear in the context of the qubit-oscillator system 

\beq
\label{eq:loop_feynman}
\setlength{\unitlength}{1cm}
\thicklines
\begin{picture}(19,0)
\put(2.5,-.0){\line(1,0){0.73}}
\put(2.5,-.0){\vector(1,0){0.5}}
\put(5.4,-.0){\line(1,0){0.73}}
\put(5.7,-.0){\vector(1,0){0.25}}

\put(4.3,-0.3){\vector(1,0){0.1}}
\put(4.3,0.3){\vector(1,0){0.1}}
\put(4.3,-0.5){\vector(1,0){0.1}}
\put(4.3,0.5){\vector(1,0){0.1}}
\qbezier(3.45,-.0)(4.25,1.)(5.18,-0.0)
\qbezier(3.45,-.)(4.25,-1.0)(5.18,-0.0)
\qbezier(3.45,-.)(4.25,0.6)(5.18,-0.0)
\qbezier(3.45,-.)(4.25,-0.6)(5.18,-0.0)
\put(4.25,-0.15){$\vdots$}
\put(3.2,-0.1)
{${\color{amber}\boldsymbol{\circledcirc}}$}
\put(2.25,-.1){${\color{forestgreen}\boldsymbol{\oplus}}$}
\put(6.1,-0.1){${\color{forestgreen}\boldsymbol{\ominus}}.$}
\put(5.15,-0.1){${\color{amber}\boldsymbol{\circledast}}$}
\end{picture}
\eeq
Here, we are considering a loop with $(n-1)$ internal lines, which would arise from the pair of $n$-squeezing operations at intermediate times. One can already see how the two-point propagator gets a loop contribution that depends on time integrals, which would be the ones that lead to UV divergences in higher-dimensional QFTs and the need of renormalization. The difference with respect to the tadpoles that appear in Eq.~\eqref{eq:Z_interacting} is that, for the particular type of $n$-squeezing vertices in the potential~\eqref{eq:n_squeezing}, these loop insertions require an even number of vertices, and will thus scale with $|\lambda_0|^{2n}$. Moreover, in a fully-fledged QFT
the internal lines would depend on the external momentum, which reflects the fact that these contributions are more similar to the sunset diagram of Eq.~\eqref{eq:Z_interacting}. Indeed, due to the directionality of the lines, there are actually no tadpoles if this interaction were to be considered in a full QFT context. If we now allow for two extra applications of the squeezing operations at two other times, we could obtain a further insertion of this loop diagram in one of the internal lines. This process carries on as we consider further applications of the vertices, and would connect to the notion of the self-energy in QFTs. 

In order to properly connect to QFTs, one would need to allow for systems that are composed by more than a single oscillator, such that the propagation of the excitations also takes place along a spatial direction.  A device with $N$ bosonic modes and programmable couplings between them could be considered as a discretization of a bosonic QFT, where the lattice would be determined by the coupling connectivity or, equivalently, the adjacency matrix of the graph defined by the mode-mode couplings. When this graph corresponds to a certain Bravais lattice, the local potentials~\eqref{eq:n_squeezing} would then impose certain rules on the momentum carried by the excitations, connecting directly to an interacting bosonic lattice field theory. Measuring the renormalization of a propagator with Feynman loops like those in Eq.~\eqref{eq:loop_feynman} would require generalizing the qubit-based tomographic reconstruction discussed in the previous section to a pair of qubits that couple to a specific pair of bosonic modes, and each of them is subjected to a different state-dependent force, increasing the number of parameters in the point estimation. 
In this context, it appears to be more reasonable to use a power expansion only in terms of the sources, as the different contributions obtained from the estimation would correspond to the full propagators with the physical parameters of the QFT, not the bare ones.

The results presented in this work suggest that one could benchmark these more complex experiments with analogous perturbative expansions in terms of  Feynman diagrams, including all the subtleties of a lattice-regularized QFTs. Let us note that the experiment, however, would measure the full generating functional of a QFT, going in this way beyond perturbation theory. In fact, one could address non-perturbative phenomena that go beyond such Feynman diagram analysis. It is by exploring these regimes of real-time dynamics in vacuum or thermal QFTs where one expects to find instances of quantum advantages for problems that are relevant in physics. 
For Gauge theories or fermionic QFTs, however, this perspective would require suitable generalizations in which the external sources couple to gauge-invariant composite operators, Wilson loops or local fermion bilinears, rather than directly to the elementary fields, considerably raising the experimental complexity.

\textbf{Author contributions.--} A. B. proposed the original idea and made the analytical calculations with input from all authors.  S. V. carried out the numerical work and discussed the results with all authors. S. S., O. B., G. Ara. and G.Aa. provided experimental and theoretical input, respectively. All authors participated in the discussion of the results and in the writing and revision of the manuscript.

\acknowledgements

We would like to thank S.~Hands, R. Srinivas and, D. J. Webb for useful discussions. A.B. warmly thanks D.~M.~Lucas for all the support during the course of this project.
S.V. and A.B. acknowledge support from PID2021-127726NB- I00 (MCIU/AEI/FEDER, UE), from the Grant IFT Centro de Excelencia Severo Ochoa CEX2020-001007-S, funded by MCIN/AEI/10.13039/501100011033, from the CSIC Research Platform on Quantum Technologies PTI-001, and from the European Union’s Horizon Europe research and innovation programme under grant agreement No 101114305 (“MILLENION-SGA1” EU Project). G.Aa. acknowledges support from STFC Consolidated Grant ST/T000813/1. G.Ar. acknowledges support from Wolfson College, Oxford. S.S. and O.B. acknowledge support from the US Army Research Office (W911NF-20-1-0038) and the UK EPSRC Hub in Quantum Computing and Simulation (EP/T001062/1).

\appendix

\section{ Path integral and vacuum persistence amplitude}
\label{app:vpa}

In this appendix, we consider the standard derivation of the path-integral representation for the generating functional $Z[J^{*},J]$ in QFT~\cite{ryder_1996}, and how it can be adapted to the $D=0+1$ dimensional vacuum persistence amplitude~\eqref{eq:gen_func}. We recall that $Z[J^{*},J]$ can be interpreted as the partition function of the system upon a Wick rotation to imaginary time, underlying the chosen notation. This connection underlies the Monte Carlo sampling of QFT path integrals, which has actually allowed for the steady progress of the lattice approach to QFTs such as the standard model, culminating in e.g. the recent verification of hadron masses from first principles~\cite{durr2009abinitio}. On the other hand, this approach is limited by the sign problem\cite{PhysRevLett.94.170201} when considering problems that involve real-time dynamics, which motivates the search for alternatives such as quantum simulators~\cite{https://doi.org/10.1002/andp.201300104,Zohar,doi:10.1080/00107514.2016.1151199,Banuls2020,Carmen_Banuls_2020,doi:10.1098/rsta.2021.0064,Klco_2022, https://doi.org/10.48550/arxiv.2204.03381,Bauer:2023qgm,halimeh2023coldatom}. 
 The generating functional has a path integral representation, which has the following derivation in the holomorphic representation~\cite{Faddeev1976,Fradkin:2021zbi}. We discretize  time with a non-zero step $\Delta t$, and apply a Trotter decomposition of the time-evolution operator. After using an over-complete basis of normalized coherent states obtained by acting with the normal-ordered displacement operator $\ket{\alpha(t)}=:\!D(\alpha(t))\!\!:\!\!\ket{0_b}$ with $\alpha(t)\in\mathbb{C}$ for $\Delta t\to 0$, we find that the generating functional reads
\beq
\label{eq:path_int}
 Z[J^{*},J]=\int_{({\rm bc})}\!\!{\rm D}\alpha^*{\rm D}\alpha\,\ee^{\ii S[\alpha^*,\alpha]}.
\eeq
Here, ${(\rm bc)}$ stands for the boundary conditions over the paths describing all possible evolutions of the system which, for the vacuum persistence amplitude, are constrained to be periodic such that $\alpha(t_0)=\alpha(t_{\rm f})=0$. In the above path integral,  we have introduced the classical action as a functional of $\alpha(t),\alpha^*\!(t)$, namely
\beq
\label{eq:action}
S[\alpha^*,\alpha]=\!\int_t \Bigl(\alpha^*(t)(\ii\partial_{t}-\omega_b)\alpha(t)-V\!\big(t,\alpha^*\!(t),\alpha(t)\big)\!\Bigr),
\eeq
where $V(t,\alpha^*,\alpha)=V_\lambda(\alpha^*,\alpha)-{J}(t)\alpha^*-{J}^*(t)\alpha$, and we have used a short-hand notation for the time integrals $
\int_t=\int_{t_0}^{t_{\rm f}}{\rm d}t
$.
In the absence of interactions, the path integral is a Gaussian functional integral that can be readily evaluated
\beq
\label{eq:free_gen_functional}
{Z}_0[J^{*},J]=Z_0[0,0]\,{\rm exp}\left\{-\int_{t_1}\int_{t_2} J^{*}\!(t_1)G_0(t_1-t_2)J(t_2)\right\},
\eeq
where $Z_0[0,0]$ is a normalization constant that does not depend on the sources, and one can see that only the free propagator $G_0(t_1-t_2)$~\eqref{eq:free_prop} appears. One can now use 
 standard manipulations in variational calculus to find a differential equation for the full generating functional in the presence of arbitrary interactions. In fact, it  has the compact closed solution
\beq
\label{eq:gen_functional_series}
{Z}[J^{*},J]=\ee^{-\ii\bigintssss_{t}V_{\rm \lambda}\!\big(-\ii\delta_ {J(t)},-\ii\delta_ {J^*\!(t)}\big)}{Z}_0[J^{*},J].
\eeq

Here, the exponential of the time-integral over the interaction potential can be evaluated at increasing orders of the interaction strength $\lambda$ by performing a Taylor series, and we note that all the subtleties of the time-ordering no longer appear as $\delta_{J(t)}\delta_{J^*(t')}=\delta_{J^*\!(t')}\delta_{J(t)},\forall t,t'$.
This series expansion of ${Z}[J^{*},J]$ has a pictorical representation in terms of all possible Feynman diagrams in which a product of source functions $J(t_1),\cdots, J(t_n)$ act at any possible intermediate times $t_1,\cdots, t_n$, pulling excitations out of the vacuum that subsequently propagate  as dictated by the free Green's functions $G_0(t'_j-t_j)$. As a consequence of the vertex $\lambda$, at each new instant of time within the set $\{t'_j\}$, there can be further scattering events that affect the free propagation of the excitations via the so-called vacuum effects, each of which as a neat representation in terms of a Feynman diagram. Time evolution carries then to a further set of times, $\{t''_j\}$, where another vertex $\lambda$ accounts for additional scatterings. This procedure is repeated $m$ times, after which  the remaining excitations are absorbed by  sink functions, turning the resulting state back to the vacuum until the final evolution time $t_{\rm f}$ is reached.

A paradigmatic case is that of polynomial interactions 
\beq
\label{eq:pol_int}
V_\lambda(a^\dagger,a)\propto\frac{\lambda_0}{n!} (a+a^\dagger)^n,
\eeq
 with a real coupling $\lambda_0\in\mathbb{R}$, being the quartic case $n=4$ a low-dimensional version of the $\lambda\phi^4$ QFT~\cite{ryder_1996,2005ftrg.book_Amit}. In that context, it is customary to work with real functions $J(t_i)=J^*\!(t_i)$ that act as both sinks and sources, and can be both depicted with the same symbol $ \color{forestgreen}{\boldsymbol{\times}}$
 in the diagrams. Additionally, since the interactions only depend on one of the mode quadratures $\phi=(a+a^\dagger)/\sqrt{2\omega}$, it is simpler to work in the quadrature basis with the following Green's function 
 \beq
 \label{eq:quad_prop}
 G_0(\omega)\mapsto \tilde{G}_0(\omega)=\frac{\ii}{\omega^2-\omega_b^2+\ii\epsilon},
 \eeq
 which would connect to the propagator of a self-interacting real scalar field in the $D=d+1$ dimensions. Hence, $\tilde{G}_0(t_i-t_j)=\bra{0}\mathcal{T}\{\phi(t_i)\phi(t_j)\}\ket{0}=\boldsymbol{-\!\!-}$ is represented by an undirected line joining the two different times, as it contains the creation of an excitation, but also the absorption of an excitation. This is reflected by the presence of two poles at $\omega=\pm\omega_b$ as $\epsilon\to 0^+$ in the propagator~\eqref{eq:quad_prop}, contrasting with Eq.~\eqref{eq:free_prop}. 
 
In this case, we must modify the expressions for the generating functional, both for the full $
{Z}[J^{*},J]\mapsto {Z}[J]={\rm exp}\{-\ii\int_{t}V_{\rm \lambda}\!(-\ii\delta_ {J(t)})\}{Z}_0[J^{*},J]$,  and free one $
{Z}_0[J,J^*]\mapsto \tilde{Z}_0[J]=\tilde{Z}_0[0]\,{\rm exp}\big\{-\half \int_{t_1}\!\int_{t_2}\, J(t_1)\tilde{G}_0(t_1-t_2)J(t_2)\big\}$, which can be easily checked fulfill Wick's theorem~\cite{PhysRev.80.268} for the propagators. The scattering of the bosonic excitations under the quartic potential can be expressed in terms of a diagrammatic expansion with time-independent interaction vertices $\lambda_0={\color{amber}\boldsymbol{\bullet}}$ that can act at any time by creating/annihilating 4 bosonic excitations, and thus connects to 4 lines. In particular, one finds 
 \begin{widetext}
 \beq
  \label{eq:Z_interacting} 
 \frac{\tilde{Z}[J]}{\tilde{Z}[0]}=\bigg( 1\hspace{-8ex}
 \setlength{\unitlength}{1cm}
\thicklines
\begin{picture}(19,0)
\put(1.25,0.0){$+\frac{\ii}{4}$}
\put(1.9,.1){\line(1,0){0.8}}
\put(1.75,0.01){$\color{forestgreen}{\boldsymbol{\times}}$}
\put(2.3,0.3){\circle{0.4}}
\put(2.21,0.01){$\color{amber}\bullet$}
\put(2.84,0.){$+\frac{1}{8}$}
\put(2.55,0.01){$\color{forestgreen}{\boldsymbol{\times}}$}
\put(3.45,0.01){$\color{forestgreen}{\boldsymbol{\times}}$}
\put(3.6,0.1){\line(1,0){0.8}}
\put(4,0.3){\circle{0.4}}
\put(4,0.69){\circle{0.4}}
\put(3.91,0.01){$\color{amber}\bullet$}
\put(3.91,0.41){$\color{amber}{\bullet}$}
\put(4.25,0.01){$\color{forestgreen}{\boldsymbol{\times}}$}
\put(4.65,0.){$+\frac{1}{8}$}
\put(5.3,0.1){\line(1,0){1.1}}
\put(5.17,0.01){$\color{forestgreen}{\boldsymbol{\times}}$}
\put(5.6,0.30){\circle{0.4}}
\put(6.1,0.30){\circle{0.4}}
\put(5.51,0.01){$\color{amber}\bullet$}
\put(6.02,0.01){$\color{amber}\bullet$}
\put(6.25,0.01){$\color{forestgreen}{\boldsymbol{\times}}$}
\put(6.6,0.){$+\frac{1}{12}$}
\put(7.4,0.1){\line(1,0){1}}
\put(7.25,0.01){$\color{forestgreen}{\boldsymbol{\times}}$}
\put(7.9,0.1){\circle{0.55}}
\put(8.08,0.01){$\color{amber}\bullet$}
\put(7.54,0.01){$\color{amber}\bullet$}
  \put(8.25,0.01){$\color{forestgreen}{\boldsymbol{\times}}$}
\put(8.5,0.0){$-\frac{\ii}{4!}$}
\put(9.22,0.1){\line(1,0){0.8}}
\put(9.61,-0.3){\line(0,1){0.8}}
\put(9.52,0.01){$\color{amber}\bullet$}
\put(9.1,0.01){$\color{forestgreen}{\boldsymbol{\times}}$}
\put(9.86,0.01){$\color{forestgreen}{\boldsymbol{\times}}$}
\put(9.465,0.41){$\color{forestgreen}{\boldsymbol{\times}}$}
\put(9.465,-0.41){$\color{forestgreen}{\boldsymbol{\times}}$}
\put(10.2,0.0){$-\frac{1}{12}$}
\put(10.95,0.1){\line(1,0){1.05}}
\put(11.61,-0.3){\line(0,1){0.8}}
\put(11.3,0.29){\circle{0.4}}
\put(11.52,0.01){$\color{amber}\bullet$}
\put(11.2,0.01){$\color{amber}\bullet$}
\put(10.8,0.01){$\color{forestgreen}{\boldsymbol{\times}}$}
\put(11.85,0.01){$\color{forestgreen}{\boldsymbol{\times}}$}
\put(11.465,0.41){$\color{forestgreen}{\boldsymbol{\times}}$}
\put(11.465,-0.41){$\color{forestgreen}{\boldsymbol{\times}}$}
\put(12.15,0.0){$-\frac{1}{32}$}
\put(12.9,.25){\line(1,0){0.8}}
\put(13.3,0.45){\circle{0.4}}
\put(13.21,0.16){$\color{amber}{\bullet}$}
\put(13.55,0.16){$\color{forestgreen}{\boldsymbol{\times}}$}
\put(12.9,-.08){\line(1,0){0.8}}
\put(12.75,0.16){$\color{forestgreen}{\boldsymbol{\times}}$}
\put(12.75,-0.165){$\color{forestgreen}{\boldsymbol{\times}}$}
\put(13.3,-0.26){\circle{0.4}}
\put(13.21,-0.165){$\color{amber}{\bullet}$}
\put(13.55,-0.165){$\color{forestgreen}{\boldsymbol{\times}}$}
\put(14.05,0.0){$-\frac{1}{16}$}
\put(14.9,.35){\line(1,0){0.8}}
\put(14.9,-.24){\line(1,0){0.8}}
\put(14.75,0.26){$\color{forestgreen}{\boldsymbol{\times}}$}
\put(15.3,0.06){\circle{0.6}}
\put(15.21,0.26){$\color{amber}{\bullet}$}
\put(15.55,0.26){$\color{forestgreen}{\boldsymbol{\times}}$}
\put(14.75,-0.33){$\color{forestgreen}{\boldsymbol{\times}}$}
\put(15.21,-0.33){$\color{amber}{\bullet}$}
\put(15.55,-0.33){$\color{forestgreen}{\boldsymbol{\times}}$}
\put(15.9,0.01){$+\cdots\bigg) $$\mathlarger{\frac{\tilde{Z}_0[J]}{\tilde{Z}_0[0]}},$}
\end{picture}
\eeq
 \end{widetext}
 where one assumes integrals over all possible intermediate times at which the sources and the vertices act. Using the above conventions, namely $\color{forestgreen}{\boldsymbol{\times}}$ =   $J(t_i)$, $\color{amber}{\bullet}$ = $\lambda_0$ and $\boldsymbol{-\!\!-}=G_0(t_i-t_j)$, any of these Feynman diagrams can be expressed as an specific integral at different orders of the sources and vertices. The first three diagrams contain the so-called tadpole diagrams, whereas the fourth one is the so-called sunset diagram, which has a very different effect in the higher-dimensional QFT, where one must also integrate over the spatial position of all sources and vertices, and momentum can run inside the internal loops. In the present $D=0+1$ dimensional case, the points of the sources and vertices only indicate different times, which must be integrated over. The last four diagrams describe scattering events during the propagation of two excitations, and the dots would include higher-order processes both in terms of the vertices and the sources.

\section{Characteristic functional for resonant sources}
\label{app:exact}

In this Appendix, we provide the exact expression for the characteristic functional of the $n=2$ squeezed state~\eqref{eq:squeezed} for harmonic sources~\eqref{eq:rwa_sources} in the resonant regime $\Delta=0$. In this case, the real-time problem is highly simplified, as the sourced 
evolution operator during the qubit-oscillator coupling is $U_J(t,t_0)\approx{\rm exp}\{-\ii (t-t_0)(\Omega\eta\ii\ee^{\ii\Delta\varphi}a^\dagger+{\rm H.c.})\}$. Therefore,  according to Eq.~\eqref{eq:charact_func}, one recovers the characteristic function with a parameter $\xi$ that simply grows linearly with the total time according to Eq.~\eqref{eq:xi_dispalcement}. The evaluation of the characteristic function within the family of Gaussian states then becomes very simple~\cite{barnett_1998}. For instance, for the vacuum state $\ket{\psi_b}=\ket{0}$, one obtains a complex-valued Gaussian 
\beq
\label{eq:free_char_function}
\chi_0(\xi^*,\xi)=\ee^{-\half\xi^*\xi},
\eeq
which is symmetric in the phase space defined by both quadratures of the bosonic mode. Indeed, by evaluating the normalised free characteristic functional in Eq.~\eqref{eq:free_charact} for the specific source and sink functions in Eq.~\eqref{eq:rwa_sources}, we find
\beq
\log\chi_0[J^*,J]=-\ii\frac{|J_0|^2}{\Delta}(t_{\rm f}-t_0)\Bigl(1-\ee^{\ii\frac{\Delta}{2}(t_{\rm f}-t_0)}{\rm sinc}\Bigl(\frac{\Delta}{2}(t_{\rm f}-t_0)\Bigr)\!\!\Bigr),
\eeq
where we have introduced the detuning $\Delta$ from the oscillator frequency $\omega_b$. Letting $\Delta\to 0$, we get $\chi_0[J^*,J]\approx\chi_0(\xi^*,\xi)$ in light of Eq.~\eqref{eq:xi_dispalcement}, and thus a perfect agreement with the above Gaussian~\eqref{eq:free_char_function}.

Let us now consider the resonant characteristic function of the squeezed state, which can be easily evaluated using $S^{\dagger}_2\!({\zeta})a S_2^{\phantom{\dagger}}\!({\zeta})=a\cosh r -a^{\dagger}\sinh r\ee^{\ii\theta}$, leading to 
\beq
\label{eq:char_squeezed}
\chi(\xi^*\!,\xi)=\chi_0\Bigl(\xi^*\!{\rm ch}r+\xi{\rm sh}r\ee^{-\ii\theta},\xi{\rm ch}r+\xi^*\!{\rm sh}r\ee^{\ii\theta}\Bigr).
\eeq
A power expansion of this characteristic function that can be used as a benchmark of the diagrammatic expansion reads
\beq
\label{eq:char_squeezed_pert_exact}
\chi=\!\bigg(\!\!1-r\,{\rm Re}\{\xi^2\ee^{-\ii\theta}\}-{r^2}|\xi|^2+\frac{r^2}{2}\big({\rm Re}\{\xi^2\ee^{-\ii\theta}\}\big)^{\!\!2}\!+\dots\!\!\!\bigg)\ee^{-\frac{|\xi|^2}{2}}\!\!\!\!,
\eeq
where we have left the free part~\eqref{eq:free_char_function} factorised to parallel the functional approach. 
 As discussed in Sec.~\ref{eq:gaussian_integrals}, each of these terms agree with the Feynman diagrams in Eqs.~\eqref{eq:fd_1},~\eqref{eq:fd_3} and \eqref{eq:fd_4} at each specific order. 

We can carry this comparison to the finite-temperature regime, starting from the free characteristic functional~\eqref{eq:free_charact_finite_T} for a non-zero mean boson number $n_B$. Using the harmonic sink and sources in Eq.~\eqref{eq:rwa_sources}, and performing the corresponding integrals, we find in this case
\begin{widetext}
\beq
\log\chi_0[J_1^*,J_1]=-\ii\frac{|J_0|^2}{\delta}(t_{\rm f}-t_0)\Bigl(1-\ee^{\ii\frac{\delta}{2}(t_{\rm f}-t_0)}{\rm sinc}\Bigl(\frac{\delta}{2}(t_{\rm f}-t_0)\Bigr)\!\!\Bigr) (1+n_B)+\ii\frac{|J_0|^2}{\delta}(t_{\rm f}-t_0)\Bigl(1-\ee^{-\ii\frac{\delta}{2}(t_{\rm f}-t_0)}{\rm sinc}\Bigl(\frac{\delta}{2}(t_{\rm f}-t_0)\Bigr)\!\!\Bigr) n_B,
\eeq
where we have performed the corresponding time integrals. Letting $\delta\to 0$, we find that $\chi_0[J_1^*,J_1]=\chi_0(\xi^*,\xi)^{1+2n_B}={\rm exp}\{-\half\xi^*\xi(1+2n_B)\}$, where the $\xi$ is given by Eq.~\eqref{eq:xi_dispalcement}. This again shows a perfect agreement with the exact characteristic function of a thermal state~\cite{barnett_1998}, which is again a Gaussian in phase space, but now has a temperature-dependent variance.

Once again, we can go beyond the free case, and also compare to the exact expression of the characteristic function of a squeezed thermal state~\cite{PhysRevA.47.4474}, namely
\beq
\label{eq:char_squeezed_thermal}
\chi(\xi^*\!,\xi)=\left(\chi_0\Bigl(\xi^*\!{\rm ch}r+\ee^{-\ii\theta}\xi{\rm sh}r,\xi{\rm ch}r+\ee^{\ii\theta}\xi^*\!{\rm sh}r\Bigr)\right)^{\!1+2n_B}.
\eeq
 This admits the following power series in the source and vertex couplings
\beq
\label{eq:char_squeezed_pert}
\chi(\xi^*,\xi)=\!\bigg(1-r\,{\rm Re}\{\xi^2\ee^{-\ii\theta}\}(1+2n_B)-{r^2}|\xi|^2(1+2n_B)+\frac{r^2}{2}\big({\rm Re}\{\xi^2\ee^{-\ii\theta}\}\big)^{\!\!2}(1+2n_B)^2\!+\dots\bigg)\ee^{-(1+2n_B)\frac{|\xi|^2}{2}}\!\!.
\eeq
\end{widetext}
 As discussed in Sec.~\ref{sec:finte_temp}, each of these terms agree with the Feynman diagrams in Eqs.~\eqref{eq:fd_1_thermal}-\eqref{eq:fd_4_thermal} that contribute to that specific order. 
 
\section{Error analysis of maximum likelihood estimation}\label{sec:error_analysis}

As explained in the main text, the maximum likelihood estimation is affected by two sources of error. As a consequence of the limited number of shots performed at each measurement point $\boldsymbol{g}_k$ we have some statistical error. Additionally, we have the systematic error produced by the fact that our model is not exactly the characteristic distribution, but a truncated expansion of the characteristic distribution. In this appendix we show how this inaccuracies in the measurements and the model propagate into the estimate \eqref{eq:ml_estimate} and give rise to Eq.~\eqref{eq:c_estimation_normal_dist}.

Let us start from the maximum likelihood cost function \eqref{eq:ML_cost_function}. The minimum of the cost function satisfies
\begin{equation}
\label{eq:cost_function_minimum}
\partial_{\boldsymbol{\theta}} \mathsf{C}_{\rm ML}= -\sum_{m,k} N_{\boldsymbol{g}_k} \tilde{f}(m|\boldsymbol{\theta}_\star, \boldsymbol{g}_k) \frac{\partial_{\boldsymbol{\theta}} \bar{p}(m|\boldsymbol{\theta}, \boldsymbol{g}_k)}{\bar{p}(m|\boldsymbol{\theta}, \boldsymbol{g}_k)} = 0.
\end{equation}
Here, we are omitting the sum over the measurement basis $s$ for the sake of notation clarity. If we have more than one measurement basis, we should be summing over the basis whenever we have a sum over the measurement points $k$ in the following equations. Since $\tilde{f}(m|\boldsymbol{\theta}_\star, \boldsymbol{g}_k) = \bar{p}(m|\boldsymbol{\theta}, \boldsymbol{g}_k) + \Delta f(m | \boldsymbol{\theta}, \boldsymbol{g}_k)$ with $\Delta f(m | \boldsymbol{\theta}, \boldsymbol{g}_k)$ small, the minimum of the cost function is slightly displaced from the true minimum $\boldsymbol{\theta}_\star=(-1,-1,1/2 )$ to $\boldsymbol{\theta}_\star + \Delta\boldsymbol{\theta}$. Taylor expanding Eq.~\eqref{eq:cost_function_minimum} around $\boldsymbol{\theta}_\star$ to first order we have
\begin{widetext}
  \begin{equation}
  \partial_i \mathsf{C}_{\rm ML} \approx -\sum_{m, k} N_{\boldsymbol{g}_k} \left[ \bar{p}(m|\boldsymbol{\theta}, \boldsymbol{g}_k) + \Delta f(m | \boldsymbol{\theta}, {\boldsymbol{g}_k})\right] \left[ \left.\frac{\partial_i \bar{p}(m|\boldsymbol{\theta}, \boldsymbol{g}_k)}{\bar{p}(m|\boldsymbol{\theta}, \boldsymbol{g}_k)}\right|_{\boldsymbol{\theta}=\boldsymbol{\theta}_\star} \right. \left. + \sum_j \left.\partial_j\frac{\partial_i \bar{p}(m|\boldsymbol{\theta}, \boldsymbol{g}_k)}{ \bar{p}(m|\boldsymbol{\theta}, \boldsymbol{g}_k)}\right|_{\boldsymbol{\theta}=\boldsymbol{\theta}_\star}\Delta \theta_j \right] = 0,
\end{equation}
where $\partial_i$ denotes $\partial/\partial \theta_i$. Keeping only first-order terms in $\Delta \boldsymbol{\theta}$ and $\Delta f_{\boldsymbol{g}_k}$ and simplifying we obtain
\begin{equation}
\sum_{j, k} \Delta \theta_j N_{\boldsymbol{g}_k} [I_{\boldsymbol{g}_k} (\boldsymbol{\theta}_\star ) ]_{ij}= \sum_{m,k} N_{\boldsymbol{g}_k} \Delta f(m | \boldsymbol{\theta}_\star,\boldsymbol{g}_k) \left. \frac{\partial_i \bar{p}(m|\boldsymbol{\theta}, \boldsymbol{g}_k)}{ \bar{p}(m|\boldsymbol{\theta}, \boldsymbol{g}_k)}\right|_{\boldsymbol{\theta}=\boldsymbol{\theta}_\star},
\end{equation}
where we have defined the matrix 
\begin{equation}
  [I_{\boldsymbol{g}_k} (\boldsymbol{\theta}_\star ) ]_{ij} = \sum_m \bar{p}(m | \boldsymbol{\theta}_\star, \boldsymbol{g}_k) \partial_i\log \bar{p}(m|\boldsymbol{\theta}, \boldsymbol{g}_k) \partial_j\log \bar{p}(m|\boldsymbol{\theta}, \boldsymbol{g}_k) |_{\boldsymbol{\theta}=\boldsymbol{\theta}_\star},
\end{equation}
\end{widetext}
which is the Fisher information at measurement point ${\boldsymbol{g}_k}$. Taking into account that $\Delta f(-1 | \boldsymbol{\theta}_\star, \boldsymbol{g}_k) = - \Delta f(+1 | \boldsymbol{\theta}_\star, \boldsymbol{g}_k)$ and $\bar{p}(-1 | \boldsymbol{\theta}_\star, \boldsymbol{g}_k) = 1-\bar{p}(+1 | \boldsymbol{\theta}_\star, \boldsymbol{g}_k)$ and defining the matrices 
\begin{align}
  &I_{ij} = \sum_{k} N_{\boldsymbol{g}_k} [I_{\boldsymbol{g}_k}(\boldsymbol{\theta}_\star)]_{ij},\\ 
  &F_{ik} = N_{\boldsymbol{g}_k} \frac{\left.\partial_i \bar{p}(+1 | \boldsymbol{\theta}, \boldsymbol{g}_k) \right|_{\boldsymbol{\theta}=\boldsymbol{\theta}_\star}}{ \bar{p}(+1 | \boldsymbol{\theta}_\star, \boldsymbol{g}_k)(1-\bar{p}(+1 | \boldsymbol{\theta}_\star,\boldsymbol{g}_k))},
\end{align}
we arrive at the expression
\begin{equation} \label{eq:delta_c_delta_f_relation}
\Delta \theta_j = \sum_{i,k} I^{-1}_{ij} F_{ik} \Delta f(+1|\boldsymbol{\theta}_\star, \boldsymbol{g}_k),
\end{equation}
which relates fluctuations $\Delta f(+1|\boldsymbol{\theta}_\star, \boldsymbol{g}_k)$ to fluctuations in the estimation $\Delta \theta_j$. When $N_{\boldsymbol{g}_k}$ is sufficiently large, the binomial distribution $\tilde{f}(m|\boldsymbol{\theta}_\star, \boldsymbol{g}_k)$ can be approximated by the normal distribution 
\begin{equation}
  \mathcal{N}\left[\mu=p(+1|\boldsymbol{\theta}_\star, \boldsymbol{g}_k), \sigma^2 = \sigma^2_{\tilde{f},k} \right],
\end{equation}
with $\sigma^2_{\tilde{f},k}= p(+1|\boldsymbol{\theta}_\star, \boldsymbol{g}_k)p(-1|\boldsymbol{\theta}_\star, \boldsymbol{g}_k)/N_{\boldsymbol{g}_k}$ the variance of $\tilde{f}(m|\boldsymbol{\theta}_\star, \boldsymbol{g}_k)$. Since $\Delta f(m | \boldsymbol{\theta}, \boldsymbol{g}_k) = \bar{p}(m|\boldsymbol{\theta}, \boldsymbol{g}_k) - \tilde{f}(m|\boldsymbol{\theta}_\star, \boldsymbol{g}_k)$ we have that
\begin{align}
  \Delta f(+1 | \boldsymbol{\theta}_\star, \boldsymbol{g}_k) \sim \mathcal{N} \left[\mu=\Delta p_{k}|_{\mathrm{sys}}, \sigma^2=\sigma^2_{\tilde{f},k}\right],
\end{align}
with $\Delta p_{k}|_{\mathrm{sys}} = p(+1 | \boldsymbol{\theta}_{\!\star}, \boldsymbol{g}_k) - \bar{p}(+1|{\boldsymbol{\theta}}_\star,\boldsymbol{g}_k)$ the difference between the true observed distribution and the truncated distribution. Under transformation \eqref{eq:delta_c_delta_f_relation} $\Delta \boldsymbol{\theta}$ behaves as
\begin{equation}
\Delta \boldsymbol{\theta} \sim \mathcal{N}\left[\mu=\Delta \boldsymbol{\theta}_{\rm sys}=I^{-1}F \Delta \boldsymbol{p}|_{\rm sys},\ \Sigma_{\boldsymbol{\theta}} = I^{-1}\right],
\end{equation}
where we have used that $I^{-1} F \mathrm{diag}(\sigma^2_{\tilde{f},k})F^T (I^{-1})^T=I^{-1}$. Thus, the systematic error displaces $\hat{\boldsymbol{\theta}}_{\rm F}$ from the true value by an amount $\Delta \boldsymbol{\theta}_{\rm sys}=I^{-1}F \Delta \boldsymbol{p}|_{\rm sys}$ and the shot noise produces a normal statistical error with covariance matrix $\Sigma_{\boldsymbol{\theta}}=I^{-1}$ which scales as $1/N$, with $N=\sum_k N_{\boldsymbol{g}_k}$.

\bibliography{bibliography}
\bibliographystyle{apsrev4-2}

\end{document}